%% file: main.tex
\documentclass[sigconf, nonacm]{acmart}

\newcommand\vldbdoi{XX.XX/XXX.XX}
\newcommand\vldbpages{XXX-XXX}
\newcommand\vldbvolume{14}
\newcommand\vldbissue{1}
\newcommand\vldbyear{2020}
\newcommand\vldbauthors{\authors}
\newcommand\vldbtitle{\shorttitle} 
\newcommand\vldbavailabilityurl{https://github.com/neemashahbazi/Fair-Count-Min/}
\newcommand\vldbpagestyle{plain} 

\input{header}

\input{mycolors-header}

\usepackage{bbm, dsfont}

\begin{document}

\title{\prob: Frequency Estimation under Equal Group-wise Approximation Factor}

%
\settopmatter{authorsperrow=3}
\author{Nima Shahbazi}
\affiliation{%
  \institution{University of Illinois Chicago}
}
\email{nshahb3@uic.edu}

\author{Stavros Sintos}
\affiliation{%
  \institution{University of Illinois Chicago}
}
\email{stavros@uic.edu}

\author{Abolfazl Asudeh}
\affiliation{%
  \institution{University of Illinois Chicago}
}
\email{asudeh@uic.edu}

\begin{abstract}
Frequency estimation in streaming data often relies on sketches like Count-Min (CM) to provide approximate answers with sublinear space. However, CM sketches introduce additive errors that disproportionately impact low-frequency elements, creating fairness concerns across different groups of elements. We introduce Fair-Count-Min, a frequency estimation sketch that guarantees equal expected approximation factors across element groups, thus addressing the unfairness issue. We propose a column partitioning approach with group-aware semi-uniform hashing to eliminate collisions between elements from different groups. We provide theoretical guarantees for fairness, analyze the price of fairness, and validate our theoretical findings through extensive experiments on real-world and synthetic datasets. Our experimental results show that Fair-Count-Min achieves fairness with minimal additional error and maintains competitive efficiency compared to standard CM sketches.
\end{abstract}
\maketitle
\pagestyle{\vldbpagestyle}
\begingroup\small\noindent\raggedright\textbf{PVLDB Reference Format:}\\
\vldbauthors. \vldbtitle. PVLDB, \vldbvolume(\vldbissue): \vldbpages, \vldbyear.\\
\href{https://doi.org/\vldbdoi}{doi:\vldbdoi}
\endgroup
\begingroup
\renewcommand\thefootnote{}\footnote{\noindent
This work is licensed under the Creative Commons BY-NC-ND 4.0 International License. Visit \url{https://creativecommons.org/licenses/by-nc-nd/4.0/} to view a copy of this license. For any use beyond those covered by this license, obtain permission by emailing \href{mailto:info@vldb.org}{info@vldb.org}. Copyright is held by the owner/author(s). Publication rights licensed to the VLDB Endowment. \\
\raggedright Proceedings of the VLDB Endowment, Vol. \vldbvolume, No. \vldbissue\ %
ISSN 2150-8097. \\
\href{https://doi.org/\vldbdoi}{doi:\vldbdoi} \\
}\addtocounter{footnote}{-1}\endgroup

\ifdefempty{\vldbavailabilityurl}{}{
\vspace{.3cm}
\begingroup\small\noindent\raggedright\textbf{PVLDB Artifact Availability:}\\
The source code, data, and/or other artifacts have been made available at \url{\vldbavailabilityurl}.
\endgroup
}

\input{introduction}

\input{preliminary}
\input{solution}
\input{efficientImplementation}
\input{analysis}

\input{experiment}
\input{related_work}
\input{conclusion}
\balance
\newpage

\bibliographystyle{ACM-Reference-Format}
\bibliography{ref}
\appendix
\section*{APPENDIX}
\input{Appendix-PoF}
\input{Appendix-exp}

\end{document}

%% file: header.tex
\usepackage{flushend}
\usepackage[a-1b]{pdfx}   
\usepackage{balance}

\usepackage{color}
\usepackage{tcolorbox}
\usepackage[labelfont=bf]{caption}
\usepackage{subcaption}
\usepackage{hyperref}
\usepackage{makecell}
\usepackage{fixltx2e}
\usepackage{rotating,multirow}
\usepackage{etoolbox,xspace}
\usepackage{algorithmicx}
\usepackage{algorithm}
\usepackage{fontawesome}
\usepackage{algpseudocode}
\usepackage{wrapfig}
\usepackage{scalerel} 
\usepackage{booktabs}
\usepackage{multirow}
\usepackage{siunitx}
\usepackage{amsmath}
\usepackage{colortbl}   
\usepackage{xcolor}     
\usepackage{todonotes}

\usepackage{pgfplots}

\algtext*{EndWhile}
\algtext*{EndIf}
\algtext*{EndFunction}
\algtext*{EndFor}
\algrenewcommand\algorithmicrequire{\textbf{Input:}}
\algrenewcommand\algorithmicensure{\textbf{Output:}}

\usepackage{amsthm}

\usepackage{tabularx}
\usepackage{ragged2e}  
\usepackage{booktabs}  
\usepackage{textcomp}
\usepackage{url}
\usepackage{comment}
\usepackage{enumitem}
\usepackage[simplified]{pgf-umlcd}
\usepackage{arydshln}   
\usepackage{tikz}
\usetikzlibrary{shapes.geometric, arrows.meta, positioning, calc}

\newcolumntype{Y}{>{\RaggedRight\arraybackslash}X}

\newtheorem{definition}{Definition}
\newtheorem{theorem}{Theorem}

\newenvironment{proof}{\paragraph{\sc Proof.}}{\hfill$\square$}

\newtcolorbox{defbox}{
  colback=orange!10,
  colframe=orange!20,
  arc=2mm, 
  fonttitle=\bfseries,
  boxrule=0mm,
  boxsep=1mm,
  left=0mm,
  right=0mm,
  top=0mm,
  bottom=0mm
}

\newcounter{example}
\renewcommand{\theexample}{\arabic{example}}

\newtcolorbox{examplebox}{
  colback=blue!10,
  colframe=blue!20,
  arc=2mm, 
  fonttitle=\bfseries,
  boxrule=0mm,
  boxsep=1mm,
  left=0mm,
  right=0mm,
  top=0mm,
  bottom=0mm
}

\newenvironment{example}{
  \refstepcounter{example}
  \begin{examplebox}{}
    \noindent {\bf Example \theexample:} 
  }
  {\end{examplebox}} 

\newcommand{\squishlist}{
 \begin{list}{$\bullet$}
  { \setlength{\itemsep}{0pt}
     \setlength{\parsep}{1pt}
     \setlength{\topsep}{1pt}
     \setlength{\partopsep}{0pt}
     \setlength{\leftmargin}{1em}
     \setlength{\labelwidth}{1em}
     \setlength{\labelsep}{0.5em} } }

\newcommand{\squishend}{
  \end{list}
}

\definecolor{low}{rgb}{1, 0.8, 0.8}   
\definecolor{mid}{rgb}{1, 1, 0.8}     
\definecolor{high}{rgb}{0.8, 1, 0.8}  
\definecolor{DarkGreen}{rgb}{0.0, 0.8, 0.0}

\newcommand{\downarrowgreen}{\textcolor{DarkGreen}{$\downarrow$}}
\newcommand{\uparrowred}{\textcolor{red}{$\uparrow$}}

\definecolor{americanrose}{rgb}{1.0, 0.01, 0.24}
\definecolor{airforceblue}{rgb}{0.36, 0.54, 0.66}
\definecolor{ao(english)}{rgb}{0.0, 0.5, 0.0}
\definecolor{ao}{rgb}{0.0, 0.0, 1.0}

\usepackage{xspace}
\newcommand{\nima}[1]{\textcolor{orange}{Nima: #1}}

\newcommand{\stitle}[1]{\vspace{2mm}\noindent{\bf #1:}\xspace}

\usepackage{xspace}

\newcommand{\Gee}{\mathcal{G}}
\newcommand{\gee}{\mathbf{g}}
\newcommand{\dee}{\mathcal{D}}

\newcommand{\uu}{\mathcal{U}}

\newcommand{\ee}{\mathbb{E}}
\newcommand{\eps}{\varepsilon}

\renewcommand{\marginpar}[2][]{}

\newcommand{\prob}{\textsc{Fair-Count-Min}\xspace}

\newcommand{\lf}{{\tt low-frequency}\xspace}
\newcommand{\hf}{{\tt high-frequency}\xspace}

\newcommand{\census}{\small \textsc{census}\xspace}
\newcommand{\google}{\small \textsc{google n-grams}\xspace}
\newcommand{\synthetic}{\small \textsc{synthetic}\xspace}

%% file: mycolors-header.tex
\definecolor{slategray}{RGB}{112, 128, 144}
\definecolor{mutedteal}{RGB}{96, 130, 133}
\definecolor{deepindigo}{RGB}{75, 0, 130}
\definecolor{olivegreen}{RGB}{85, 107, 47}
\definecolor{dustyrose}{RGB}{133, 96, 111}
\definecolor{softblue}{RGB}{70, 130, 180}          
\definecolor{forestgreen}{RGB}{34, 139, 34}        
\definecolor{mutedpink}{RGB}{200, 112, 147}        
\definecolor{burntred}{RGB}{165, 42, 42}           
\definecolor{darkyellow}{RGB}{184, 134, 11}        
\definecolor{warmtaupe}{RGB}{139, 119, 101}        
\definecolor{dustylavender}{RGB}{150, 123, 182}    
\definecolor{steelblue}{RGB}{70, 130, 180}         

%% file: introduction.tex
\section{Introduction}\label{sec:intro}

Estimating item frequencies in data streams is a fundamental problem in computer science, where given a universe of elements, the objective is to count the appearance of each within the stream.
However, due to memory limitations and processing time in streaming settings, exact counting is often impractical. Consequently, approximating the counts using techniques such as the Count-Min (CM) sketch has become widely adopted. 
The CM estimates the frequency of elements by randomly hashing the elements into a smaller set of buckets, while adding up the frequencies of all elements in each bucket.
The frequency count of each bucket is then returned as an upper bound of the frequency of elements it contains.

The CM sketches offer space-efficient frequency estimations with small error guarantees.
Nevertheless, {\em the error guarantees are additive}. Such errors may be negligible for popular elements with a high frequency in the data stream.
However, as further elaborated in Example~\ref{eg:1}, the unpopular elements with low frequencies in the stream are disproportionately impacted under such guarantees.
That is because even small additive errors can cause a significant relative increase in the count estimation of low-frequency elements. 
This imbalance gives rise to a fairness issue -- an aspect overlooked in the existing work.

To address this issue, in this paper, we introduce {\bf Fair-Count-Min} (FCM), a frequency estimation sketch with the equal group-level approximation factor guarantee, ensuring that the relative count-estimation increase is equal for various groups, irrespective of their popularity in the data stream.
To the best of our knowledge, FCM is the first frequency estimation sketch with provable {\em multiplicative error} guarantees in its estimation.

Our key idea in the design of the FCM is the partitioning of the sketch buckets strategy based on a group-aware, semi-uniform hashing scheme that prevents collisions between elements of different groups. 
Then, we prove that, carefully selecting the number of buckets allocated to each group, our FCM sketch guarantees an equal expected approximation factor across the groups, i.e., equal ratios of the true frequency to the estimated one.
The FCM sketch applies to binary and non-binary grouping of the elements, while the grouping can be based on element popularity or any other grouping strategy, e.g., based on the demographic groups.

FCM does not increase the memory requirement of the CM.
Furthermore, FCM and CM have the same time complexities for update and query operations. In other words, the time taken to look up the frequency estimation of an element and the time to increase the frequency of an element in the data stream are the same in CM and FCM.

In a general setting, where multiple hash functions are used in the FCM, identifying the proper number of buckets allocated to each group requires solving an equation, for which we propose efficient exact and approximation algorithms.

Furthermore, we theoretically analyze the price of fairness (PoF) of FCM. Interestingly, we show that achieving fairness may even reduce the total additive error for a specific case when the CM sketch uses a single random hash function, while experimentally demonstrating that the PoF is small for the general settings.

In addition to theoretical analysis, we perform extensive experiments on multiple real-world and synthetic datasets to evaluate the performance of our proposed approach in practice.
In summary, our experiments verify (a) while CM is usually not fair, FCM achieves fairness in all experiments, (b) the price of fairness is negligible, and (c) FCM has the same memory and time efficiency as CM.

\newpage
\stitle{Summary of contributions}
\begin{itemize}[leftmargin=*]
\item We propose a novel notion of \emph{group fairness} in the frequency estimation context, defined by the requirement that the approximation factor (the \emph{multiplicative} overestimation error) remains equal across different groups. This contrasts with traditional approaches based on {\em additive} error bounds, which tend to disproportionately impact unpopular elements with low frequency.
Using this notion of fairness, we introduce {\em Fair-Count-Min} (FCM), the frequency estimation sketch that guarantees group fairness.
    
\item We introduce a \emph{column-partitioning} technique based on semi-uniform hash functions to develop FCM. This method assigns groups to disjoint sets of hash buckets, thereby eliminating inter-group collisions and promoting fairness in estimation.
Our proposed method, FCM, is agnostic to both the grouping strategy and the underlying distribution of the data elements.
Furthermore, FCM does not increase the memory and time requirements of the regular count-min sketch.

\item We theoretically prove that FCM is fair, i.e., it ensures an equal expected approximation factor across groups. Furthermore, we analyze the \emph{price of fairness} and show that it is typically small—and in some cases, can even be negative.

\item We design both exact and approximate algorithms to efficiently compute the optimal allocation of buckets per group.

\item We empirically evaluate FCM using real-world and synthetic datasets. The results demonstrate that FCM successfully achieves group fairness while incurring only a minor additional additive error, and it preserves the time and space efficiency characteristic of standard CM sketches.
\end{itemize}
\stitle{Paper organization}
The rest of the paper is organized as follows. In Section~\ref{sec:pre}, we provide necessary preliminaries, introduce the fairness challenge in CM sketches, and formally define the group-fair frequency estimation problem. Section~\ref{sec:col} details our FCM sketch, including its construction via group-aware semi-uniform hashing and fairness guarantees. Section~\ref{sec:efficient} presents efficient algorithms for computing bucket allocations across groups. In Section~\ref{sec:price}, we theoretically analyze the price of fairness. Section~\ref{sec:exp} reports our experimental results, comparing FCM to standard CM and Row-Partitioning baselines. Finally, we conclude with a discussion of implications and future work.

%% file: preliminary.tex
\section{Preliminary}\label{sec:pre}
\subsection{Data Model}

Let $\dee$ be a data stream consisting of elements (or events), each belonging to a type from a finite universe $\uu=\{e_1, e_2, \dots, e_n\}$.
Let $N$ be the number of elements in the data stream $\dee$.
We use the function $f:\uu\rightarrow [N]$ to refer to the frequency of each element type in $\dee$. That is, $f(e)$ is the number of times an element of type $e$ appears in $\dee$.
Formally,

\[f(e) = \sum_{x \in \mathcal{D}} \mathbbm{1}(x = e),\]  
where $\mathbbm{1}(x = e)$ is the indicator function that equals 1 if $x$ is of type $e$ and 0 otherwise. 

\subsection{Count-Min Sketch}
Frequency estimation is a fundamental problem in streaming data processing. Given a data stream $\dee$, the goal is to efficiently estimate the frequency $f(e)$ of each element type $e$. It is easy to see that finding an exact solution to the above problem requires $\Theta(n)$ space, as it involves maintaining $n$ counters for each element type. 
To overcome this limitation, frequency estimation data structures (a.k.a sketches) have been designed to provide approximate answers to such queries while using sub-linear space. In addition to being space-efficient, these sketches provide probabilistic error guarantees through tunable parameters and enable constant-time update and query operations. 

Count-Min (CM) is among the most widely used sketches for frequency estimation. It represents the frequency estimates using a two-dimensional array $CM$ of $d$ rows and $w$ columns. The sketch employs $d$ independent hash functions $\mathsf{h}_1, \mathsf{h}_2, \dots, \mathsf{h}_d$, where each $\mathsf{h}_i:\uu\xrightarrow{}[w]$ maps an element type to a column index (called {\em bucket} or {\em bin} in the rest of the paper) in row $i$. These hash functions distribute element types across different buckets to reduce the impact of hash collisions. When a new element of type $e$ arrives in the stream, the sketch updates its counts as follows:  
\[
CM[i, \mathsf{h}_i(e)] \gets CM[i, \mathsf{h}_i(e)] + 1, \quad \forall i \in [d].
\]  
Each row records a count for $e$, though potential overestimations may occur due to hash collisions. To estimate the frequency of an element type\footnote{In the rest of the paper, we use ``element $e$'' and ``element type $e$'' interchangeability to refer to elements of type $e$.} $e$, the sketch returns:  
\[
\hat{f}(e) = \min_{i=1}^{d} CM[i, \mathsf{h}_i(e)].
\]  
Taking the minimum across multiple rows mitigates over-counting errors. As previously mentioned, the estimation error in CM sketch occurs due to hash collisions. For each element type $e$ the estimated frequency $\hat{f}(e)$ is always an upper bound on the true frequency $f(e)$:
\begin{equation}\label{eq:error}
\hat{f}(e)=f(e)+\eps_A(e),
\end{equation}

where the additive error $\eps_A(e)$ corresponds to the combined frequency of all other elements that hash into the same bucket as $e$. In expectation, the additive error $\eps_A(e)$ is bounded by $\frac{N}{w}$. Formally:
\begin{equation}\label{eq:exp_error}
\ee[\hat{f}(e)]=f(e)+\frac{\sum_{\forall e^{\prime}\neq e, \; \mathsf{h}(e^{\prime})=\mathsf{h}(e)}f(e^{\prime})}{w}\leq f(e)+\frac{N}{w}
\end{equation}

Several variations of the CM sketch exist, including the CM sketch with conservative updates~\cite{mazziane2022analyzing}, the Count-Mean-Min sketch~\cite{deng2007new}, and other related frequency estimation structures such as the Count sketch~\cite{charikar2002finding} and Spectral Bloom Filters~\cite{cohen2003spectral}. In this study, we focus exclusively on the CM sketch, leaving the exploration of these alternatives for future work.

\subsection{Fairness: Equal Expected Approximation Factor}
Existing CM sketches provide an additive upper-bound error $\eps_A$ on the frequency estimation of the elements.
However,
while $\eps_A$ may be a negligible error for the popular, high-frequency elements, it can be intolerable for the unpopular, low-frequency ones.
In other words, the significance of the frequency estimation error is relative to the frequency value.
To further clarify this using a toy example, let us consider Example~\ref{eg:1}.

\begin{example}\label{eg:1}
    {\it \small
    Consider a CM sketch with a hash function $h$ that maps the events of types $\uu=\{e_1, e_2, \dots, e_n\}$ to $w=200$ bins. Let $N=10000$ be the number of elements in the data stream. Hence, the additive estimation error is bounded by $\eps_A\leq 50$.
    Suppose the frequencies of the element types $e_1$ and $e_2$ are $f(e_1)=1000$ and $f(e_2)=10$.
    As a result, the expected frequency estimation is bounded by $1050$ for $e_1$, resulting in a small error of only $5\%$ of $f(e_1)$. On the other hand, this error is as high as $500\%$ for $e_2$. 
    }
\end{example}

From Example~\ref{eg:1}, it is clear that the additive error does not provide a strong guarantee independent of the element frequencies. Specifically, providing large estimation errors relative to small frequency values, it is {\em unfair} for low-frequency element types.
This motivates us to instead promote {\bf approximation factor} as a stronger frequency estimation guarantee.

\begin{definition}[Approximation Factor ]\label{def:approx}
    The approximation factor of count-min sketch CM for an element type $e$ is $\alpha(e)$, if its estimation $\hat{f}(e)$ of the frequency of $e$ satisfies the following condition:
    \[
        \frac{f(e)}{\hat{f}(e)} = \alpha(e).
    \] 
\end{definition}


Using Definition~\ref{def:approx}, we define a notion of {\em group fairness} provides equal multiplicative frequency estimation errors for various element groups.

Let each element type $e$ be associated with a group $g\in \Gee$, where $\Gee=\{\gee_1,\gee_2,\dots,\gee_{\ell}\}$.
For example, each group $\gee_i$ might contain element types with comparable popularity.
Then, a count-min sketch is called fair if it satisfies Definition~\ref{def:fairness}.

\begin{definition}[Group-Fair Count-Min]\label{def:fairness} 
A count-min sketch is group-fair iff:
\[
\forall \gee_i, \gee_j \in \Gee, \quad \ee_{e \in \gee_i}\big[\alpha(e)\big] = \ee_{e' \in \gee_j}\big[\alpha(e')\big]
\].
\end{definition}

Without loss of generality 
and to ease the explanations, 
we use the binary groups $l$ (\lf) and $h$ (\hf) for explaining our sketches and drawing the analyses. 

Our results readily hold for non-binary and arbitrary grouping of element types, as we shall further discuss in Section~\ref{sec:gen}. 

\subsection{Problem Formulation}

With the necessary terms and notations defined, we now formally define our problem of interest:

\begin{definition} Given a data stream of elements $\dee$ drawn from a universe of element types $\uu$ where each element type belongs to a group $g\in\Gee$, design a group-fair count-min sketch.
\end{definition}

\subsection{Solution Overview}
A main issue that causes the unbounded multiplicative error in traditional count-min sketches is that high-frequency elements can collide with low-frequency ones, adding a major overestimation to their counts.

One way to reduce the chance of such collisions is by significantly increasing the number of bins $w$ and/or the value of $d$, the number of independent hash functions used for the estimation, i.e., the number of rows in the sketch.
This, however, has two major issues: first, 
given a low-frequency element $e$ to ensure 
it is unlikely that a high-frequency element collides with it in at least one of the $d$ rows, the size of the sketch increases in the order of $n$, the number of element types -- losing its purpose.
Second, even after increasing the size of the sketch, it cannot provide a guarantee of a multiplicative error as low and high-frequency elements can still collide.

Instead, to design group-fair min-count sketches, our goal is to {\em make it impossible for elements in different groups to collide}.

Specifically, we ensure our goal by designing group-aware ``semi-uniform'' hash schemes that isolate low and high-frequency elements in Section~\ref{sec:col}. 
Proving that our sketch satisfies Definition~\ref{def:fairness}, we study its price of fairness in Section~\ref{sec:price}.

%% file: solution.tex
\begin{figure}[!tb]
    \centering
    \begin{tikzpicture}
        \fill[mutedpink] (1,0) rectangle (5,1);
        \fill[softblue] (5,0) rectangle (8,1);
        \draw (1,0) rectangle (2,1); \node at (1.5,0.5) {$1$};
        \draw (2,0) rectangle (3,1); \node at (2.5,0.5) {$2$};
        \draw (3,0) rectangle (4,1); \node at (3.5,0.5) {$\cdots$};
        \draw (4,0) rectangle (5,1); \node at (4.5,0.5) {$w_l$};
        \draw (5,0) rectangle (6,1); \node at (5.5,0.5) {$w_l+1$};
        \draw (6,0) rectangle (7,1); \node at (6.5,0.5) {$\cdots$};
        \draw (7,0) rectangle (8,1); \node at (7.5,0.5) {$w$};

        \draw[<->, thick] (1,1.3) -- (5,1.3) node[midway, above] {$w_l$};
        \draw[<->, thick] (5,1.3) -- (8,1.3) node[midway, above] {$w_h = w - w_l$};
    \end{tikzpicture}
    \caption{Illustration of a column-partitioning based group-fair min-count with one row, i.e., one hash function $\mathsf{h}(.)$.}
    \label{fig:col-one}
\end{figure}
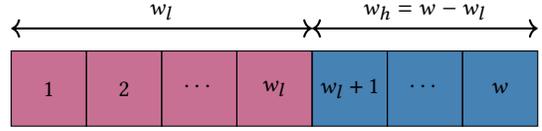

\section{Fair-Count-Min using Group-Aware Semi-Uniform Hashing}\label{sec:col}
Our first idea for generating group-fair count-min is to utilize ``semi-uniform'' hash functions that separate element types of low and high frequency. 
Specifically, reserving $w_l$ bins for the \lf group $l$ and $w_h=w-w_l$ for the \hf group $h$, the semi-uniform hash function $\mathsf{h}(.)$ assigns 
each low-frequency element to the first $w_l$ bins and the others to the remaining $w_h$ bins (Figure~\ref{fig:col-one}):
\[
\mathsf{h}: \begin{cases}
    l \rightarrow [w_l]\\
    h \rightarrow  w_l+[w_h]
\end{cases}
\]

The key question we shall answer in the rest of this section is whether there exists a value $w_l$ (hence $w_h=w-w_l$) for which a count-min sketch based on the semi-uniform hash scheme becomes group fair.

In the following, first, we focus on the case where $d=1$, i.e., the sketch is based on only one hash function $h$.
Next, we extend our analysis to $d=2$ and finally to the general values of $d$.

\subsection{$d=1$}

First, let us derive the expected approximation factor for an element $e$ in the \lf group $l$.
Let $n_l$ be the number of element types in $l$.

Let $C_i$ be the total frequency counts in each low-frequency bucket $i\in [w_l]$. That is, \[C_i = \sum _{e_j:\mathsf{h}(e_j)=i} f(e_j)\]

For every element $e_j\in l$, 
the value of the bucket it hashed to is returned as its frequency estimation. That is,
\(\hat{f}(e_j)=C_{\mathsf{h}(e_j)}\).
Therefore, following the approximation factor definition,
\[\hat{f}(e_j) = \frac{f(e_j)}{\alpha(e_j)} = C_{\mathsf{h}(e_j)}.\]

Hence,
\begin{align}
    \alpha(e_j) =  \frac{f(e_j)}{C_{\mathsf{h}(e_j)}}
\end{align}

Therefore,
\begin{align}\label{eq:temp1}
 \nonumber   \ee \left[\alpha_l\right]
    &= \frac{1}{n_l} \sum_{e_j\in l} \alpha(e_j) = \frac{1}{n_l} \sum_{e_j\in l} \frac{f(e_j)}{C_{\mathsf{h}(e_j)}}\\
\nonumber &=\frac{1}{n_l} \sum_{i=1}^{w_l}\sum_{e_j:\mathsf{h}(e_j)=i} \frac{f(e_j)}{C_{i}}  \hspace{7mm} \text{\tt\scriptsize  //breaking the calculation per cell}\\
\nonumber   & = \frac{1}{n_l} \sum_{i=1}^{w_l} \frac{1}{C_i} \sum_{e_j:\mathsf{h}(e_j)=i} f(e_j) \hspace{7mm} \text{\tt\scriptsize  //factoring out the common term}\\
\nonumber   & = \frac{1}{n_l} \sum_{i=1}^{w_l} \frac{1}{C_i} C_i  \hspace{16mm} \text{\tt\scriptsize  //replacing sum of frequencies with $C_i$} \\
            &= \frac{1}{n_l} \sum_{i=1}^{w_l} 1
                = \frac{w_l}{n_l}
\end{align}

Similarly, 
the expected approximation factor for the group $h$ can be computed as
\[
\ee\big[\alpha_h\big] = \frac{w_h}{n_h} = \frac{w-w_l}{n-n_l}
\]

To satisfy group fairness (Definition~\ref{def:fairness}), we need to ensure that the expected group fairness for the two groups is the same. That is, $\ee\big[\alpha_l\big] = \ee\big[\alpha_h\big]$.

Hence, to ensure fairness, $w_l$ is selected as follows:
\begin{align} \label{eq:wlAllocation}
    \frac{w_l}{n_l} = \frac{w-w_l}{n-n_l} \Rightarrow ~ w_l = \frac{n_l}{n} w
\end{align}
In other words, according to Equation~\ref{eq:wlAllocation}, if the number of bins allocated to each group is proportional to their size, i.e., $\frac{n_l}{n}$, is sufficient to ensure fairness.
Formally,

\begin{theorem}\label{thm:1}
A Count-Min sketch 
with a group-aware semi-uniform hash function $\mathsf{h}(.)$ is group-fair, if the number of bins $l$ allocated to each group is proportional to the ratio of element types from that group. That is,  
$w_l = \frac{n_l}{n} w, \forall \gee_l\in \mathcal{G}$.
\end{theorem}

Before moving to the larger values of $d$, i.e., count-min sketches with more number of hash functions, let us take a closer look at Theorem~\ref{thm:1} and what it indicates.
A regular min-count sketch hashes the $n$ element types uniformly across the $[w]$ bins (irrespective of which group they belong to).
As a result, the expected number of element types in each bucket is $\frac{n}{w}$.
According to Theorem~\ref{thm:1}, $w_l = \frac{n_l}{n} w, \forall l\in \mathcal{G}$. 
As a result, the number of elements in each bucket for each group $l$ is:
\begin{align}\label{eq:temp2}
    \frac{n_l}{w_l} = \frac{n_l}{\frac{n_l}{n} w} = \frac{n}{w}
\end{align}

Interestingly, this means to ensure fairness, we need to {\em allocate enough bins to each group such that expected number of element types hashed to each bucket remains unchanged}, i.e., $\frac{n}{w}$.
This helps us in extending $d$ to more than one row by {\em equalizing the expected number of element types hashed in each bin} across various groups to ensure fairness.



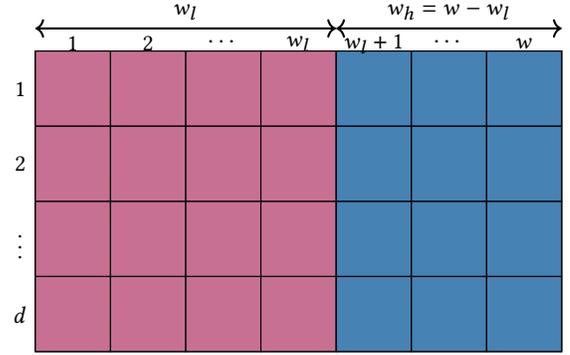
\begin{figure}[!tb]
    \centering
    \begin{tikzpicture}
        \fill[mutedpink] (1,0) rectangle (5,4);
        \fill[softblue] (5,0) rectangle (8,4);
        
        \draw (1,0) rectangle (2,4); \node at (1.5,4.1) {$1$};
        \draw (2,0) rectangle (3,4); \node at (2.5,4.1) {$2$};
        \draw (3,0) rectangle (4,4); \node at (3.5,4.1) {$\cdots$};
        \draw (4,0) rectangle (5,4); \node at (4.5,4.1) {$w_l$};
        \draw (5,0) rectangle (6,4); \node at (5.5,4.1) {$w_l+1$};
        \draw (6,0) rectangle (7,4); \node at (6.5,4.1) {$\cdots$};
        \draw (7,0) rectangle (8,4); \node at (7.5,4.1) {$w$};

        \draw[<->, thick] (1,4.3) -- (5,4.3) node[midway, above] {$w_l$};
        \draw[<->, thick] (5,4.3) -- (8,4.3) node[midway, above] {$w_h = w - w_l$};

        \draw (1,3) rectangle (8,4); \node at (0.8,3.5) {$1$};
        \draw (1,2) rectangle (8,3); \node at (0.8,2.5) {$2$};
        \draw (1,1) rectangle (8,2); \node at (0.8,1.5) {$\vdots$};
        \draw (1,0) rectangle (8,1); \node at (0.8,0.5) {$d$};
    \end{tikzpicture}
    \caption{Illustration of a column-partitioning based group-fair min-count with $d$ rows.}
    \label{fig:col-d}
\end{figure}

\subsection{General Value of $d$}\label{sec:d}

Having established that a count-min sketch with a single group-aware semi-fairness hash function ($d=1$) ensures group fairness, we now generalize this result to the case of $d$ hash functions.

Similar to $d=1$, our goal is to partition the $w$ columns across various groups such that the group fairness requirement is satisfied, i.e., all groups have the same expected approximation factor.

We recall from Theorem~\ref{thm:1} that achieving group fairness requires equalizing the expected number of element types hashed into the minimum-count bucket across $d$ rows for all groups. 

When $d=1$, the estimated frequency of an element is the frequency count of the bucket it is hashed into.
On the other hand, when $d>1$, a frequency estimation $\hat{f}(e)$ is the {\em minimum} of the frequency counts of the buckets $e$ is hashed to at each row. 

Let us define the size of a bucket as the number of element types hashed to it. 
For an element $e_j\in l$ and a row $i$, let $\ell=\mathsf{h}_i(e_j)$. 
Let the random integer variable $X_i$ be the size of the bucket of the element $e$ at row $i$ , i.e., $|\{e\in l | \mathsf{h}_i(e)=\ell\}|$.
The elements are hashed uniformly at random and independently into the bins. As a result, the probability that a random element is hashed into the bucket $\ell$ is $\frac{1}{w_l}$.
As a result, $X_i$ follows the binomial distribution \(X_i \sim Bin\left(n_l,\frac{1}{w_l}\right)\):

\begin{gather*}
   \Pr(X_i=x) = \binom{n_l}{x} \cdot \left(\frac{1}{w_l}\right)^{x} \cdot \left(\frac{w_l-1}{w_l}\right)^{n_l - x}
\end{gather*}

Let $\mu_l$ be the expected frequency for group $l$.
Then, 
the expected frequency count of each bucket is $\mu_l$ times the size of the bucket.
Following this argument, we simplify our analysis 
by equalizing the expected number of element types hashed into the {\em minimum-size} bucket across $d$ rows for all groups. 

Fix an element $e_j$ in the group $l$.
Let $X_1,\cdots, X_d$ be the random variables reflecting the sizes of the $d$ buckets $e_j$ is hashed to across various rows, while each $X_i$ following the Binomial distribution 
\[X_1,\cdots,X_d \sim Bin\left(n_l,\frac{1}{w_l}\right)\]

Define the random variable $Y_j$ as the minimum of $X_1$ to $X_d$. That is,
\[
    Y \sim \min(X_1, \cdots, X_d).
\]
We aim to find the expected value of $Y$:
\[
\ee[Y] = \sum_{x=1}^n x P(Y=x) 
\]
Since the domain of the random variable $Y$ is the nonnegative integers \(\{1, 2, \cdots\}\), the expected value can alternatively be computed using the following equation~\cite{feller1991introduction}.
\[
    \ee[Y] = \sum_{x=1}^n P(Y\geq x) 
\]

\begin{gather*}
\Pr(Y \geq x) = \Pr(\min(X_1,\cdots, X_d) \geq x) = \Pr(X_1 \geq x, \cdots, X_d \geq x).
\end{gather*}

Since $X_1,\cdots, X_d$ are independent and identically distributed, the probability is computed as:

\begin{align*}
\Pr(Y \geq x) &= \Pr(X_1 \geq x, \cdots, X_d \geq x) \\
    &= \prod_{i=1}^d \Pr(X_i \geq x) = \Pr(X \geq x)^d
\end{align*}

where \(\Pr(X=x) = \binom{n_l}{x} \cdot \left(\frac{1}{w_l}\right)^{x} \cdot \left(\frac{w_l-1}{w_l}\right)^{n_l - x}\).
Hence, the probability $\Pr(X \geq x)$ is:
\[
    \Pr(X \geq x) = \sum_{i=x}^{n_l} \binom{n_l}{i} \left(\frac{1}{w_l}\right)^i \left(\frac{w_l-1}{w_l}\right)^{n_l - i}.
\]

Putting everything together, we obtain the expected value of $Y$ as:

\begin{align}\label{eq:expectedmd}
 \nonumber \ee[Y] &= \sum_{x=1}^{n_l} \Pr(X \geq x)^d\\
       &= \sum_{x=1}^{n_l} \left[\sum_{i=x}^{n_l} \binom{n_l}{i} \left(\frac{1}{w_l}\right)^i \left(\frac{w_l-1}{w_l}\right)^{n_l - i} \right]^d.
\end{align}


Finally, equalizing Equation~\ref{eq:expectedmd} for the groups $l$ and $h$, we specify the number of columns allocated to each group as $w_l$ and $w_h=w-w_l$, by solving the Equation~\ref{eq:solve} ($n_h = n-n_l$):

\begin{align}\label{eq:solve}
\nonumber
\ee_{e\in l}[Y] &= \ee_{e\in  h}[Y] \Rightarrow \\
\sum_{x=1}^{n_l} &\left[\sum_{i=x}^{n_l} \binom{n_l}{i} \left(\frac{1}{w_l}\right)^i \left(\frac{w_l-1}{w_l}\right)^{n_l - i} \right]^d\\
\nonumber
&=
\sum_{x=1}^{n-n_l} \left[\sum_{i=x}^{n-n_l} \binom{n-n_l}{i} \left(\frac{1}{w-w_l}\right)^i \left(\frac{w-w_l-1}{w-w_l}\right)^{n-n_l - i} \right]^d
\end{align}

Later in Section~\ref{sec:efficient}, we shall provide an efficient algorithm for solving Equation~\ref{eq:solve}.



\begin{figure}[!tb]
    \centering
    \begin{tikzpicture}
        \fill[softblue] (1,0) rectangle (5,3);
        \fill[mutedpink] (1,3) rectangle (5,6);
        
        \draw (1,0) rectangle (2,6); \node at (1.5,6.15) {$1$};
        \draw (2,0) rectangle (3,6); \node at (2.5,6.15) {$2$};
        \draw (3,0) rectangle (4,6); \node at (3.5,6.15) {$\cdots$};
        \draw (4,0) rectangle (5,6); \node at (4.5,6.15) {$w$};

        \draw (1,5) rectangle (5,6); \node at (0.5,5.5) {$1$};
        \draw (1,4) rectangle (5,5); \node at (0.5,4.5) {$\vdots$};
        \draw (1,3) rectangle (5,4); \node at (0.5,3.5) {$d_l$};
        \draw (1,2) rectangle (5,3); \node at (0.5,2.5) {$d_l+1$};
        \draw (1,1) rectangle (5,2); \node at (0.5,1.5) {$\vdots$};
        \draw (1,0) rectangle (5,1); \node at (0.5,0.5) {$d$};

        \draw[<->, thick] (0,0) -- (0,2.98) node[midway, left, rotate=90, yshift=2mm, anchor=center] {$d_h=d-d_l$};
        \draw[<->, thick] (0,3.02) -- (0,6) node[midway, left, rotate=90, yshift=2mm, anchor=center] {$d_l$};
    \end{tikzpicture}
    \caption{Illustration of the row-partitioning baseline.}
    \label{fig:row}
\end{figure}
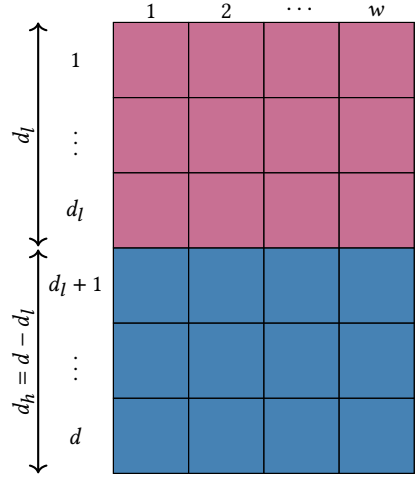

\subsection{Negative Result: Row Partitioning Baseline}\label{sec:row}
FCM follows a column partitioning approach, using the group-aware semi-uniform hashing schemes.
Alternatively, one can partition the rows by allocating a specific number of exclusive rows to each group, where the allocation of rows to each group is determined by the number of element types within that group (Figure~\ref{fig:row}). 

Following Equation~\ref{eq:expectedmd}, the expected size of the minimum-size bucket across the $d'$ rows, $w'$ columns, and a group with $n'$ elements can be computed as, 

\begin{align}\label{eq:expectedmd2}
\mathcal{E}(n',d',w') = \sum_{x=1}^{n'} \left[\sum_{i=x}^{n'} \binom{n'}{i} \left(\frac{1}{w'}\right)^i \left(\frac{w'-1}{w'}\right)^{n' - i} \right]^{d'}
\end{align}

Let $d_l$ and $d_h=d-d_l$ be the number of rows allocated to the groups $l$ and $h$. 
Then, we can find the proper values of $d_l$ and $d_h$ that achieve fairness by solving the following equation:
\begin{align}\label{eq:rowd}
\mathcal{E}(n_l,d_l,w) = \mathcal{E}(n_h,d_h,w)
\end{align}

There, however, is a major issue making this approach unlikely to work: unlike $w$, which is a relatively large number, $d$ is a small constant (usually in the order of tens).
As a result, there are only a small, constant number of choices for $d_l\in [d]$.
This, in practice, makes it unlikely for $d$ to be divisible such that Equation~\ref{eq:rowd} is satisfied.
We shall verify this negative result in our experiments in Section~\ref{sec:exp}.

\subsection{Generalization to Multiple Arbitrary Groups}
\label{sec:gen}
So far, we have explained our techniques for the binary grouping of the elements based on their frequencies. 

Our approach works for any (arbitrary) grouping of the elements since none of our definitions, results, and analyses are based on any assumption about how elements are grouped. This is also observed in our experiments in Section~\ref{sec:exp}, where we, for example, use demographic groups to group the elements.

Extending our results to the non-binary grouping of the element types is straightforward. 
Let $\Gee=\{\gee_1, \gee_2,\cdots, \gee_\ell\}$ be the non-binary set of groups.
For every group $\gee_l\in \Gee$, transform the problem into the binary grouping by merging all other groups as the super-group ``others''. Then, for $d=1$, following Theorem~\ref{thm:1}, $w_l = \frac{n_l}{n}w$.
Similarly, for $d>1$, the value of $w_l$, $\forall \gee_l\in \Gee$, can be computed by solving Equation~\ref{eq:solve}. We propose an efficient algorithm for finding $w_l$ values when $d>1$ in Section~\ref{sec:efficient}.

%% file: efficientImplementation.tex
\newcommand{\well}{w_l}
\newcommand{\wh}{w_h}
\section{Computing the column widths}
\label{sec:efficient}

Our algorithm is straightforward; we should compute the integer value of $\well$ such that Equation~\eqref{eq:solve} holds. We note that Equation~\eqref{eq:solve} may not hold for an integer value of $\well$. In this case the goal is to compute the integer value of $\well$ such that the left side of Equation~\eqref{eq:solve} is as close as possible to the right side of Equation~\eqref{eq:solve}.
A naive implementation is to try every value of $\well$ and calculate the value of the left and right side of Equation~\eqref{eq:solve} by simply evaluate every term in the sum. Such an implementation would take $\Omega(n^3)$ time because we should try $O(n)$ values of $\well$, and for each such value the calculation of each side of Equation~\eqref{eq:solve} takes $\Omega(n^2)$ time to evaluate the summations.

We first show an exact near-linear time algorithm to compute the value of $\well$ that solves Equation~\ref{eq:solve} and then we briefly discuss an approximation algorithm that is used in our experiments to efficiently evaluate the sums of the binomial function.

\newcommand{\func}{v}
\paragraph{Efficient exact computation}
We first assume that we have two groups, the high-frequency elements and the low-frequency elements. Then we extend our method to multiple groups $\Gee=\{\gee_1,\ldots, \gee_\ell\}$.
Similarly, to the previous section,
let $\mathcal{E}(n_l,d,\well)$ be the function that represents the left side of Equation~\ref{eq:solve}. Equivalently, the right side of Equation~\eqref{eq:solve} is captured by $\mathcal{E}(n_h,d,w-\well)$. 
By definition, the function $\mathcal{E}$  computes the expected error of light (resp. heavy) elements. The values $n_l$, $n_h$, and $d$ are fixed, so the more bins allocated to light (respectively, heavy) items, the smaller the expected error. Hence, the function $\mathcal{E}(n_l,d,\well)$ is monotone non-increasing with respect to $\well$, while $\mathcal{E}(n_h,d,w-\well)$ is monotone non-decreasing with respect to $\well$.
This property leads to the observation that instead of trying all values of $\well$, we can run a binary search in the range $[1,\ldots, n]$ and compute the best value of $\well$. The overall idea of our algorithm is the following. For every value of $\well$ in the binary search, we evaluate $\mathcal{E}(n_l,d,\well)$ and $\mathcal{E}(n_h,d,w-\well)$. If $\mathcal{E}(n_l,d,\well)<\mathcal{E}(n_h,d,w-\well)$ then we try larger values of $\well$ in the binary search. Otherwise, we continue the binary search with smaller values. In the end, we return the value of $\well$ found by the binary search such that $|\mathcal{E}(n_l,d,\well)-\mathcal{E}(n_h,d,w-\well)|$ is minimized.

Even if someone naively applies binary search, the running time would be $\Omega(n^2)$ to evaluate the sums. Given a value of $\well$ we show how to compute $\mathcal{E}(n_l,d,\well)$ in near-linear time with respect to $n$. The computation of $\mathcal{E}(n_h,d,w-\well)$ is equivalent.
First, for $x=1$ we compute the sum $\sum_{i=1}^{n_l}\binom{n_l}{i}\left(\frac{1}{\well}\right)^i\left(\frac{\well-1}{\well}\right)^{n_{l-i}}$ as follows. For $i=1$, we compute $b_1=\binom{n_l}{1}$, $t_1=\frac{1}{\well}$, and $s_1=\left(\frac{\well-1}{\well}\right)^{n_l-1}$. We store $\alpha_1=b_1\cdot t_1\cdot s_1$. Then for $i\in \{2,\ldots, n_l\}$, we observe that $b_i=b_{i-1}\cdot \frac{n_l+1-i}{i}$,
$t_i=t_{i-1}\cdot \frac{1}{\well}$, $s_i=s_{i-1}\cdot\frac{\well}{\well-1}$, and we set $\alpha_i=b_i\cdot t_i\cdot s_i$. Then, we set $\beta_{n_l}=\alpha_{n_l}$ and for every $i=n_l-1,\ldots, 1$, we compute $\beta_i=\alpha_i+\beta_{i+1}$. The main observation of our algorithm is that
$\mathcal{E}(n_l,d, \well)=\sum_{x=1}^{n_l}\beta_x^d$.

In order to analyze the runtime of our algorithm, we assume a modification of the real-RAM model of computation where every basic arithmetic operation is computed in $O(1)$ time. We note that for any pair of positive integer numbers $q, k$, it holds that $\binom{q}{k}=\prod_{i=1}^{k}\frac{q+1-i}{i}$ so $\binom{q}{k}$ can be computed in $O(k)$ time. Finally, the $k$-th power of any real number can be computed in $O(\log k)$ time.

We analyze the runtime in one iteration of the binary search. By definition of our model of computation, we compute $b_1, t_1$ in $O(1)$ time and $s_1$ in $O(\log n)$ time. Then $\alpha_1$ is computed in $O(1)$ time. For every value of $i\in\{2,\ldots,n_l\}$, given $b_{i-1}, t_{i-1}, s_{i-1}$ the values of $b_i, t_i, s_{i-1}$ and hence $\alpha_i$ are computed in $O(1)$ time. So all values $\alpha_i$ for $i\in\{1,\ldots, n_l\}$ are computed in $O(n+\log n)=O(n)$ time in total. Next, we observe that given $\beta_{i+1}$, the value $\beta_i$ is computed in $O(1)$ time, so all values $\beta_i$ are computed in $O(n)$ time. Finally, the sum $\sum_{x=1}^{n_l}\beta_x^d$ is computed in $O\left(\sum_{x=1}^n\log(d)\right)=O(n\log d)$ time. The binary search is executed for $O(\log n)$ iterations, so in total the running time of our algorithm is $O(n\log(n)\log(d))$. Notice that usually, $n\gg d$ so the total running time is bounded by $O(n\log^2 n)$.

We can extend our method to multiple groups $\Gee=\{\gee_1,\ldots, \gee_\ell\}$.
For each group $\gee_j\in\Gee$, the function
$\mathcal{E}(n_j,d,w_j)$
represents the expected error of the elements in group $\gee_j$ as computed by Equation~\eqref{eq:expectedmd}. The goal is to compute $w_j$ for every $\gee_j\in\Gee$, given that $\sum_{j\in[\ell]}w_j=w$. We solve this problem, by recursively execute our algorithm for two groups (light and heavy) $\ell-1$ times.
Intuitively, if we have $\ell$ groups, we construct an instance with two colors: one that consists of the elements in the first group, and one that consists of the elements from all groups excluding the first group.
For a value $j\in\{1,\ldots, \ell-1\}$, let $W_j$ be the number of bins that should be assigned to groups $\gee_j, \gee_{j+1}\ldots, \gee_\ell$. Initially, notice that $W_1=w$. We set $n_j'=\sum_{b=j+1}^{\ell}n_b$. We assume that there exist two groups, one with $n_j$ elements and the other one with $n_j'$ elements, while the total number of bins is $W_j$. We solve this instance using our efficient implementation for two groups. Let $w_j$ be the number we compute. We set $W_{j+1}\leftarrow W_j-w_j$ and we continue recursively with $j\leftarrow j+1$. 
The overall time of our algorithm for groups $\Gee=\{\gee_1,\ldots, \gee_\ell\}$ is $O(\ell\cdot n\cdot\log^2n)$.

\paragraph{Approximate practical implementation}
While in theory, our previous algorithm works in the real-RAM model of computation, a direct calculation of $\binom{n}{i}$ might cause overflow in a practical implementation.
In practice, while we run the binary search on $\well$, instead of computing the binomial coefficients exactly, we use Stirling's approximation, i.e. for every positive integer $k$, $k!\approx \left(\frac{k}{\mathbf{e}}\right)^k\cdot \sqrt{2\pi k}$, where $\mathbf{e}$ is the Euler's number. It is known that the Gamma function $\Gamma(z)=\int_0^{\infty}t^{z-1}\mathbf{e}^{-t}dt$ satisfies $\Gamma(k)=k!$.
 Taking logarithms converts the products in the definition of the binomial coefficient into sums of logs which are easier to handle and avoid overflow. Then we approximate each $\log\Gamma(\cdot)$ term with Stirling’s expansion, $\log\Gamma(z)\approx z\log z-z+\frac{1}{2}\log(2\pi)+\frac{1}{12z}+O(z^{-3})$. In our practical implementation, we use the logarithm of $\Gamma$ function to estimate the values of $\mathcal{E}(n_l, d,\well)$ and  $\mathcal{E}(n_h,d, w-\well)$ for each value $\well$ in the binary search.


%% file: analysis.tex
\section{Price of Fairness Analysis}\label{sec:price}
Improving fairness usually comes at a cost, known as the {``price of fairness''} (aka cost of fairness) -- the reduction in the overall performance as a result of resolving unfairness~\cite{barocas2023fairness}.
After introducing FCM and its details, in this section, we study its price of fairness.
Following Equation~\ref{eq:error}, the overall performance of a CM sketch can be measured in terms of the sum of its expected additive error across all element types \(\mathcal{U}=\{e_1,\cdots, e_n\}\).
That is, 
    \begin{align}
    \mathcal{L}_{CM} = \sum_{i=1}^n \eps_A(e_i)
    \end{align}
Subsequently, we define the price of fairness (PoF) of a FCM sketch as the increase in its total expected additive error, compared to the standard CM sketch. That is,
    \begin{align}\label{eq:temp20}
    PoF = \mathcal{L}_{FCM} - \mathcal{L}_{CM} 
    \end{align}

In the following, first, we provide our price of fairness analysis for $d=1$, proving that fairness even improves the expected performance, i.e., $PoF<0$, when $d=1$.
For $d>1$, however, fairness comes at a positive (but small) cost.

\subsection{$d=1$}

Let us begin by computing $\mathcal{L}_{CM}$, the total additive error for the standard CM sketch.
By Chapter 3 of~\cite{cormode2020small}, the expected additive error of an element type $e_j\in\mathcal{U}$ is $$\eps_A(e_j) = \frac{N-f(e_j)}{w}.$$
Then, the total additive error of the standard CM sketch is computed as

\begin{align}\label{eq:temp22}
  \mathcal{L}_{CM} &= \sum_{j=1}^n \eps_A(e_j)= \sum_{j=1}^n \frac{N-f(e_j)}{w}=n\frac{N}{w}-\frac{N}{w}=(n-1)\frac{N}{w}
\end{align}

Now, let us compute $\mathcal{L}_{FCM}$, the total additive error for the FCM sketch.
\[
    \mathcal{L}_{FCM} = \sum_{j=1}^n \eps_A(e_j) = \sum_{e_j\in \gee_1} \eps_A(e_j)+\cdots + \sum_{e_j\in \gee_\ell} \eps_A(e_j)
\]
Following the same calculation as in Equation~\eqref{eq:temp22}, for every group $\gee_l=\Gee$, 
\begin{align*}
     \sum_{e_j\in \gee_l} \eps_A(e_j)&= (n_l-1)\frac{N_l}{w_l}
\end{align*}
Hence,
\[
\mathcal{L}_{FCM} = \sum_{e_j\in \gee_1} \eps_A(e_j)+\cdots + \sum_{e_j\in \gee_\ell} \eps_A(e_j)
 \nonumber = \sum_{l=1}^\ell (n_l-1)\frac{N_l}{w_l}
\]

From Theorem~\ref{thm:1}, we know, $w_l=\frac{n_l}{n}w, \forall \gee_l\in \Gee$, while $\sum_{l=1}^\ell N_l=N$.
Therefore,

\begin{align}\label{eq:temp23}
\nonumber \mathcal{L}_{FCM} &= \sum_{l=1}^\ell (n_l-1)\frac{N_l}{w_l}\\
\nonumber &=\sum_{l=1}^\ell (n_l-1)\frac{N_l}{\frac{n_l}{n}w}=\sum_{l=1}^\ell n\left(1-\frac{1}{n_l}\right)\frac{N_l}{w}\\
&=n\sum_{l=1}^\ell\frac{N_l}{w} - \sum_{l=1}^\ell\frac{n}{n_l}\cdot\frac{N_l}{w}=n\frac{N}{w} - \sum_{l=1}^\ell\frac{n}{n_l}\cdot\frac{N_l}{w}
\end{align}

Interestingly, combining Equations~\ref{eq:temp20}, \ref{eq:temp22}, and \ref{eq:temp23}, we show that $PoF<0$, for a CM with random hashing scheme:
\begin{align*}\label{eq:pof1}
PoF &= \mathcal{L}_{FCM} - \mathcal{L}_{CM}  \\`
&= n\frac{N}{w} - \sum_{l=1}^\ell\left(\frac{n}{n_l}\cdot\frac{N_l}{w}\right) - (n-1)\frac{N}{w}\\
&=\frac{N}{w}-\sum_{l=1}^\ell\frac{n}{n_l}\cdot\frac{N_l}{w}\\
&<\frac{N}{w}-\sum_{l=1}^\ell\frac{N_l}{w} \hspace{20mm} \text{\tt\scriptsize  //since $\frac{n}{n_l}>1, \forall g_l\in\Gee$}\\
&=\frac{N}{w}-\frac{N}{w}=0
\end{align*}

The negative PoF sounds counterintuitive, but as we shall further elaborate in our appendix
, it can be explained by the difference between the random distribution of the elements versus their uniform distribution to the buckets. We further show that the PoF of FCM is zero for $d=1$ when uniform hashing is used.


\subsection{$d>1$}
Next, we consider the price of fairness for multiple rows $d>1$.
The error of an element type $e_j\in\mathcal{U}$ is $$\min_{p\in[d]} \sum_{e\neq e_j:\mathsf{h}_p(e)=\mathsf{h}_p(e_j)}f(e),$$
so the expected additive error is
\begin{equation}\label{eq:expPriceFairnessd}
    \eps_A(e_j) = \ee\left[\min_{p\in[d]} \sum_{e\neq e_j:\mathsf{h}_p(e)=\mathsf{h}_p(e_j)}f(e)\right].
\end{equation}

For the standard CM sketch the total additive error is computed as $\mathcal{L}_{CM}=\sum_{j=1}^n\eps_A(e_j)$ using Equation~\eqref{eq:expPriceFairnessd} to compute the sum considering $w$ buckets.
For the FCM sketch the total additive error is computed as 
$\mathcal{L}_{FCM}=\sum_{l=1}^\ell\sum_{e_j\in \gee_l}\eps_A(e_j)$, where each sum $\sum_{e_j\in \gee_l}\eps_A(e_j)$ is computed using Equation~\eqref{eq:expPriceFairnessd} for $w_l$ buckets.
Since we do not have a closed form for the $w_l$ values (for $d>1$) from Equation~\eqref{eq:solve}, we cannot directly compute $PoF=\mathcal{L}_{CM}-\mathcal{L}_{FCM}$. Instead, we compute the PoF experimentally in real and synthetic datasets in Section~\ref{sec:pof_exp}.
Particularly, Table~\ref{tbl:negative_pof} and other experiments, including Figures~\ref{fig:pof_n_l_google} to \ref{fig:pof_n_groups_census} confirm the small PoF for $d>1$.

%% file: experiment.tex
\section{Experiments}\label{sec:exp}
In addition to the theoretical analysis, we perform extensive experiments across a range of settings to validate the fairness as well as the efficiency of our proposed sketch. Consistent with the theoretical guarantees established in previous sections, the experimental results confirm the effectiveness and efficiency of our sketch in practical scenarios.
\subsection{Experiments Setup}
The experiments were conducted on an Apple M2 Pro processor, 16 GB memory, running macOS. The algorithms were implemented in Python 3.12.
\subsection{Datasets}
For evaluation, we used two real-world datasets along with several synthetic ones. To assess the scalability of our proposed sketch in large-scale scenarios, we selected datasets that are sufficiently large to reflect real-world streaming environments. 

\stitle{Census} This dataset contains a one percent sample of the {\em Public Use Microdata Samples} person records derived from the complete 1990 census dataset. It includes around 2.5 million records with 68 categorical attributes. We use this dataset for experiments involving arbitrarily defined groups. Entity types are defined by concatenating the values of \texttt{iYearsch}, \texttt{dIndustry}, and the grouping attribute. For binary group experiments, the \texttt{iSex} attribute is used to define the groups resulting in 430 element types. To extend the grouping to $\ell$ groups, we utilize other attributes with higher cardinalities.

\stitle{Google N-Grams} This dataset is a large-scale collection of n-gram frequency counts extracted from texts in books digitized through the Google Books project. It has been used in related research to evaluate frequency estimation sketches. In our evaluations, we use a sample of 2-grams that begin with {\em a\_} to generate a stream of approximately 1.25 million element types, with a total frequency of around 5 million. We utilize this dataset for experiments in which groups are defined based on the frequency of elements associated with a given element type. If an element's frequency is below a specified threshold, it is assigned to group $l$; otherwise, it is assigned to group $h$. This approach is further extended to partition elements into $\ell$ frequency-based groups.

\stitle{Synthetic}
Finally, we generate several synthetic datasets in which the frequency of element types follows different distributions, including Gaussian, Zipfian, Exponential, and Uniform. We apply a frequency-based grouping to these datasets, similar to the approach used for the {\google} dataset. Each dataset contains $n=20,000$ element types, with the total frequency determined by the parameters of the underlying distribution. In our experiments on unfairness and the price of fairness, we report results for the case where frequencies are drawn from a Zipfian distribution.

\subsection{Baselines}
To demonstrate the effectiveness of our approach, we compare it against two baseline methods. The first baseline is the standard {\bf Count-Min} sketch, which we use across all of our experiments. The second baseline is the {\bf Row-Partitioning} approach discussed in Section~\ref{sec:row}, which, as expected, does not consistently achieve equal group-wise approximation factors in practice. We use this baseline in a subset of our experiments to illustrate the negative result of row-partitioning.

\subsection{Evaluation Plan}
We evaluate our proposed sketch based on three metrics 1) unfairness, 2) price of fairness and 3) efficiency. For each metric, we examine the impact of varying four parameters: the number of element types in the disadvantaged group $n_l$, the width of the sketch $w$, the depth of the sketch $d$, and the number of groups $\ell$. Due to space constraints and the confirmed similarity of results across different datasets, for certain experiments, we present results for a single dataset per setting, with the full results available in
appendix. Finally, we repeat each experiment five times and report the average values for each setting.

\input{plots1}

\subsection{Unfairness}
In this experiment, we measure the impact of varying certain parameters on the unfairness of the sketch. Unfairness is measured as the difference between the expected approximation factor of the disadvantaged and advantaged groups, i.e., $\ee[\alpha_l]-\ee[\alpha_h]$. In the multi-group setting, unfairness is measured as the difference between the maximum and minimum mean approximation factors across all groups, i.e., $\min_{\forall \gee_i\in \Gee }(\ee[\alpha_{\gee_i}]) - \max_{\forall \gee_j \in \Gee }(\ee[\alpha_{\gee_j}])$. A sketch is considered group-fair if the unfairness value is zero. 

\stitle{Varying group size} We begin with the experiment that examines the impact of varying the number of element types $n_l$ in the unpopular group on unfairness. To do so, we vary the cutoff threshold on the frequency of each element: elements with frequencies below the threshold are assigned to group $l$, while those above it are assigned to group $h$. As the threshold increases, $n_l$ grows while $n_h$ decreases, keeping the total number of elements $n$ constant.
Let us now focus on Figure~\ref{fig:unfairness_n_l_google}. As $n_l$ increases, the standard CM sketch exhibits greater unfairness because the expected approximation factor for group $l$ decreases as the group grows larger, while group $h$ becomes smaller. This is due to the standard CM sketch favoring more accurate estimates for the smaller group $h$, which contains high-frequency elements, over the larger group $l$, which consists of numerous low-frequency elements. For the Row-Partitioning baseline, the optimal integral values of $d_l$ and $d_h$ selected by the algorithm gradually deviate from the optimal fractional values determined by the equality~\ref{eq:rowd}. Specifically, with $d = 5$, the sketch would ideally assign a value $4 < d_l < 5$ (closer to 5) 
to group $l$, which is not feasible. As a result, we observe a significant disparity between the approximation factors of the groups for larger values of $n_l$. On the other hand, the FCM maintains the approximation factor difference near zero, regardless of the value of $n_l$. Results conveying a similar message for the {\synthetic} dataset are also presented in Figure~\ref{fig:unfairness_n_l_synthetic}.

\stitle{Varying the number of columns} Next, we examine the impact of varying the sketch width $w$ on unfairness. An instance of this experiment is illustrated in Figure~\ref{fig:unfairness_w_google}. Since $n_l = 1.2M$ is relatively large, the approximation factors for all groups and among all approaches are initially close to zero because the values of $w$ are comparatively small. For the standard CM baseline, as $w$ increases, the approximation factor for group $h$ gradually approaches 1, while it remains relatively low for group $l$. After surpassing a certain threshold, $w$ becomes large enough for the mean approximation factor of group $l$ to also approach 1. For the Row-Partitioning baseline, as $w$ initially increases, it struggles to find suitable integral $d_l$ values to allocate rows to group $l$. However, after surpassing a certain threshold, as the sketch becomes wide enough to accommodate fewer elements per bucket, the approximation factor difference begins to decrease.
As expected, FCM maintains a zero difference between the approximation factors of the two groups, regardless of the value of $w$.

\stitle{Varying the number of rows} Next, we evaluate the impact of increasing the sketch depth $d$ on unfairness. The results are presented in Figures~\ref{fig:unfairness_d_google} and~\ref{fig:unfairness_d_census}. For the standard CM baseline, the sketch initially provides more accurate estimates (approximation factors closer to 1) for group $h$. As $d$ increases, the accuracy improves for both groups, but at a slightly faster rate for group $h$. As before, FCM maintains zero unfairness regardless of the sketch depth. For the Row-Partitioning approach, we note some interesting observations in As shown in Figure~\ref{fig:unfairness_d_census}, depending on the value of $d$ and its divisibility, the unfairness fluctuates between nearly zero and a significant value as $d$ increases. In this experiment, $n_l$ and $n_h$ are very close, so as long as the sketch can assign equal $d_l$ and $d_h$ values to each group, it performs well w.r.t keeping unfairness at zero. This occurs only when $d$ takes even values.

\stitle{Varying the number of groups} In our final experiment on unfairness, we investigate the impact of increasing the number of groups. The results are shown in Figures~\ref{fig:unfairness_n_groups_google} and~\ref{fig:unfairness_n_groups_census}. Having demonstrated that the Row-Partitioning baseline performs inconsistently and is highly sensitive to the value of $d$, we henceforth compare our approach only with the standard CM sketch. As expected, FCM maintains zero unfairness, independent of the number of groups or the grouping strategy used. However, for the standard CM, an increase in the number of groups generally leads to higher unfairness. This is particularly noticeable in the case of equi-width frequency-based grouping, where the sketch is more accurate for smaller groups with high-frequency elements and less accurate for those larger groups with low-frequency elements (Figure~\ref{fig:unfairness_n_groups_google}). In the case of arbitrary grouping, unfairness is entirely dependent on the size of each group. As shown in Figure~\ref{fig:unfairness_n_groups_census}, the number of groups is determined based on a different attribute with specific cardinality at each iteration. While the general trend is an increase in unfairness as the number of groups grows, in some iterations (e.g., 5 and 10), the unfairness values are closer, indicating that the groups are more evenly sized. In contrast, for iteration 9, one group is significantly larger or smaller than the others, resulting in higher unfairness.

\subsection{Price of Fairness}\label{sec:pof_exp}
Thus far, we have established that FCM guarantees group-fairness regardless of the parameters involved. In the next set of experiments, we examine the cost of enforcing this constraint compared to the standard CM. Following the discussion in Section~\ref{sec:price} and Equation~\ref{eq:temp20}, we measure the price of fairness in terms of the total additive error required to achieve fairness.

First, let us empirically validate the interesting result that PoF is negative for $d=1$, 
proven in Section~\ref{sec:price}. 
To do so, we use a synthetic dataset where the frequencies of groups $l$ and $h$ are drawn from Gaussian distributions with $\mu_l = 100, \sigma_l = 50$ and $\mu_h = 1000, \sigma_h = 500$, respectively, and a total of $n = 10{,}000$ element types, we vary the unpopular group size $n_l$ from 9000 to 1000. The price of fairness is computed as the difference in total additive error between FCM and standard CM sketches ($\text{PoF} = \mathcal{L}_{\text{FCM}} - \mathcal{L}_{\text{CM}}$). The results, along with the empirical and theoretical additive errors shown in Table~\ref{tbl:negative_pof}, emphasize the superiority of FCM over standard CM in terms of additive error when $d = 1$. 

However, when $d > 1$, FCM incurs a positive but small price of fairness, and is generally outperformed by standard CM. We repeated the above experiment with $d=1$, and the results in Table~\ref{tbl:negative_pof} further support our findings. 
Still, we notice that the PoF is much smaller in the real datasets compared to the synthetic dataset. That is because synthetic frequency values are generated using a Gaussian distribution, resulting in larger values with lower variance.
This is further clear when comparing Table~\ref{tbl:negative_pof} with Figure~\ref{fig:pof_n_l_google} -- the results on the {\google} dataset, where one can confirm the small gap between the total additive error of the CM (the blue line) with FCM (the orange line).
To further evaluate the price of fairness for $d>1$ across various settings, similar to the previous section, we varied (a) group size (Figure~\ref{fig:pof_n_l_google}), (b) number of columns (Figure~\ref{fig:pof_w_google}), (c) number of rows (Figures~\ref{fig:pof_d_google} and~\ref{fig:pof_d_census}), and (d) the number of groups (Figures~\ref{fig:pof_n_groups_google} and~\ref{fig:pof_n_groups_census}), and measured the total additive error for CM versus FCM.
The results across all settings were consistent, conveying a unified message: FCM achieves group-fairness at a negligible, if any, price of fairness. Due to the space limitations, extensive results are moved to the 
appendix. We confirm that we observed similar results across all settings and all other datasets.

\begin{table*}[htbp]
\small
\centering
\begin{tabular}{|c|c|c|r|r|r|r|r|r|r|r|r|}
\hline
\multicolumn{3}{|c|}{\bf Total Additive Error} & \multicolumn{9}{c|}{\textbf{$n_l (n=10{,}000)$}} \\
\cline{4-12}
 \multicolumn{3}{|c|}{} & \textbf{9000} & \textbf{8000} & \textbf{7000} & \textbf{6000} & \textbf{5000} & \textbf{4000} & \textbf{3000} & \textbf{2000} & \textbf{1000}\\
\hline
 & \multirow{2}{*}{theoretical} & \textbf{CM}  & 19,047,019 & 28,054,816 & 37,081,001 & 46,092,289 & 55,121,050 & 64,147,002 & 73,186,882 & 82,323,243 & 91,265,426 \\
\multirow{5}{*}{\rotatebox{90}{$d=1$}} &                            & \textbf{FCM} & 18,985,224 & 28,028,987 & 37,003,399 & 46,050,674 & 55,115,538 & 64,089,586 & 73,165,549 & 82,249,909 & 91,228,281 \\
\cline{2-12}
 & \multirow{3}{*}{empirical}  & \textbf{CM}  & 19,039,343 & 28,085,806 & 37,053,427 & 46,096,489 & 55,113,194 & 64,234,176 & 73,227,052 & 82,283,374 & 91,328,074 \\
 &                             & \textbf{FCM} & 18,991,509 & 27,982,573 & 37,002,246 & 46,071,394 & 55,086,890 & 64,133,815 & 73,168,189 & 82,230,649 & 91,276,271 \\
\cline{3-12}
 &                  & \textbf{PoF} &  \downarrowgreen  -47,834   &\downarrowgreen -103,233   &\downarrowgreen  -51,181   &\downarrowgreen  -25,095   &\downarrowgreen  -26,304   &\downarrowgreen -100,361   &\downarrowgreen  -58,863   &\downarrowgreen  -52,725   &\downarrowgreen  -51,803   \\
\hline
 \multirow{3}{*}{\rotatebox{90}{$d=5$}}& \multirow{3}{*}{empirical}  & \textbf{CM}&  7,964,348& 11,572,548& 16,458,026& 22,023,181& 28,305,699& 34,442,308& 40,959,703& 47,230,290& 54,257,770  \\
 &                             & \textbf{FCM} &  11,695,556& 17,114,933& 22,666,115& 28,053,739& 33,856,154& 39,448,942& 44,913,211& 50,089,698& 55,893,637  \\
\cline{3-12}
 &                  & \textbf{PoF} &  \uparrowred 3,731,208     &\uparrowred 5,542,385    &\uparrowred 6,208,089     &\uparrowred 6,030,558    &\uparrowred 5,550,455    &\uparrowred 5,006,634    &\uparrowred 3,953,508    &\uparrowred 2,859,408    &\uparrowred 1,635,867    \\
\hline

\end{tabular}
\caption{comparison of CM and FCM w.r.t additive errors and the price of fairness across different $n_l$ values for $d=1$ and $d=5$.}
\label{tbl:negative_pof}
\end{table*}

\input{plots2}

\subsection{Efficiency}
Having established the efficacy of the FCM sketch, we now turn our attention to its efficiency. A frequency-estimation sketch involves two main stages: (1) construction and (2) query time. In the following, we evaluate the impact of varying key parameters on the time required for each stage and provide a comparison with the standard CM sketch:
\subsubsection{Construction Time}
We define the construction time as the duration required to initialize the sketch and insert the entire stream into it. In summary, the construction time of FCM is comparable to that of the standard CM. For both sketches, the construction time is independent of the disadvantaged group size $n_l$ (Figure~\ref{fig:construction_n_l_google}), the sketch width $w$ (Figure~\ref{fig:construction_w_google}), and the number of groups $\ell$ (Figure~\ref{fig:construction_n_groups_google}); varying these parameters has no impact on construction time. The construction time for both sketches grows linearly with the sketch depth $d$, as expected, since it corresponds to the number of times each element is inserted into the sketch. This result is presented in Figure~\ref{fig:construction_d_google}.
\subsubsection{Query Time}
Query time is defined as the time required for the sketch to perform an estimate operation for a given element. The query time results exhibit similar patterns to the construction times, with FCM and standard CM showing comparable values and trends. The query times are independent of the disadvantaged group size $n_l$ (Figure~\ref{fig:query_n_l_google}), the sketch width $w$ (Figure~\ref{fig:query_w_google}), and the number of groups $\ell$ (Figure~\ref{fig:query_n_groups_google}). Similarly, the query time for both sketches also increases linearly with the sketch depth $d$, as shown in Figure~\ref{fig:query_d_google}.

\subsection{Validating Group Columns Allocation}
In our final experiment, we evaluate 
experimentally validate our mathematical analysis in Section~\ref{sec:d}, using a Monte Carlo method for estimating the value of $w_l$ -- the number of columns allocated to the group $l$.

We note that instead of the mathematical derivation, one could develop the following Monte Carlo method based on repeated sampling to estimate the value of $w_l$: 

Consider $n_l$, $n_h$, $w$, $d$, and the parameters of arbitrary distributions from which the frequencies of groups $l$ and $h$ are drawn. For each group $\gee \in \{l, h\}$, we draw $n_\gee$ samples from the corresponding distribution to serve as the frequencies of the elements. As a result, we obtain a list of $n$ tuples of the form {\tt (frequency, group)}. Next, we perform a binary search over the range of $w$ to find the optimal value of $w_l$ that minimize the empirical difference in approximation factors between the two groups, where the empirical difference is evaluated by inserting the sampled frequencies into a FCM sketch with 1 row and measuring the approximation factors for each group. This procedure is repeated $d$ times (once for each sketch row), and we compute the average of the resulting $w_l$ values. Finally, we repeat the entire process over multiple iterations and report the overall average of $w_l$.

\begin{figure}[h]
    \centering
    \begin{minipage}[t]{0.8\linewidth}
    \centering
    \includegraphics[width=\textwidth]{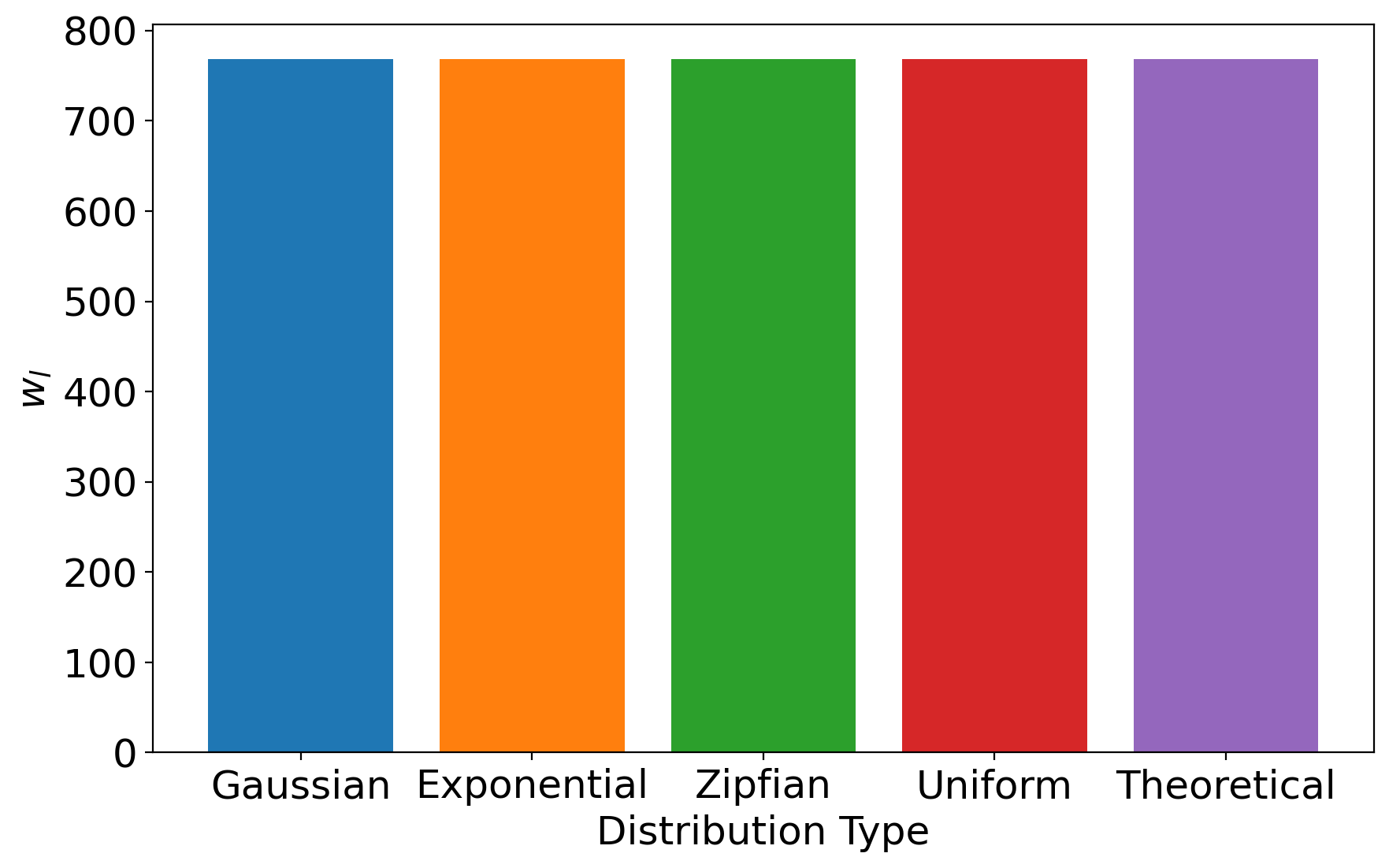}
    \vspace{-2.5em}
    \caption{Monte Carlo-based approach to finding $w_l$ vs. theoretical value. $w_l$ only depends on $n_l$ and is independent of the distribution of frequencies of each group.}
    \label{fig:mc_comparison}
\end{minipage}
\end{figure}

We use this Monte Carlo method to estimate $w_l$ under a range of synthetic distributions. 
We note from our analysis that the value of $w_l$ does not depend on the distribution of data. Hence, we expect to observe similar results from different runs of the Monte Carlo estimation.
We then compare the number of columns assigned to each group by this method to the theoretical values derived from the analysis in Section~\ref{sec:d}. 
 The results are shown in Figure~\ref{fig:mc_comparison}. The values of $w_l$ were identical across different frequency distributions and, as expected, depended solely on the values of $n_l$ and $n_h$.

%% file: plots1.tex
\begin{figure*}[!tb] 
\centering
    \includegraphics[width=0.75\textwidth]{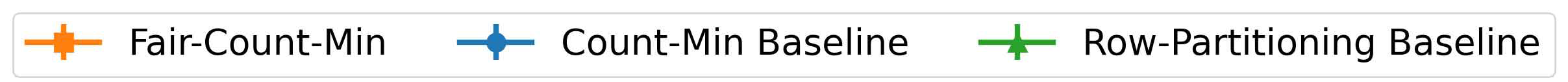}
    \begin{minipage}[t]{0.32\linewidth}
        \centering
        \includegraphics[width=\textwidth]{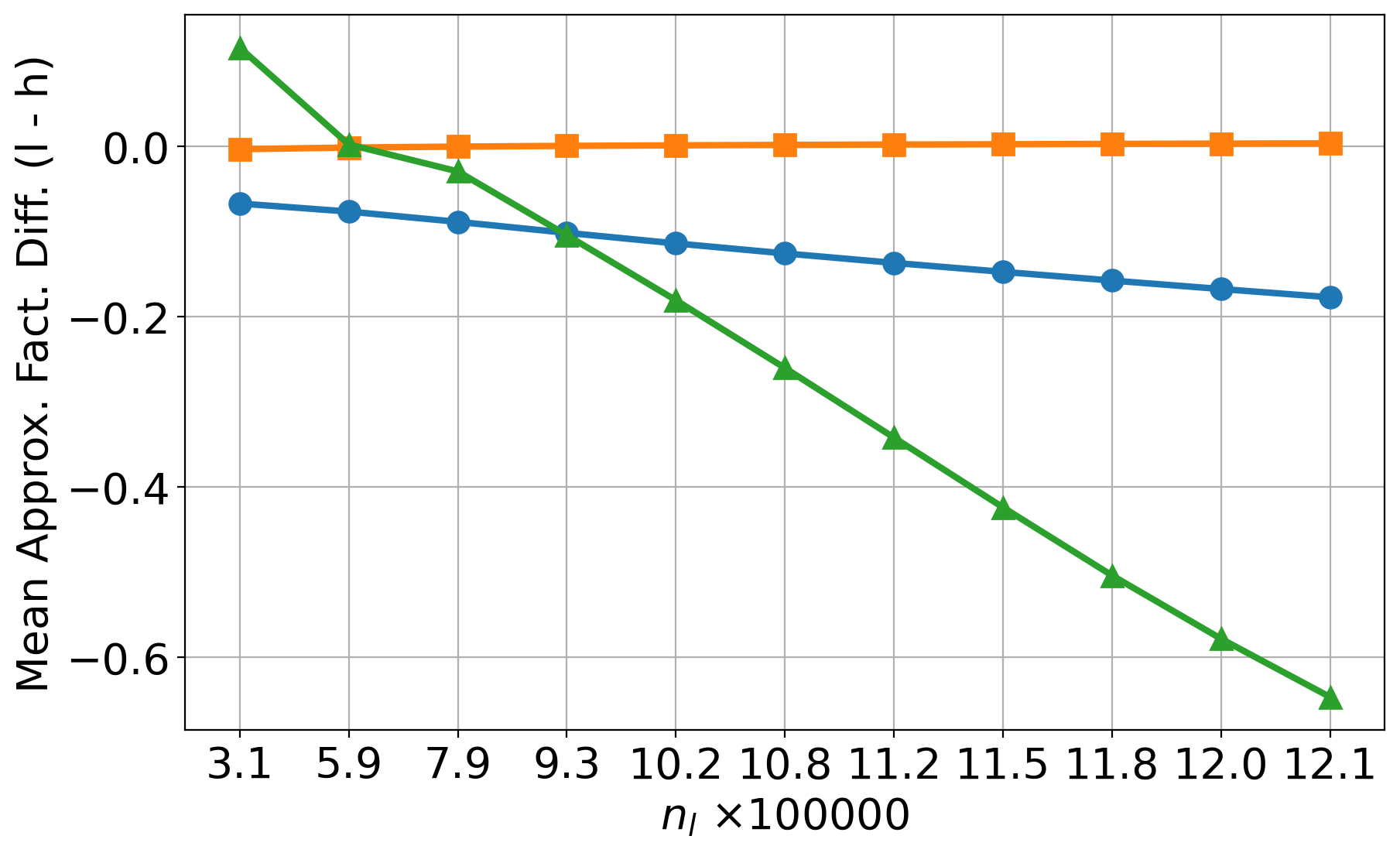}
        \vspace{-1.5em}
        \caption{effect of varying disadvantaged group size $n_l$ on unfairness, \google, $w=65536, d=5$.}
        \label{fig:unfairness_n_l_google}
    \end{minipage}
    \hfill
    \begin{minipage}[t]{0.32\linewidth}
        \centering
        \includegraphics[width=\textwidth]{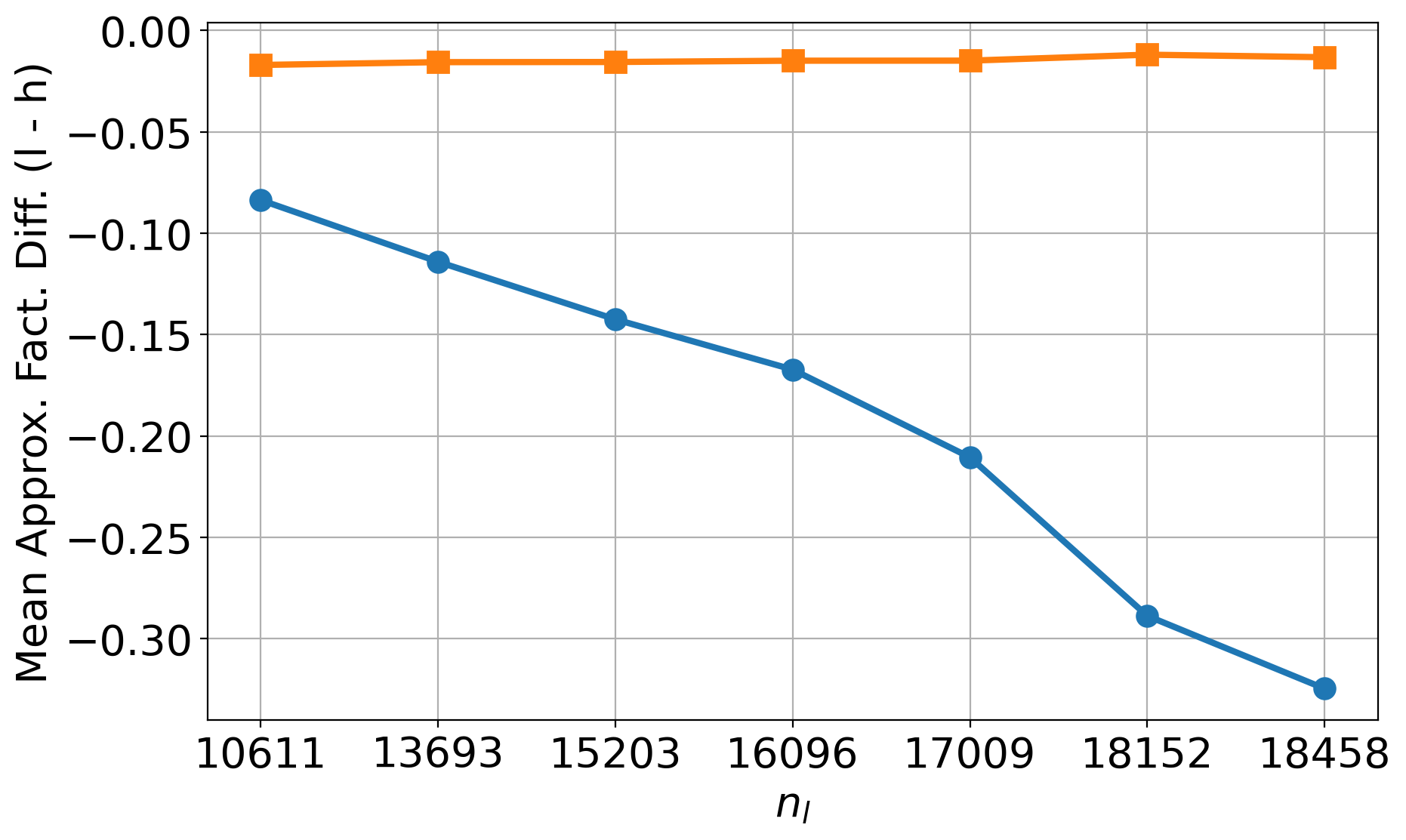}
        \vspace{-1.5em}
        \caption{effect of varying disadvantaged group size $n_l$ on unfairness, \synthetic, $n=20K,w=512,d=10$.}
        \label{fig:unfairness_n_l_synthetic}
    \end{minipage}
    \hfill
    \begin{minipage}[t]{0.32\linewidth}
        \centering
        \includegraphics[width=\textwidth]{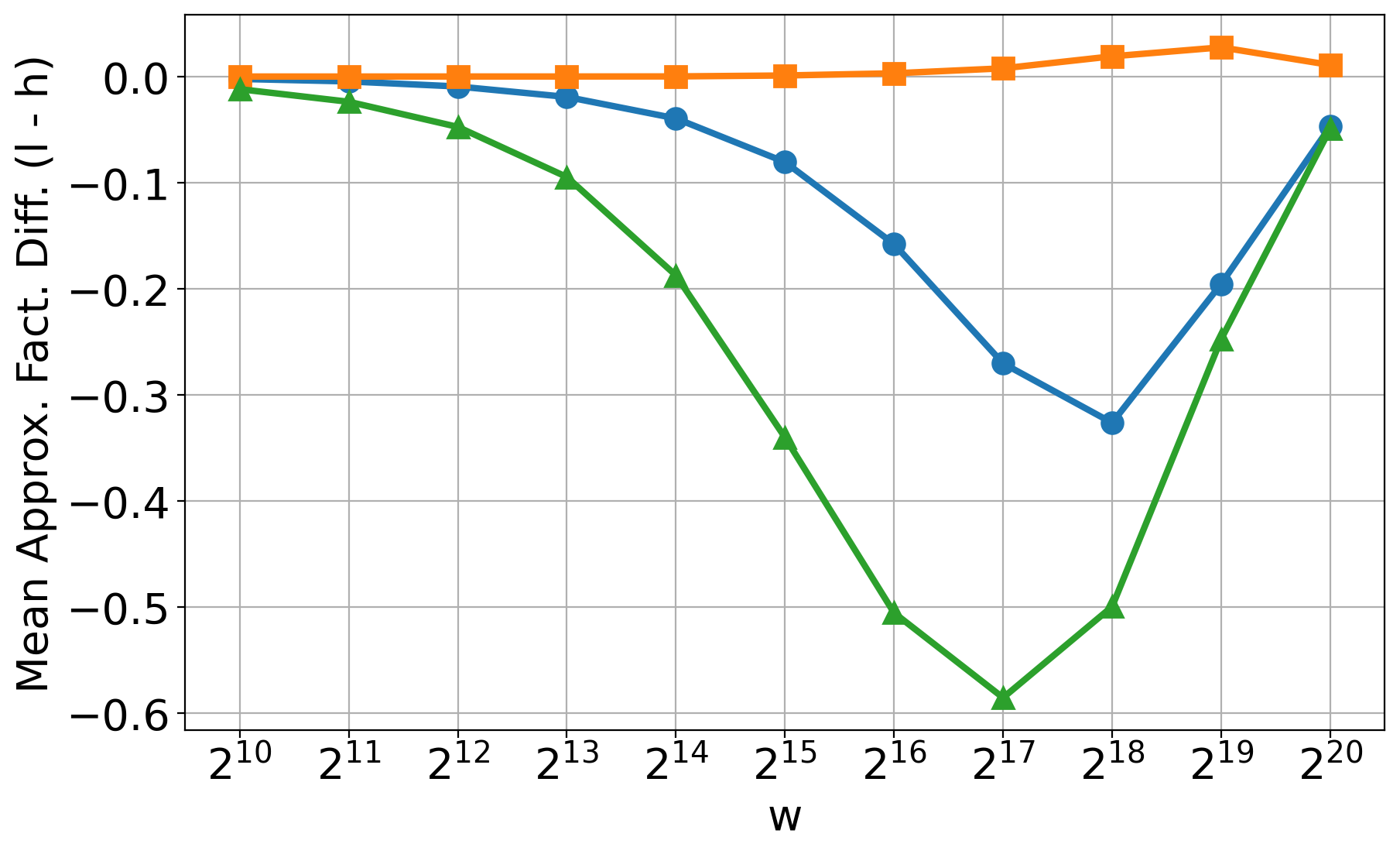}
        \vspace{-1.5em}
        \caption{effect of varying sketch width $w$ on unfairness, \google, $n=1.2M, d=5$.}
        \label{fig:unfairness_w_google}
    \end{minipage}
    \hfill
    \begin{minipage}[t]{0.32\linewidth}
        \centering
        \includegraphics[width=\textwidth]{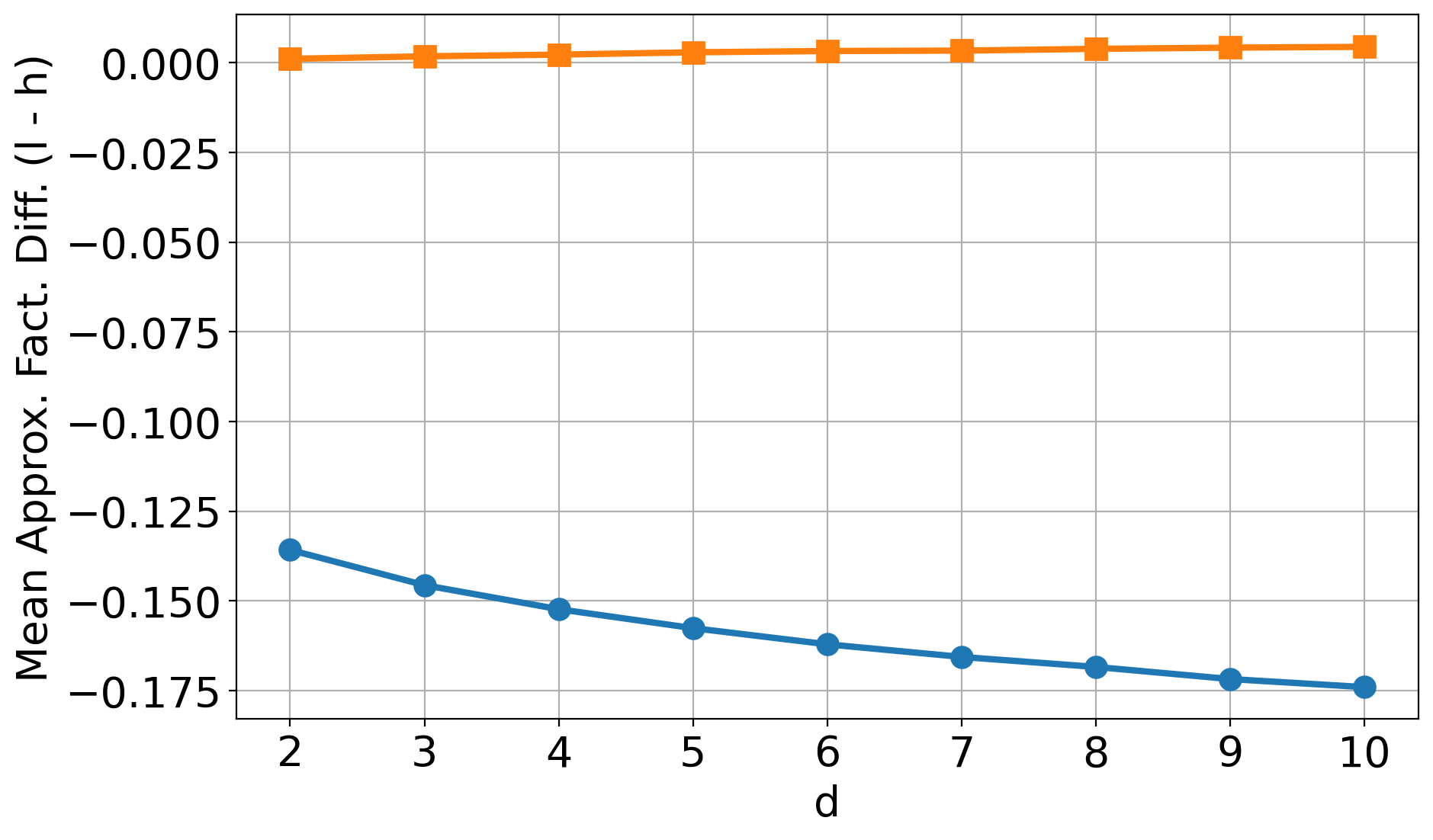}
        \vspace{-1.5em}
        \caption{effect of varying sketch depth $d$ on unfairness, \google, $n=1.2M, w=65536$.}
        \label{fig:unfairness_d_google}
    \end{minipage}
    \begin{minipage}[t]{0.32\linewidth}
        \centering
        \includegraphics[width=\textwidth]{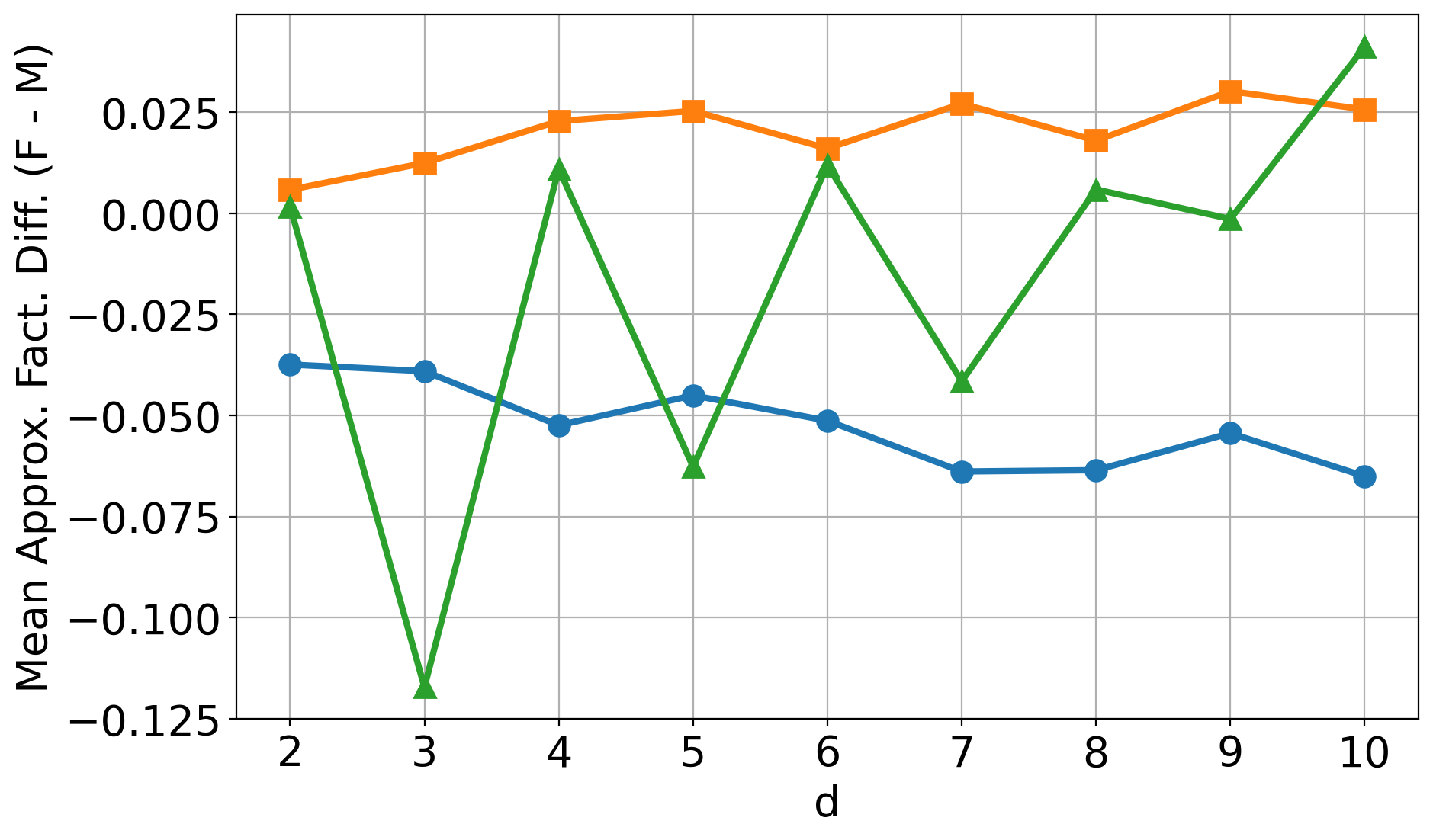}
        \vspace{-1.5em}
        \caption{effect of varying sketch depth $d$ on unfairness, \census $n=430, w=64$.}
        \label{fig:unfairness_d_census}
    \end{minipage}
    \hfill
    \begin{minipage}[t]{0.32\linewidth}
        \centering
        \includegraphics[width=\textwidth]{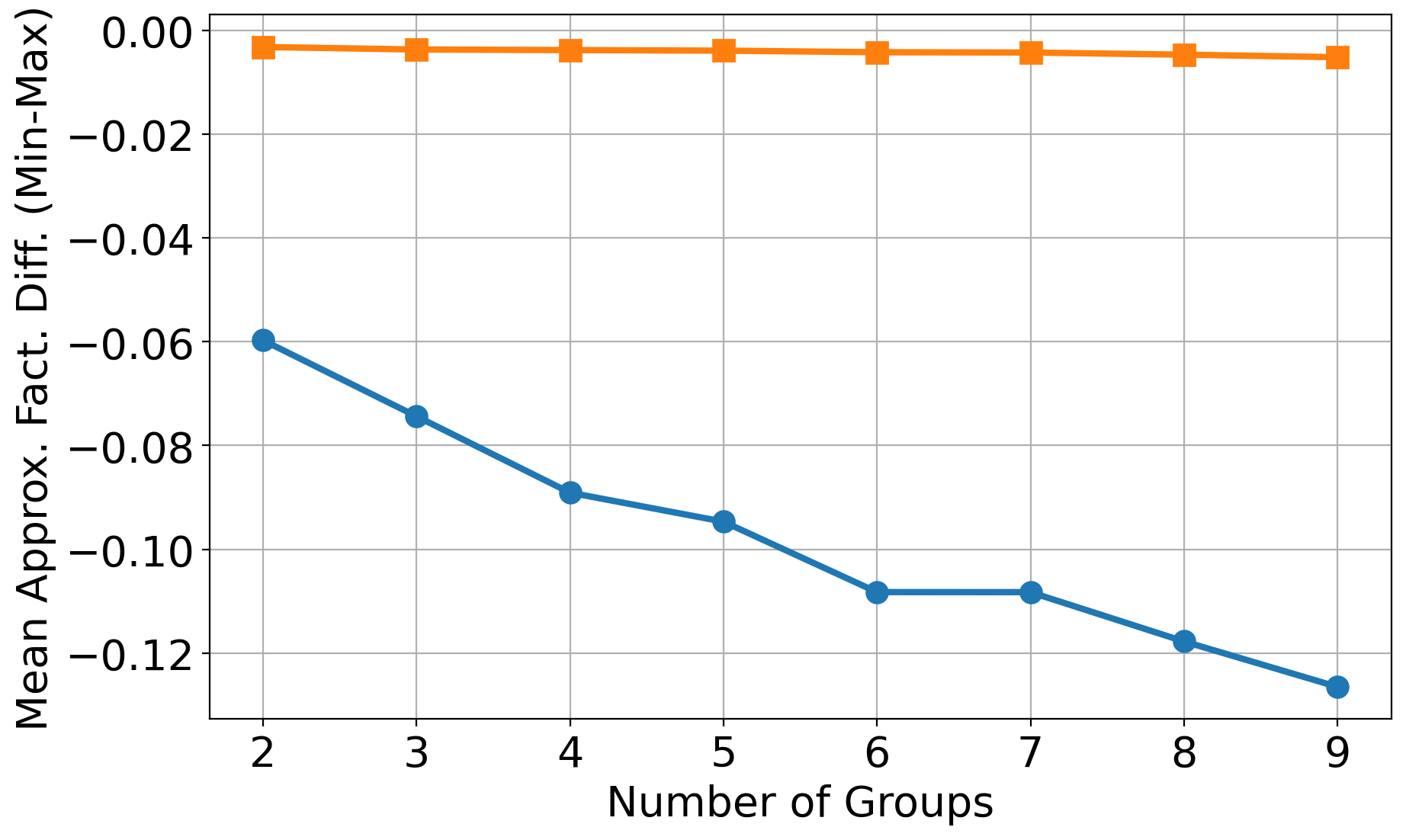}
        \vspace{-1.5em}
        \caption{effect of varying number of groups $\ell$ on unfairness, \google, $n=1.2M,w=65536,d=5$.}
        \label{fig:unfairness_n_groups_google}
    \end{minipage}
    \hfill
    \begin{minipage}[t]{0.32\linewidth}
        \centering
        \includegraphics[width=\textwidth]{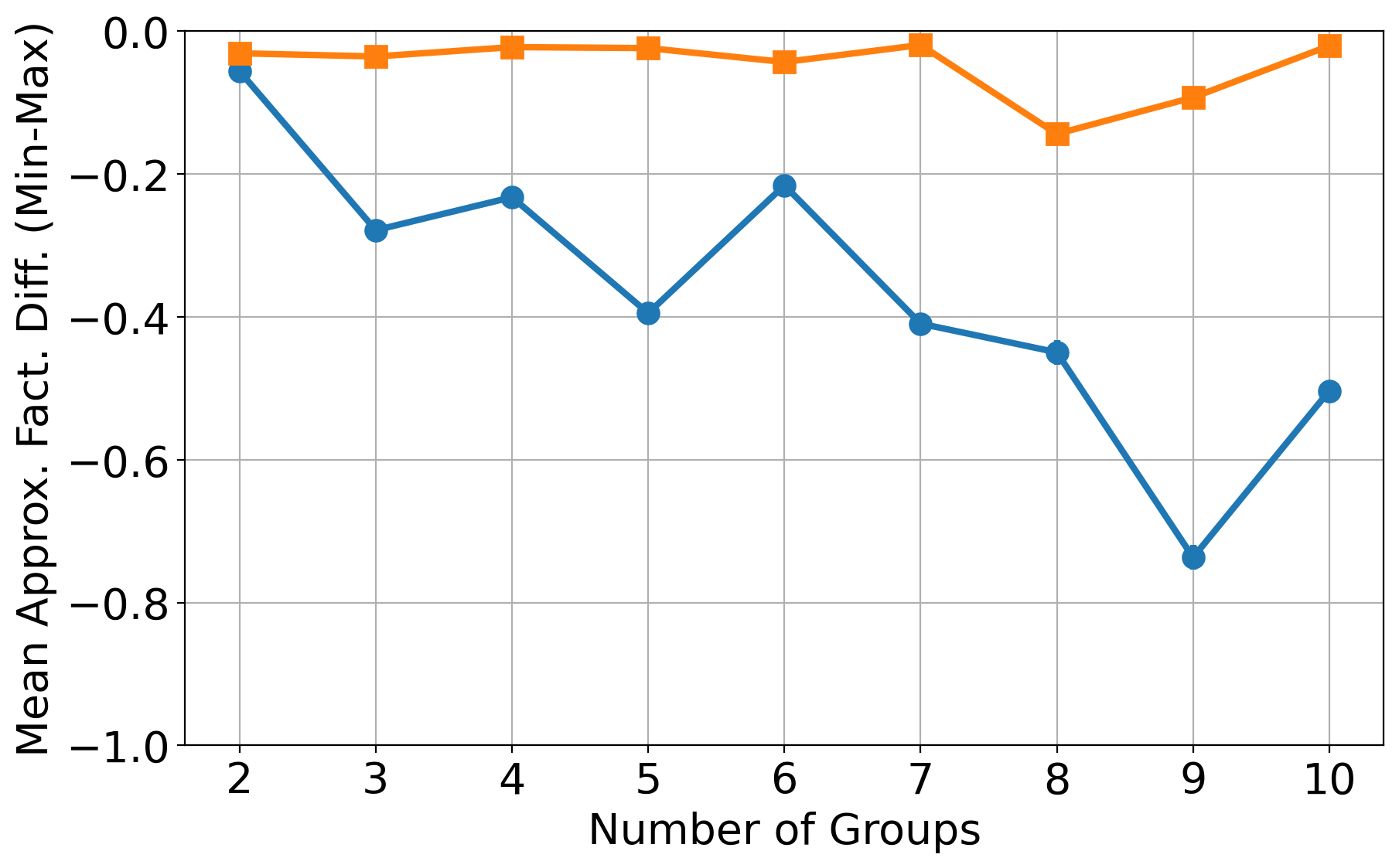}
        \vspace{-1.5em}
        \caption{effect of varying number of groups $\ell$ on unfairness, \census, $n=430,w=64,d=10$.}
        \label{fig:unfairness_n_groups_census}
    \end{minipage}
    \begin{minipage}[t]{0.32\linewidth}
        \centering
        \includegraphics[width=\textwidth]{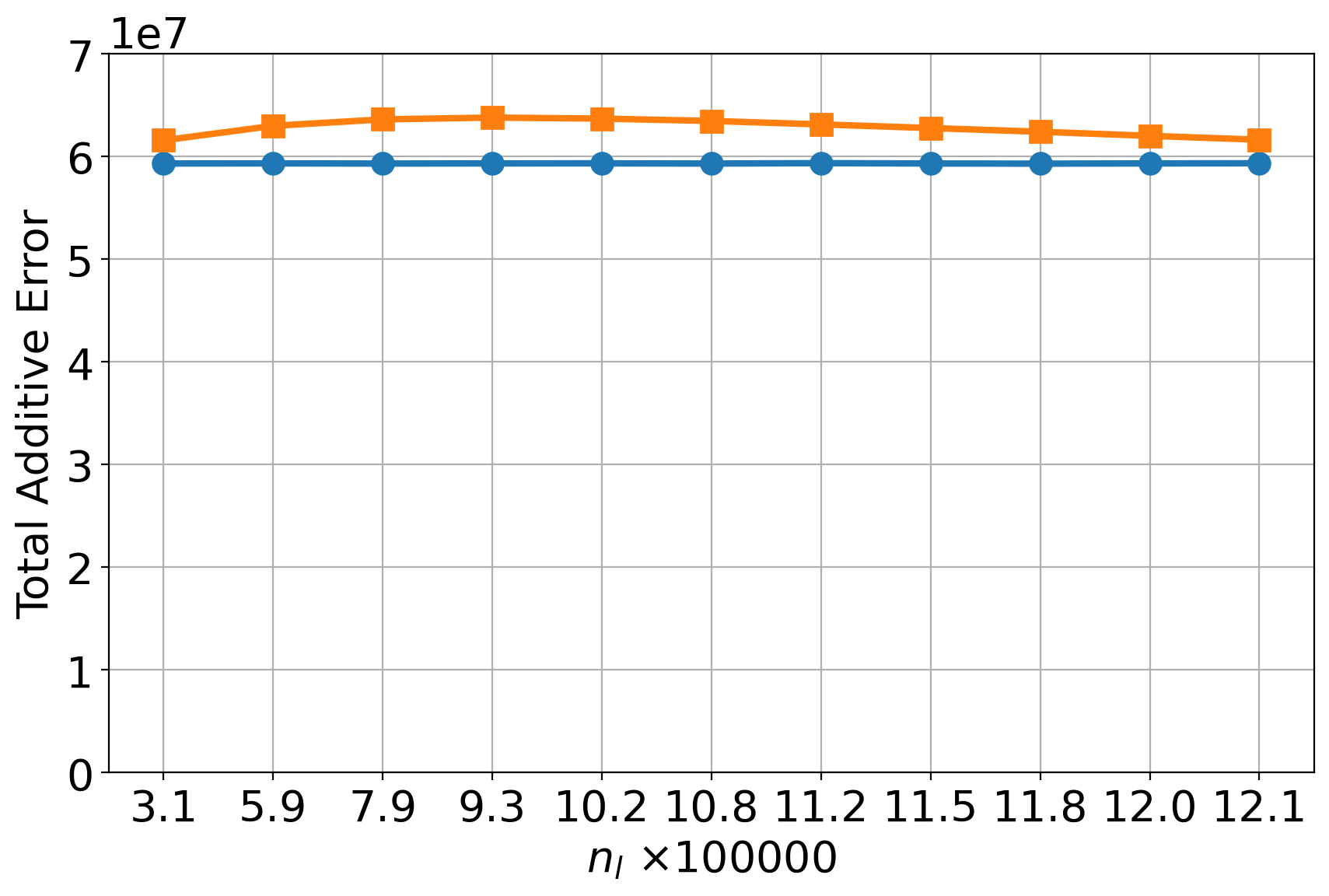}
        \vspace{-1.5em}
        \caption{effect of varying disadvantaged group size $n_l$ on price of fairness, \google, $w=65536, d=5$.}
        \label{fig:pof_n_l_google}
    \end{minipage}
    \hfill
    \begin{minipage}[t]{0.32\linewidth}
        \centering
        \includegraphics[width=\textwidth]{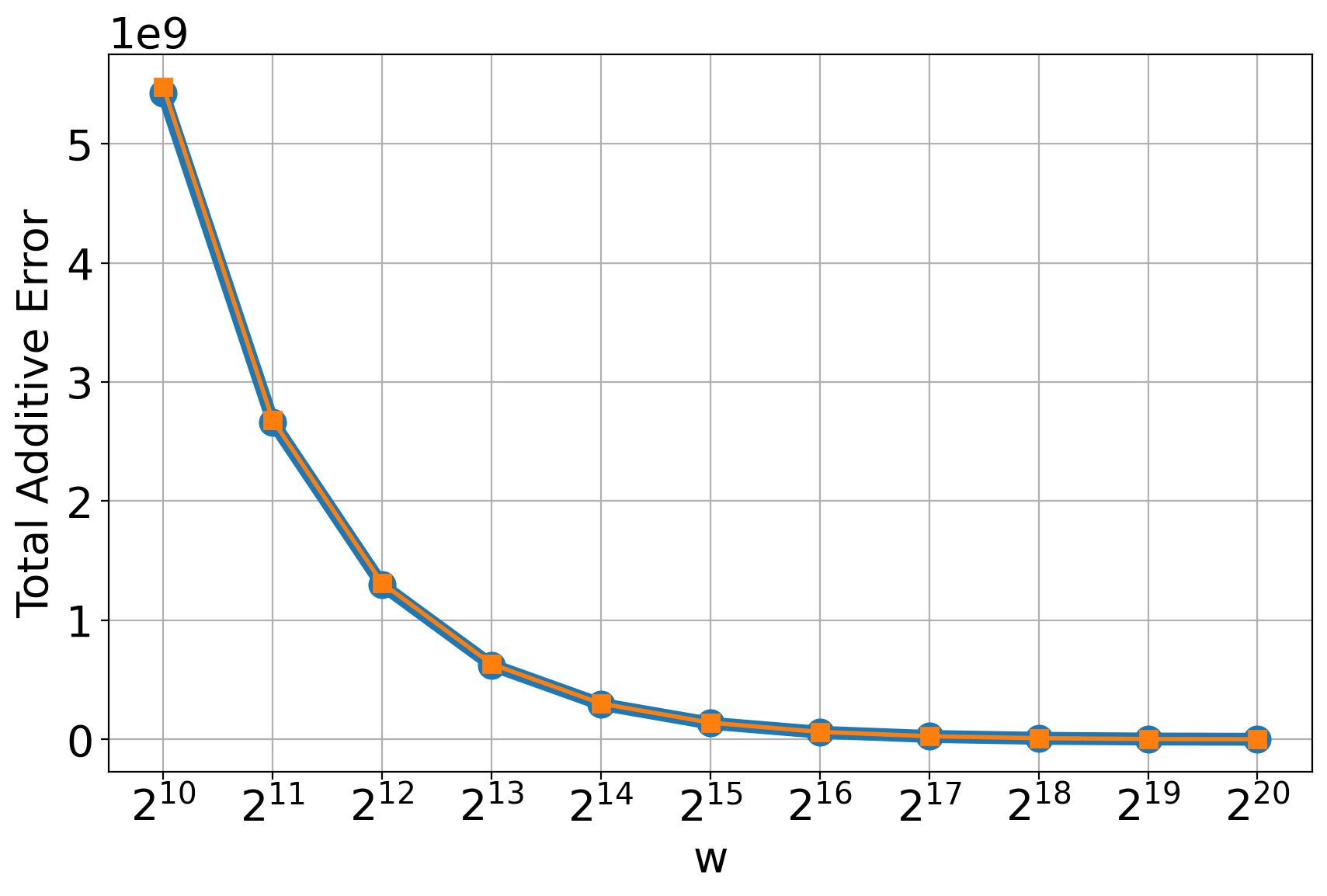}
        \vspace{-1.5em}
        \caption{effect of varying sketch width $w$ on price of fairness, \google, $n=1.2M, d=5$.}
        \label{fig:pof_w_google}
    \end{minipage}
    \begin{minipage}[t]{0.32\linewidth}
        \centering
        \includegraphics[width=\textwidth]{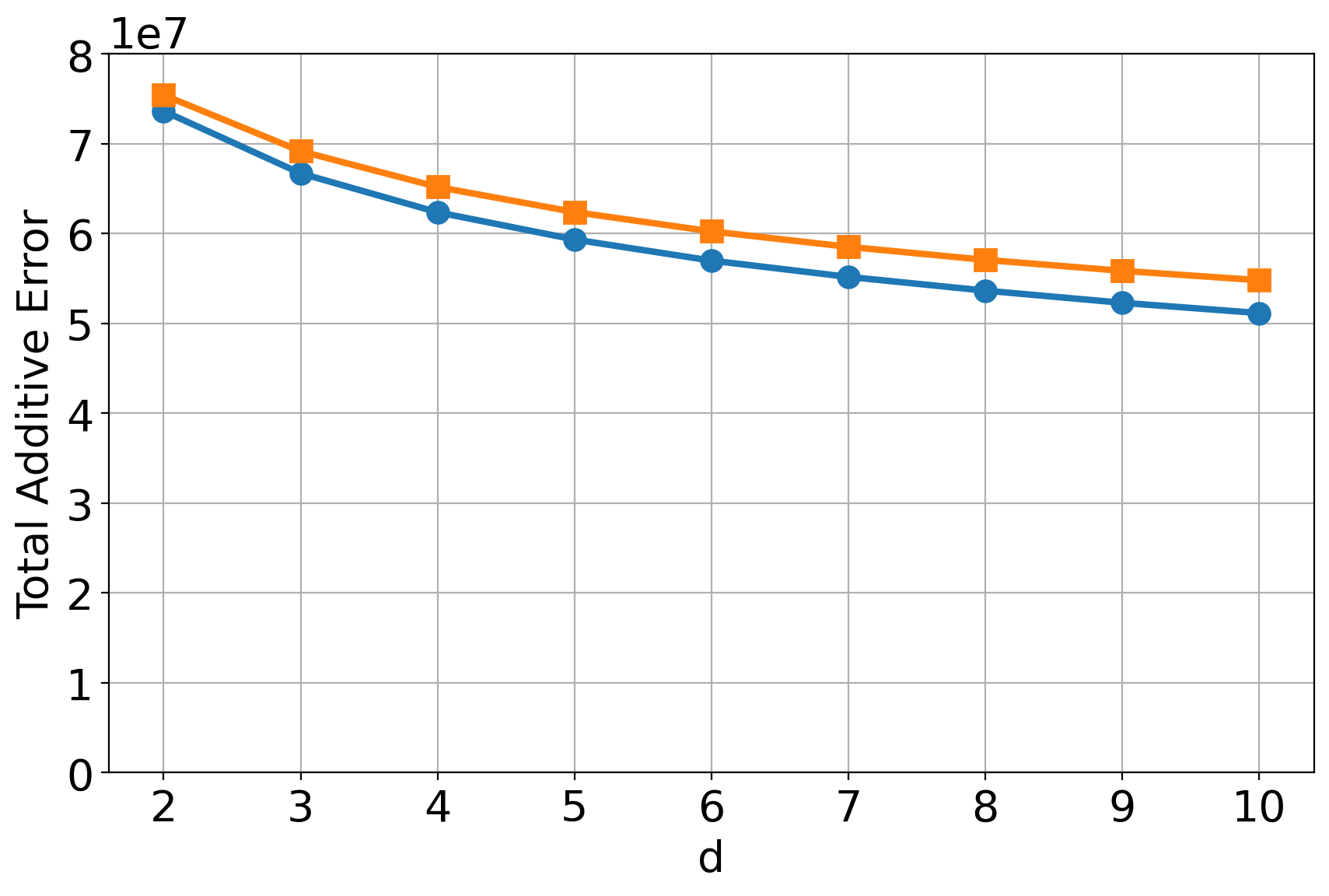}
        \vspace{-1.5em}
        \caption{effect of varying sketch depth $d$ on price of fairness, \google, $n=1.2M, w=65536$.}
        \label{fig:pof_d_google}
    \end{minipage}
    \hfill
    \begin{minipage}[t]{0.32\linewidth}
        \centering
        \includegraphics[width=\textwidth]{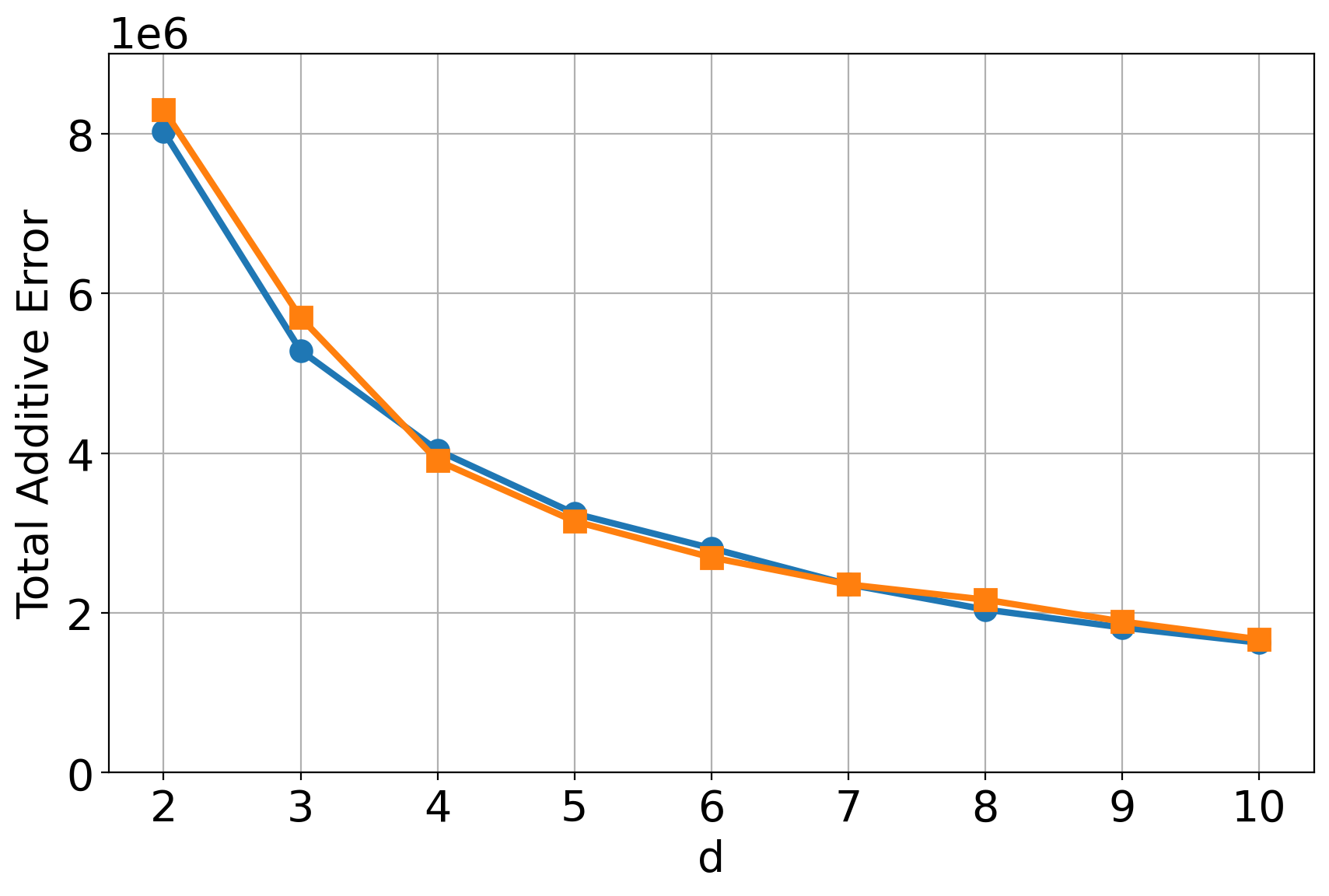}
        \vspace{-1.5em}
        \caption{effect of varying sketch depth $d$ on price of fairness, \census, $n=430, w=64$.}
        \label{fig:pof_d_census}
    \end{minipage}
    \hfill
    \begin{minipage}[t]{0.32\linewidth}
        \centering
        \includegraphics[width=\textwidth]{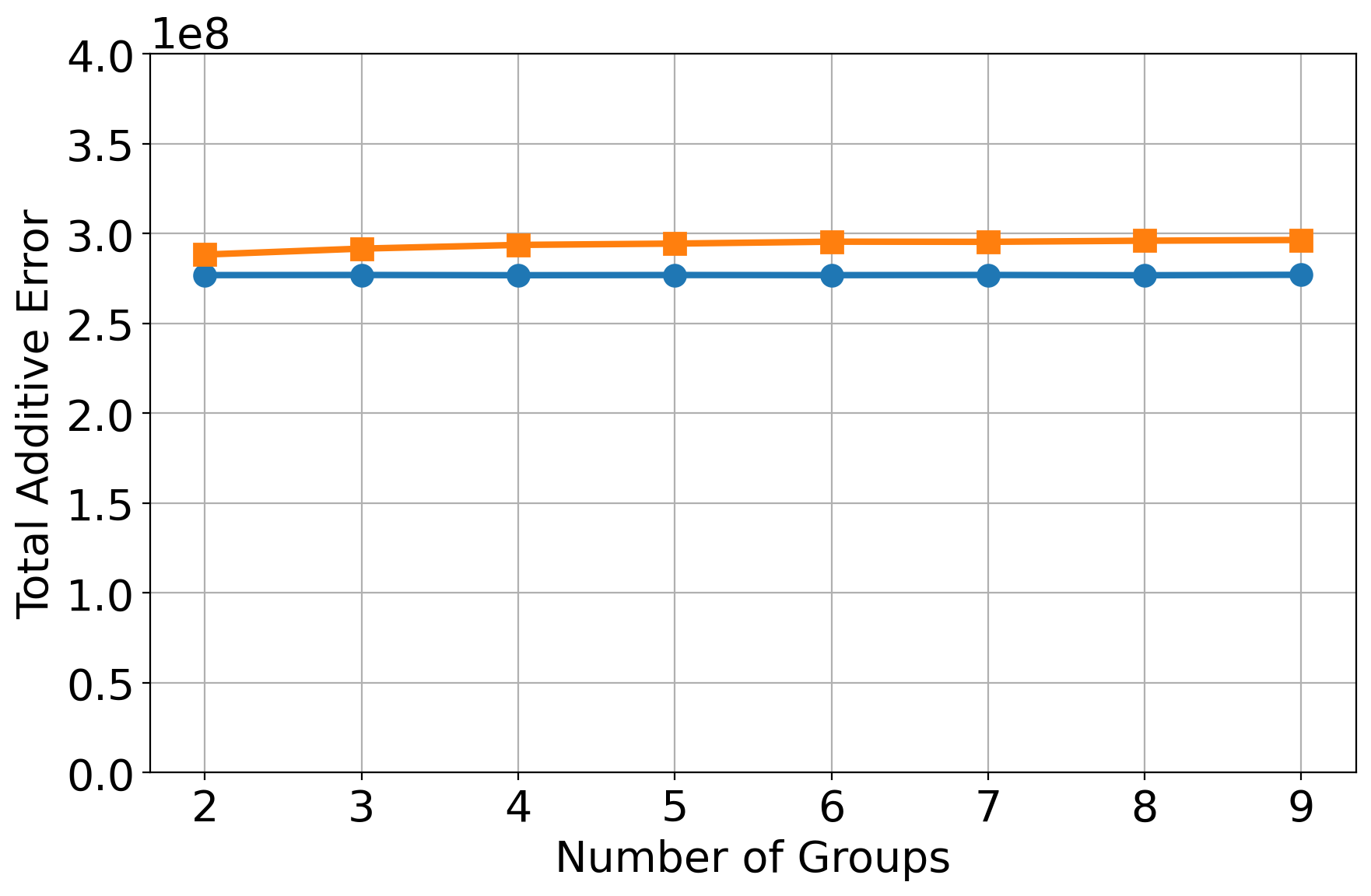}
        \vspace{-1.5em}
        \caption{effect of varying number of groups $\ell$ on price of fairness, \protect\google, $n=1.2M,w=65536,d=5$\protect\footnotemark.}
        \label{fig:pof_n_groups_google}
    \end{minipage}
\end{figure*}

\footnotetext{For this experiment, we used a larger sample of 18 million elements as the stream to ensure that all groups contained a meaningful number of element types. This accounts for the larger additive error values observed in this dataset compared to the other experiments.}

%% file: plots2.tex
\begin{figure*}
    \centering
    \includegraphics[width=0.45\textwidth]{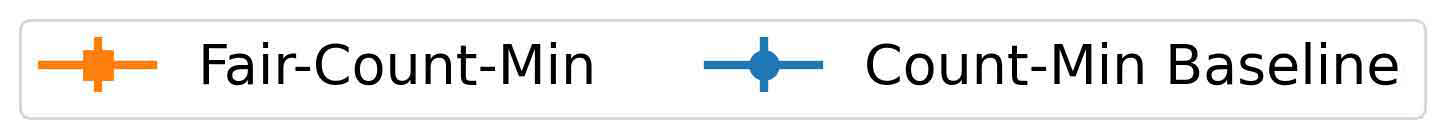}\\
    \begin{minipage}[t]{0.32\linewidth}
        \centering
        \includegraphics[width=\textwidth]{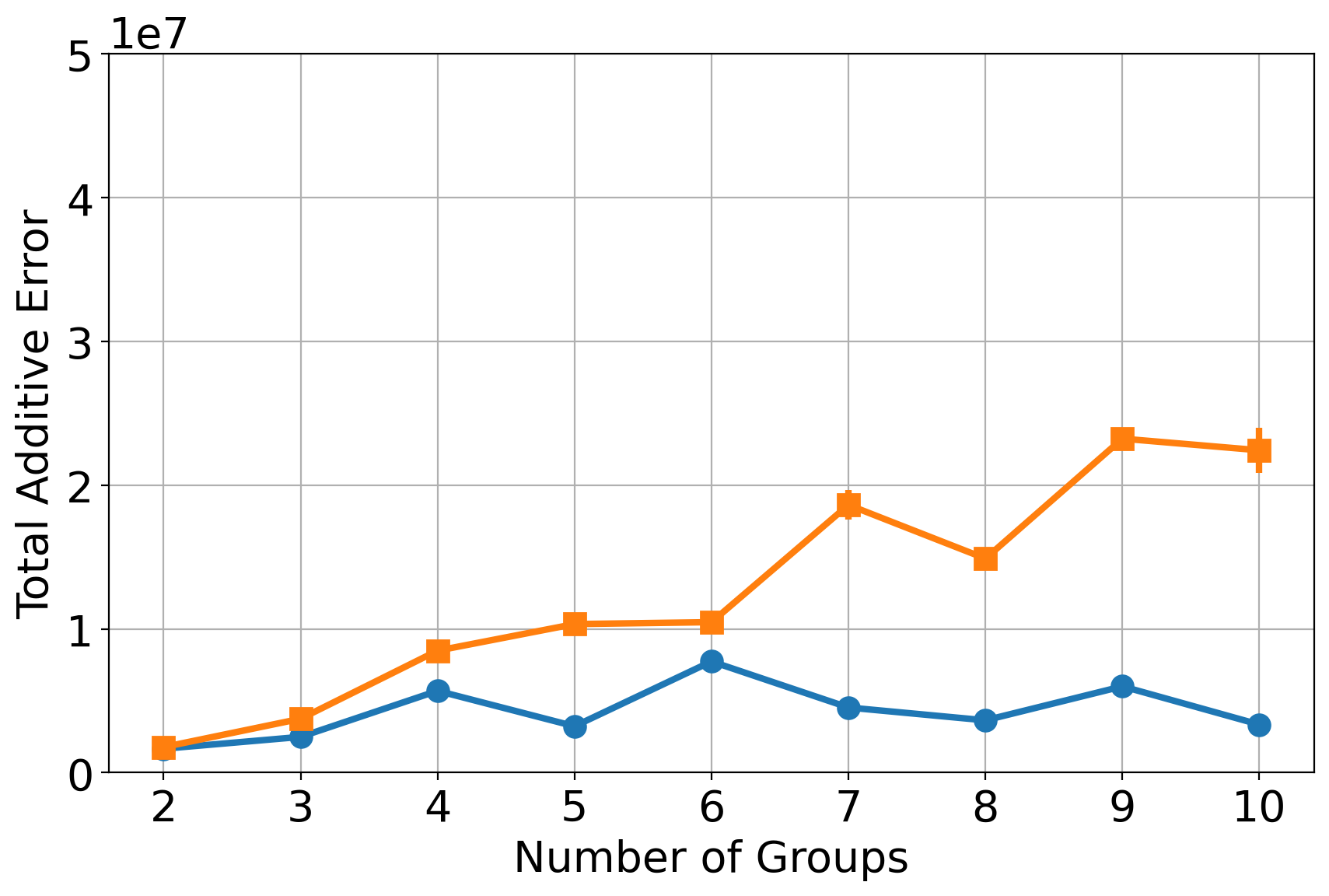}
        \vspace{-1.5em}
        \caption{effect of varying number of groups $\ell$ on price of fairness, \census, $n=430,w=64,d=10$.}
        \label{fig:pof_n_groups_census}
    \end{minipage}
    \hfill
    \begin{minipage}[t]{0.32\linewidth}
        \centering
        \includegraphics[width=\textwidth]{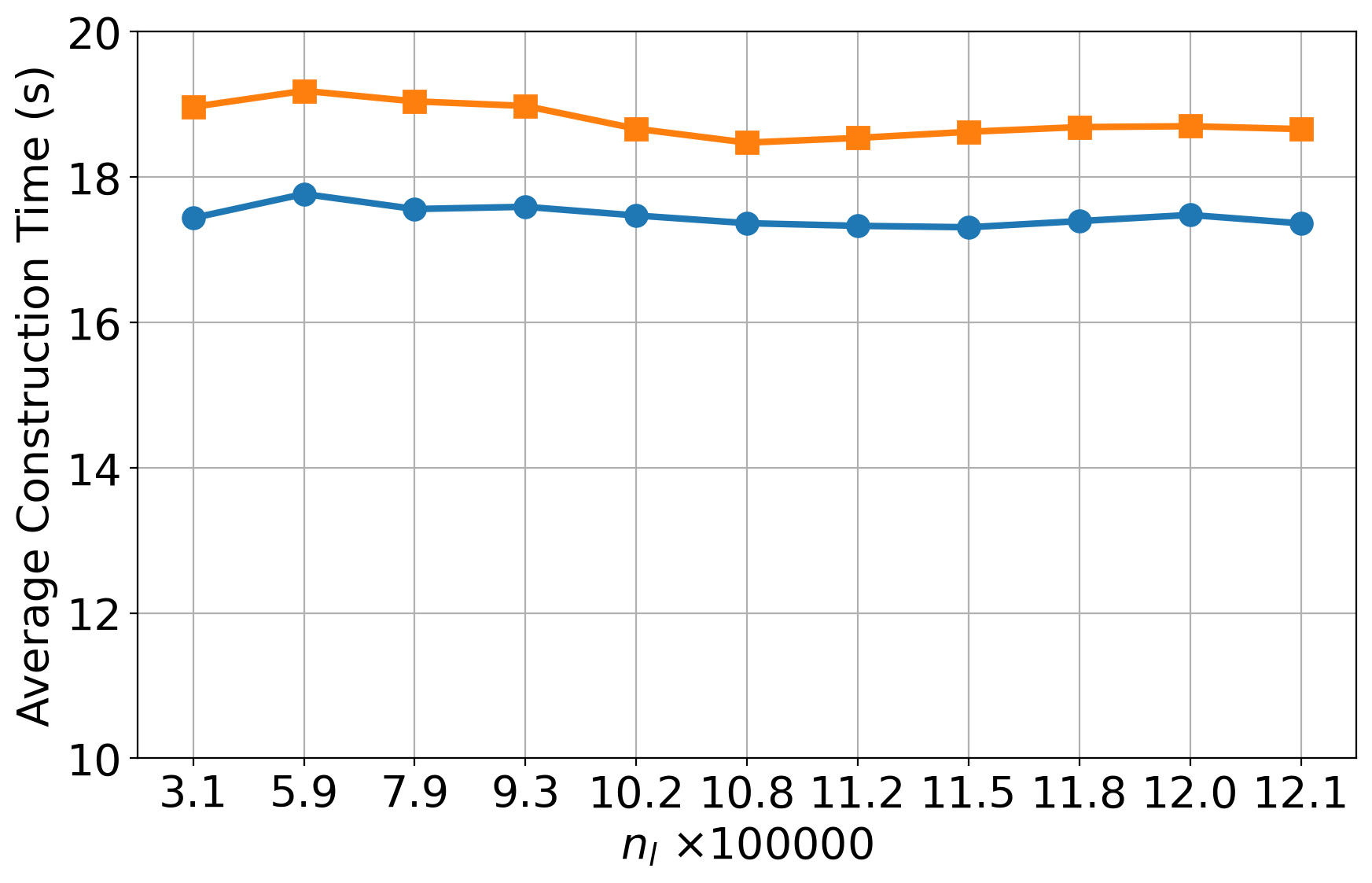}
        \vspace{-1.5em}
        \caption{effect of varying disadvantaged group size $n_l$ on construction time, \google, $n=1.2M, w=65536$.}
        \label{fig:construction_n_l_google}
    \end{minipage}
    \hfill
    \begin{minipage}[t]{0.32\linewidth}
        \centering
        \includegraphics[width=\textwidth]{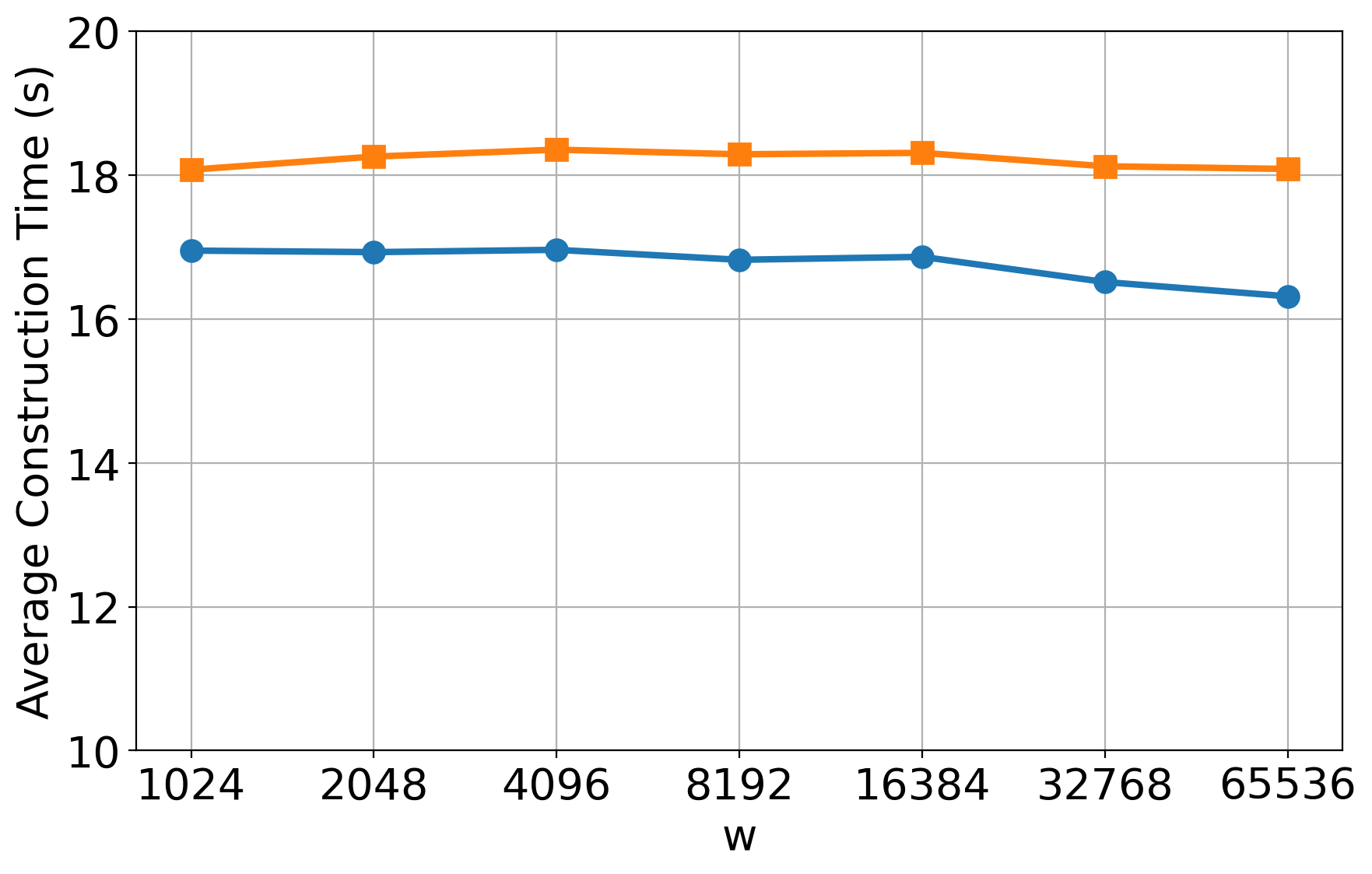}
        \vspace{-1.5em}
        \caption{effect of varying sketch width $w$ on construction time, \google, $n=1.2M, d=5$.}
        \label{fig:construction_w_google}
    \end{minipage}
    \hfill
    \begin{minipage}[t]{0.32\linewidth}
        \centering
        \includegraphics[width=\textwidth]{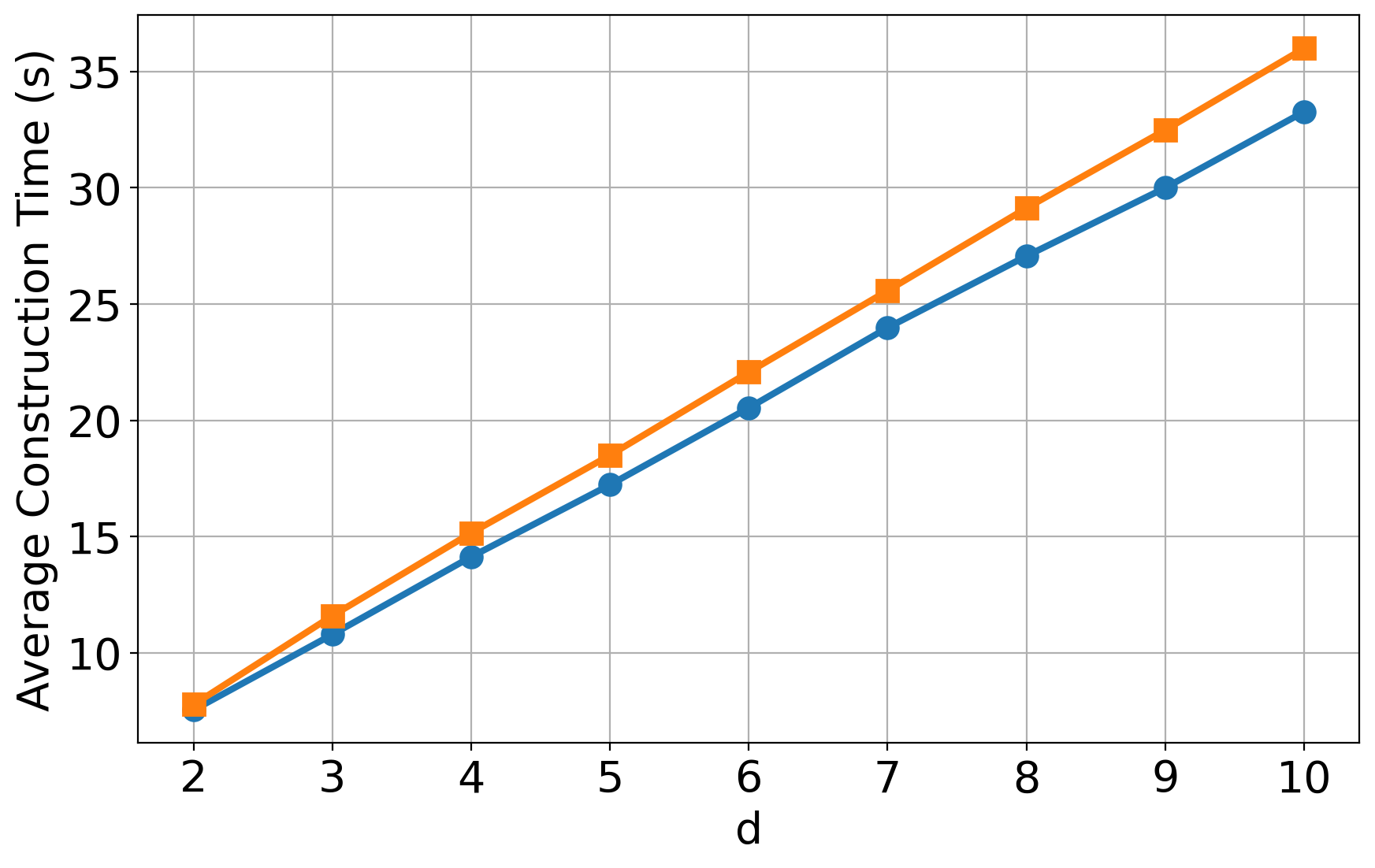}
        \vspace{-1.5em}
        \caption{effect of varying sketch depth $d$ on construction time, \google, $n=1.2M, w=65536$.}
        \label{fig:construction_d_google}
    \end{minipage}
    \hfill
    \begin{minipage}[t]{0.32\linewidth}
        \centering
        \includegraphics[width=\textwidth]{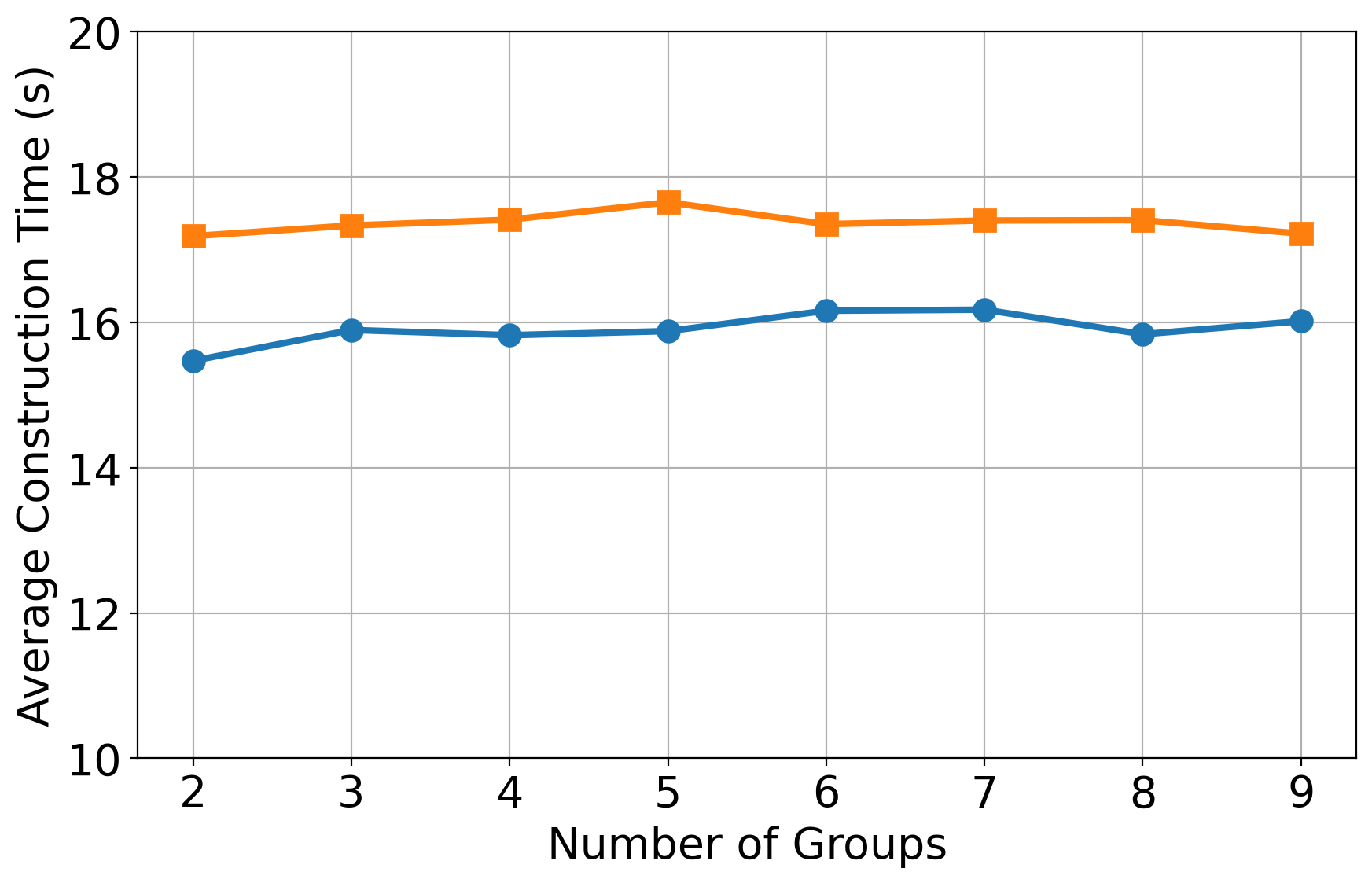}
        \vspace{-1.5em}
        \caption{effect of varying number of groups $\ell$ on construction time, \google, $n=1.2M,w=65536,d=5$.}
        \label{fig:construction_n_groups_google}
    \end{minipage}
    \hfill
    \begin{minipage}[t]{0.32\linewidth}
        \centering
        \includegraphics[width=\textwidth]{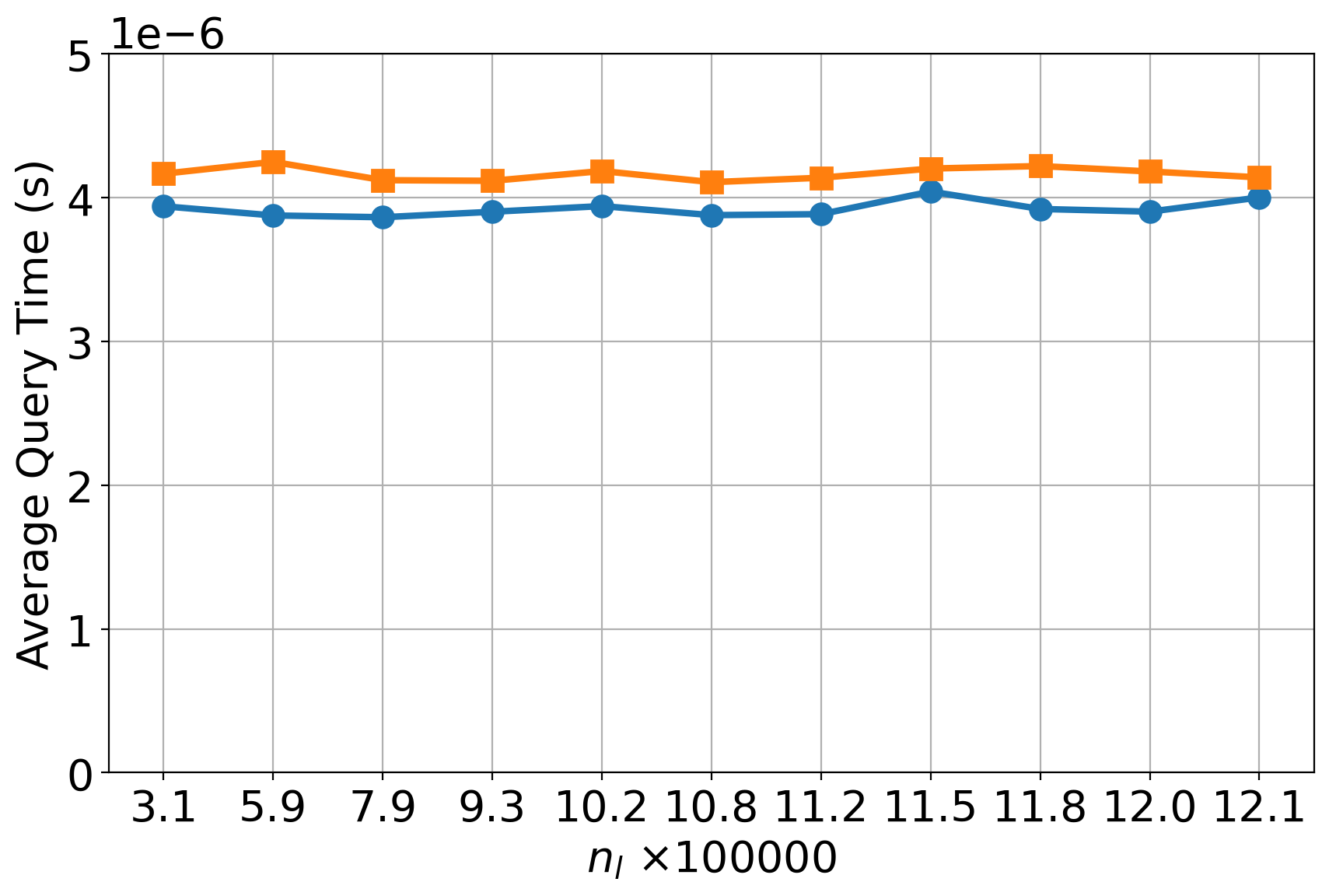}
        \vspace{-1.5em}
        \caption{effect of varying disadvantaged group size $n_l$ on query time, \google, $n=1.2M, w=65536$.}
        \label{fig:query_n_l_google}
    \end{minipage}
    \hfill
    \begin{minipage}[t]{0.32\linewidth}
        \centering
        \includegraphics[width=\textwidth]{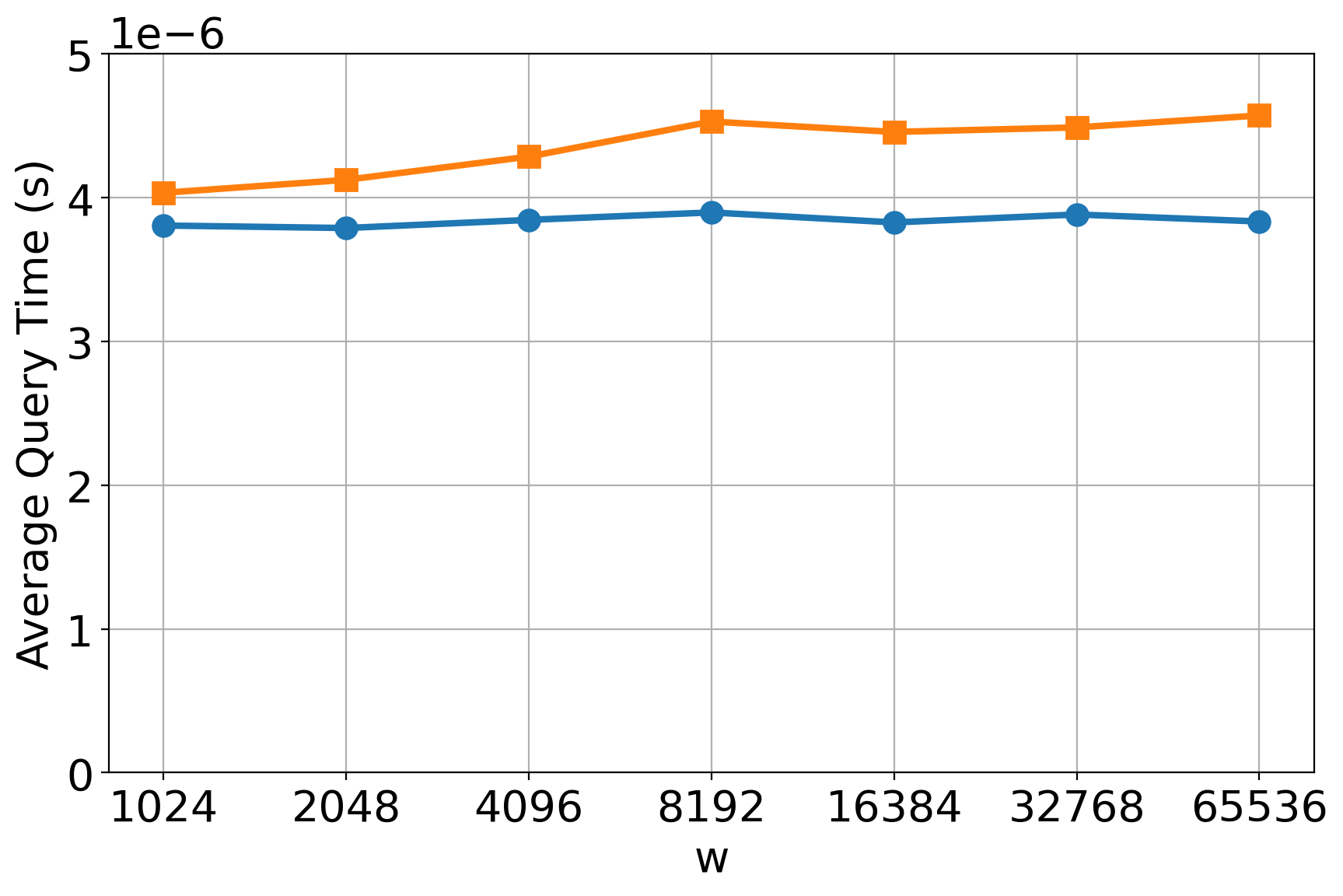}
        \vspace{-1.5em}
        \caption{effect of varying sketch width $w$ on query time, \google, $n=1.2M, d=5$.}
        \label{fig:query_w_google}
    \end{minipage}
    \hfill
    \begin{minipage}[t]{0.32\linewidth}
        \centering
        \includegraphics[width=\textwidth]{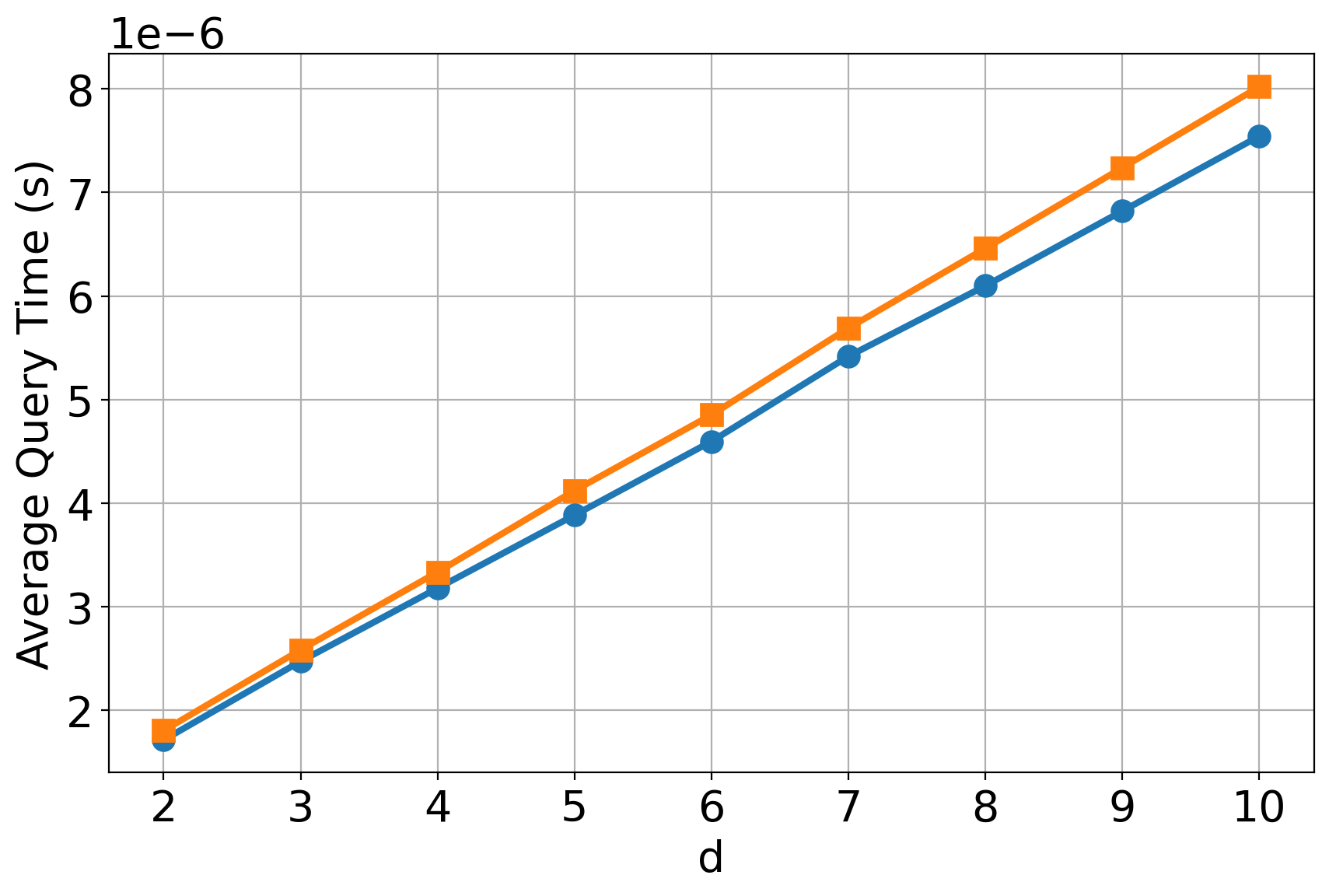}
        \vspace{-1.5em}
        \caption{effect of varying sketch depth $d$ on query time, \google, $n=1.2M, w=65536$.}
        \label{fig:query_d_google}
    \end{minipage}
    \hfill
    \begin{minipage}[t]{0.32\linewidth}
        \centering
        \includegraphics[width=\textwidth]{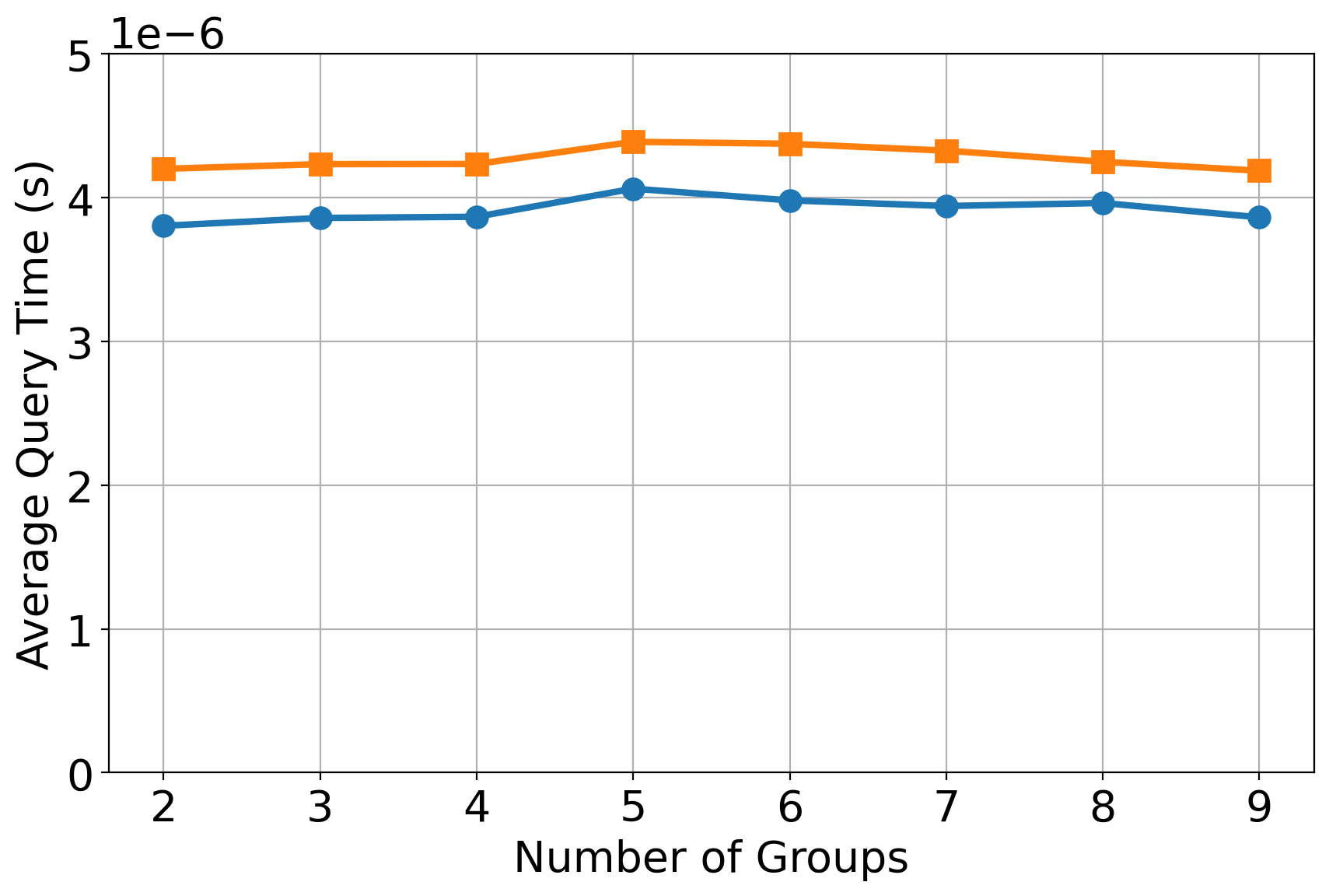}
        \vspace{-1.5em}
        \caption{effect of varying number of groups $\ell$ on query time, \google, $n=1.2M,w=65536,d=5$.}
        \label{fig:query_n_groups_google}
    \end{minipage}
\end{figure*}

%% file: related_work.tex
\section{Related Work}
\paragraph{Count-Min sketch}
The CM sketch was first described by Cormode and Muthukrishnan~\cite{Cormode2009, cormode2005improved}. It uses hash functions to map events to frequencies, but unlike a hash table uses only sub-linear space, at the expense of overcounting some events due to collisions.
Count–min sketch is an alternative to count sketch~\cite{charikar2002finding} and AMS sketch~\cite{alon1996space} and can be considered an implementation of a counting Bloom filter~\cite{fan2000summary, guo2010false} or multistage-filter~\cite{Cormode2009}.
Over the years multiple variations of min-count sketches have been proposed in the literature~\cite{pitel2015count, mazziane2024count, ting2018count, fusy2023count, rottenstreich2021avoiding} improving the estimation or the performance under different settings.
While CM sketches have been thoroughly studied there is still room for improvement.
For example, the random hashing procedure may induce substantial uncertainty in the estimation, especially for low-frequency tokens. Furthermore, it is often the case that there exists an a priori knowledge on the data, and therefore it may be desirable to incorporate such a knowledge into the estimates

Recently, machine learning is used to resolve these issues. More specifically, machine learning techniques~\cite{nguyenpartitioned, dolera2023learning, zhang2020learned, hsu2019learning, aamand2019frequency, cai2018bayesian, dolera2021bayesian} are used to learn either the hash functions or a partition of the elements into two groups (based on their frequency) to improve the estimations or the performance compared to the traditional CM sketch.
Interestingly, partitioned-learned CM sketches~\cite{nguyenpartitioned, hsu2019learning} learn a partition of the elements into low frequency and high frequency elements. While this partition is the same as the one we used in our analysis, our objective is orthogonal to their problem. They do not necessarily aim to achieve the same estimation error between low and high frequency elements. Instead, they process light and heavy elements separately trying to improve the overall estimation error.
To the best of our knowledge, none of these traditional and learned schemes can handle fair count-min estimations with theoretical guarantees.
Finally, learning has been used to obtain other data structures as well, such as $B$-trees~\cite{kraska2018case} or bloom filters~\cite{kraska2018case, vaidya2020partitioned, mitzenmacher2018model}.

\paragraph{Fairness}
Fairness in data-driven systems has been studied by various research communities but mostly in machine learning (ML)~\cite{barocas2017fairness,mehrabi2021survey,pessach2022review}.
Most of the existing work is on training a ML model that satisfies some fairness constraints. Some pioneering fair-ML efforts include~\cite{dwork2012fairness,zafar2017fairness,celis2019classification,calmon2017optimized,feldman2015certifying,hardt2016equality,kamiran2012data}.
Biases in data has also been studied extensively~\cite{stoyanovich2022responsible,shahbazi2023representation,olteanu2019social,salimi2019data,asudeh2019assessing,asudeh2021identifying} to ensure data has been prepared responsibly~\cite{salimi2019interventional,nargesian2021tailoring,salimi2020database,shetiya2022fairness}.
Recent studies of fair algorithm design include
fair
clustering~\cite{chierichetti2017fair,bera2019fair,schmidt2020fair,ahmadian2020fair,bohm2020fair,makarychev2021approximation,chlamtavc2022approximating,hotegni2023approximation},
fairness in resource allocation and facility location problem~\cite{blanco2023fairness,jiang2021rawlsian,donahue2020fairness,he2020inherent,zhou2019public,gorantla2023fair},
min cut~\cite{li2023near},
max cover~\cite{asudeh2023maximizing}, set cover~\cite{dehghankar2025fair},
game theoretic approaches~\cite{zhao2021fairness,donahue2023fairness,algaba2019shapley},
hiring~\cite{raghavan2020mitigating,aminian2023fair},
ranking~\cite{asudeh2019designing,singh2018fairness,singh2019policy,zehlike2022fairness},
recommendation~\cite{chen2023bias,li2022fairness, swift2022maximizing},
representation learning~\cite{he2020geometric},
etc.

Fairness in data structures is significantly under-studied, with the existing work being limited to ~\cite{aumuller2021fair,aumuller2022sampling,aumuller2020fair,shahbazi2024fairhash}.
Aumuller et al.~\cite{aumuller2021fair,aumuller2022sampling,aumuller2020fair} study individual fairness in near-neighbor search, while to the best of our knowledge, Shahbazi et al.~\cite{shahbazi2024fairhash} is the only work that studies group fairness in data structure design, and more specifically in hashing. Both in nearest-neighbor search and fair hashing, the objective is fundamentally different, so these techniques cannot be used for our fair-CM sketch.

%% file: conclusion.tex
\section{Conclusion}
In this paper we introduced Fair-Count-Min, a new frequency estimation framework that ensures equal expected approximation factors across different groups, addressing a key fairness limitation of traditional Count-Min sketches. By proposing a fairness-preserving design—column partitioning with group-aware hashing, our work offers strong theoretical guarantees and validates its effectiveness through extensive experiments. Future research directions include extending Fair-Count-Min integrating learned hash functions for further efficiency gains, and exploring fairness notions beyond group-wise guarantees, such as individual fairness in streaming data sketches.

%% file: Appendix-PoF.tex
\section{Price of Fairness: Randomness vs. Uniformity}

\begin{figure}[H]
    \centering
    \begin{tikzpicture}
          \begin{axis}[
              ybar,
              symbolic x coords={1,2,3,4,5,6,7,8,9,10},
              xtick=data,
              xlabel={Bucket Number},
              ylabel={Size},
              ymin=0,
              ymax=20,
              width=0.45\textwidth,
              height=0.4\textwidth,
              nodes near coords,
              enlarge x limits=0.1
            ]
            \addplot coordinates {(1,9) (2,18) (3,11) (4,8) (5,13) (6,8) (7,10) (8,7) (9,6) (10,10)};
          \end{axis}
    \end{tikzpicture}
    \caption{Illustration of a random distribution of $n=100$ elements to $w=10$ buckets.}
    \label{fig:randomdistribution}
\end{figure}
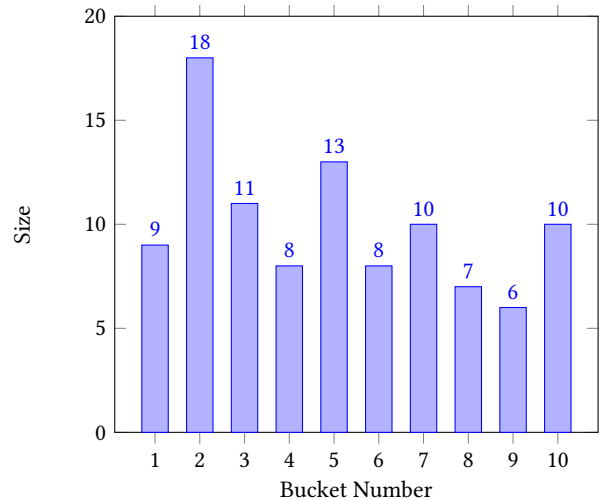

In Section~\ref{sec:price}, we observed a counterintuitive result for $d=1$, proving a negative price of fairness (PoF) for FCM, compared to a regular CM sketch that uses a random hash function.
This result can be explained by the fact that a random hash function is unlikely to {\em uniformly} distribute the elements to the buckets (an interesting related topic is {\em the occupant problem}~\cite{motwani1996randomized}).
As a result, some of the buckets will have more than $\frac{n}{w}$ elements in them. For example, Figure~\ref{fig:randomdistribution} illustrates a random hashing of $n=100$ element types into $w=10$ buckets.
While a uniform distribution would allocate $10$ elements to each bucket, the random hashing allocated up to $18$ elements in one bucket.

Elements in large buckets will suffer from a large additive error, and there are a large number of them in such buckets. Hence, such buckets significantly increase the total additive error of the CM.

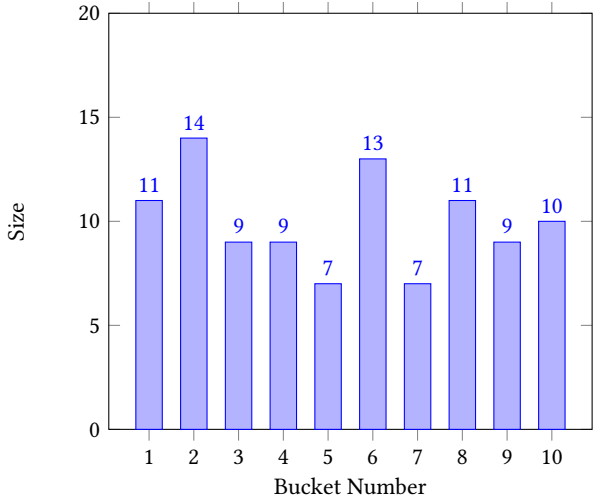
\begin{figure}[H]
    \centering
    \begin{tikzpicture}
          \begin{axis}[
              ybar,
              symbolic x coords={1,2,3,4,5,6,7,8,9,10},
              xtick=data,
              xlabel={Bucket Number},
              ylabel={Size},
              ymin=0,
              ymax=20,
              width=0.45\textwidth,
              height=0.4\textwidth,
              nodes near coords,
              enlarge x limits=0.1
            ]
            \addplot coordinates {(1,11) (2,14) (3,9) (4,9) (5,7) (6,13) (7,7) (8,11) (9,9) (10,10)};
          \end{axis}
    \end{tikzpicture}
    \caption{Illustration of a random distribution of $n=100$ elements with two groups of sizes $n_l=50$ and $n_h=50$ to $w=10$ buckets, based on FCM.}
    \label{fig:randomdistribution2}
\end{figure}

FCM increases the size-uniformity of the buckets by reserving $w_l=\frac{n_l}{n}w$ for each group $g_l$. For example, Figure~\ref{fig:randomdistribution2} illustrates the allocation of the 100 elements to the 10 buckets, by dividing (arbitrarily) the elements into two groups of size $n_l=50$, $n_h=50$. As a result, each group is allocated $5$ buckets, and the distribution of the elements is more uniform, reducing the maximum size of the buckets (in this example) to 14.
Increasing the uniformity (reducing the change of large-size buckets) helps to reduce the total additive error of FCM versus CM.

\subsection{PoF for Uniform hashing ($d=1$)}
Using a hashing scheme (such as data-informed hashmaps~\cite{kraska2018case}) that ensures uniformity by allocating exactly $\frac{n}{w}$ elements to each bucket can resolve the non-uniformity issue in the CM sketches.
In the following, we compute the PoF of SCM for $d=1$ when a uniform hashing scheme is used.

Since the hashing is uniform, the size of each bucket $i$ in the CM sketch is \[C_i=\frac{n}{w}\]
Therefore, each element $e_j$ collides with exactly \(\frac{n}{w}-1\) other elements. 
In other words, the frequency of each element contributes to the additive error of exactly \(\frac{n}{w}-1\) other elements.

As a result, the total additive error can be computed as

\begin{align}\label{eq:temp32}
 \nonumber    \mathcal{L}_{CM} &= \sum_{j=1}^n \eps_A(e_j) \\
 \nonumber    & = \sum_{j=1}^n f(e_j) \left(\frac{n}{w}-1\right)= \left(\frac{n}{w}-1\right)\sum_{j=1}^n f(e_j) \\
    &= N\left(\frac{n}{w}-1\right)
\end{align}

Now, let us compute $\mathcal{L}_{FCM}$, the total additive error for the regular fair-count-min sketch.
\[
    \mathcal{L}_{FCM} = \sum_{j=1}^n \eps_A(e_j) = \sum_{e_j\in \gee_1} \eps_A(e_j)+\cdots + \sum_{e_j\in \gee_\ell} \eps_A(e_j)
\]
Following the same calculation as in Equation~\ref{eq:temp32}, for every group $\gee_l=\Gee$, 
\begin{align*}
     \sum_{e_j\in \gee_l} \eps_A(e_j)&= N_l\Big(\frac{n_l}{w_l}-1\Big) 
\end{align*}
Hence,
\[
\mathcal{L}_{FCM} = \sum_{e_j\in \gee_1} \eps_A(e_j)+\cdots + \sum_{e_j\in \gee_\ell} \eps_A(e_j)
 \nonumber = \sum_{l=1}^\ell N_l\Big(\frac{n_l}{w_l}-1\Big)
\]

From Theorem~\ref{thm:1}, we know, $w_l=\frac{n_l}{n}w, \forall \gee_l\in \Gee$, while $\sum_{l=1}^\ell N_l=N$.
Therefore,

\begin{align}\label{eq:temp33}
\nonumber \mathcal{L}_{FCM} &= \sum_{l=1}^\ell N_l\Big(\frac{n_l}{w_l}-1\Big)\\
\nonumber&= \sum_{l=1}^\ell N_l\left( \frac{n_l}{w_l} \right) - \sum_{l=1}^\ell N_l
= \sum_{l=1}^\ell N_l\left( \frac{n_l}{w_l} \right) - N\\
\nonumber& = \left( \frac{n}{w} \right)\sum_{l=1}^\ell N_l - N \hspace{20mm} \text{\tt\scriptsize  //since $\frac{n_l}{w_l} = \frac{n}{w}, \forall \gee_l\in \Gee$}\\
& = N \left( \frac{n}{w} \right) - N = N\Big(\frac{n}{w}-1\Big)
\end{align}

Finally, combining Equations~\ref{eq:temp20}, \ref{eq:temp32}, and \ref{eq:temp33}, we get,
\begin{align}\label{eq:pof1}
PoF = \mathcal{L}_{FCM} - \mathcal{L}_{CM}  = N\Big(\frac{n}{w}-1\Big) - N\Big(\frac{n}{w}-1\Big) = 0
\end{align}

That is, the PoF is zero for $d=1$, when a uniform hashing scheme is used.

%% file: Appendix-exp.tex
\section{Additional Experiment Results}
In this section, we present comprehensive experimental results across three datasets: 1) {\google}, 2) {\census}, and 3) {\synthetic}. We also provide the absolute values of the approximation factors and additive errors for each group to facilitate a clearer understanding of the trends in unfairness and the associated price of fairness under each setting.

\input{plots_appendix}

%% file: plots_appendix.tex

\begin{figure*}[!tb] 
\centering
    \begin{minipage}[t]{0.32\linewidth}
        \centering
        \includegraphics[width=\textwidth]{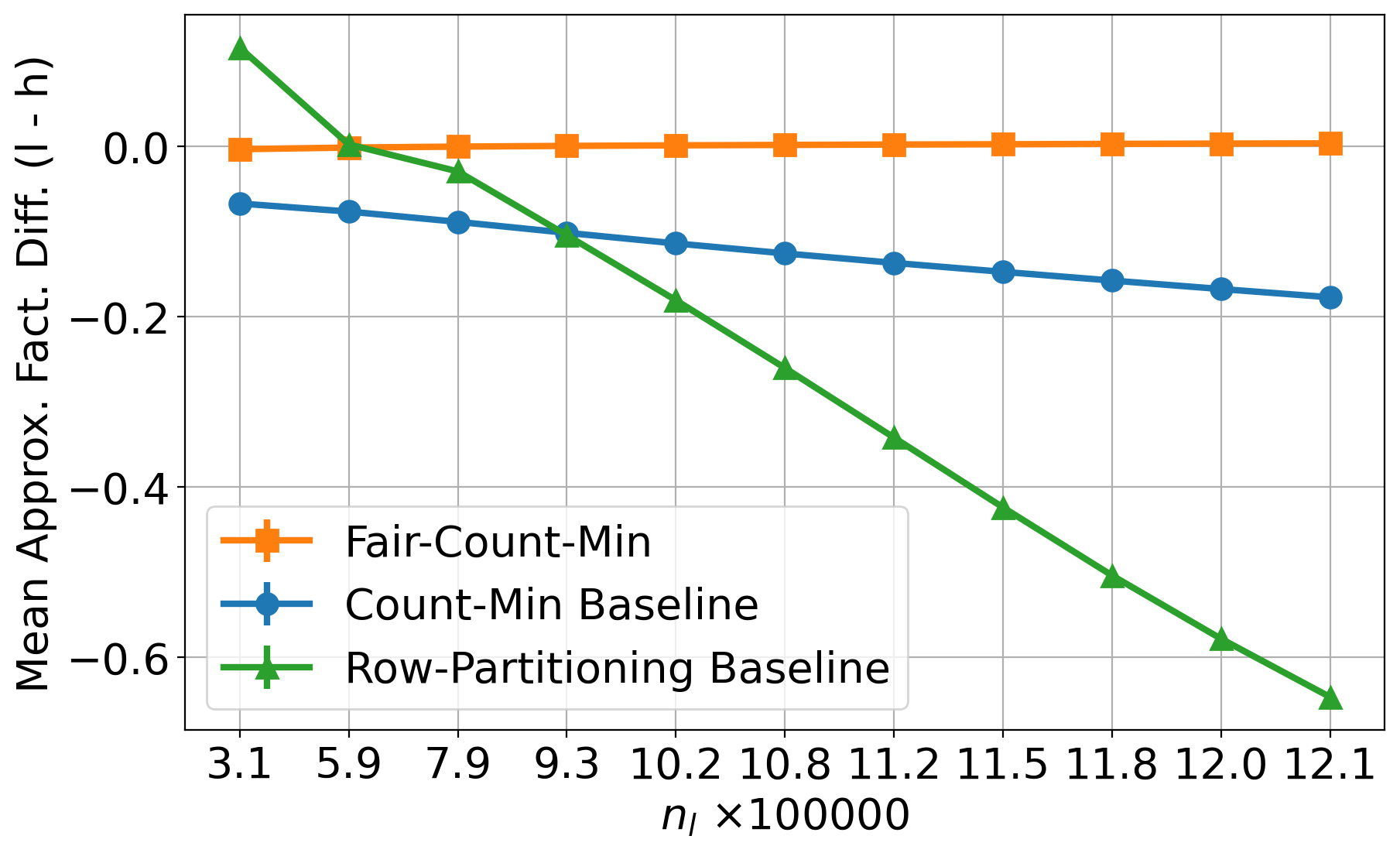}
        \vspace{-2.5em}
        \caption{effect of varying group size $n_l$ on unfairness, \google, $w=65536, d=5$.}
        \label{fig:}
    \end{minipage}
    \hfill
    \begin{minipage}[t]{0.32\linewidth}
        \centering
        \includegraphics[width=\textwidth]{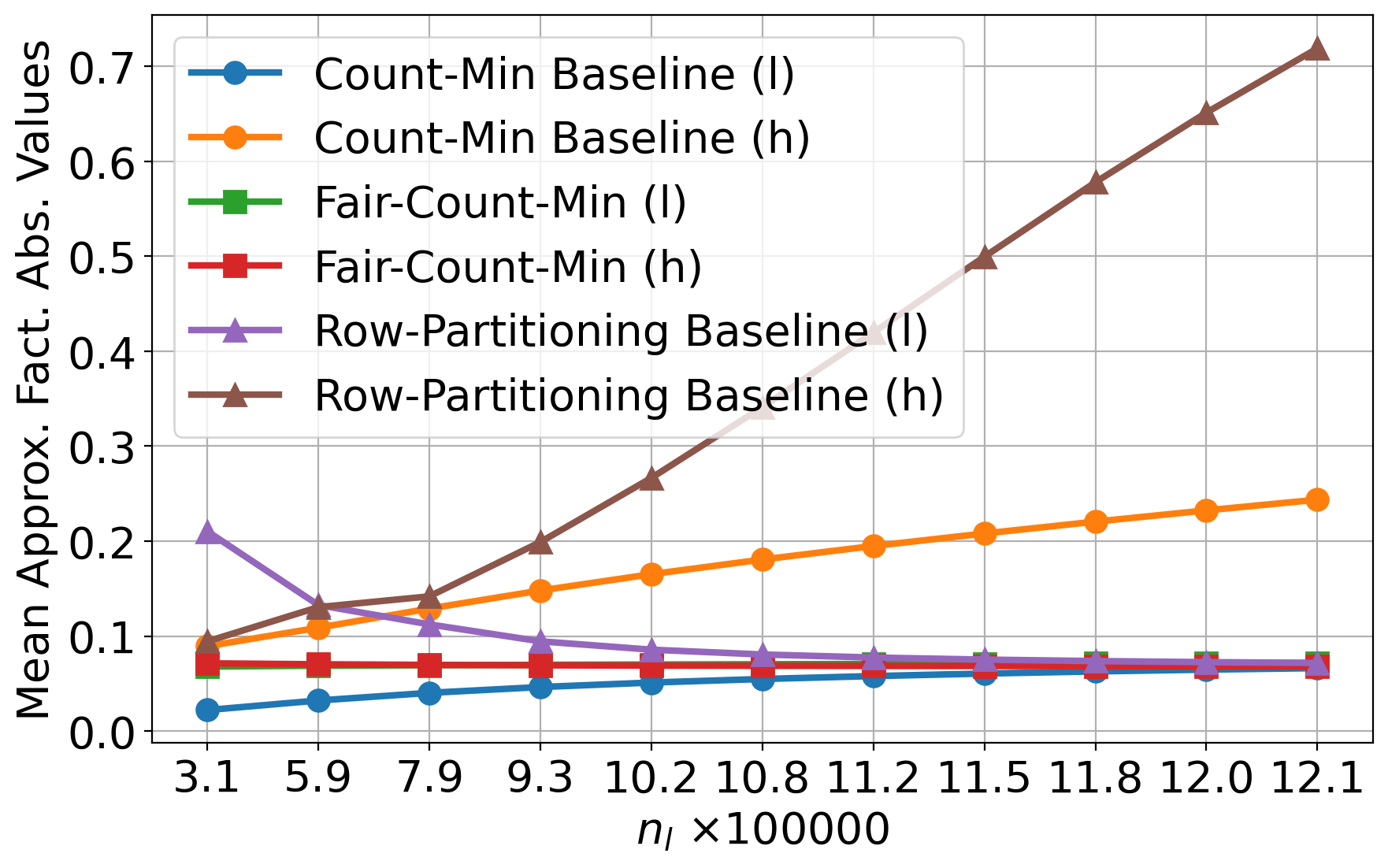}
        \vspace{-2.5em}
        \caption{effect of varying group size $n_l$ on approximation factors, \google, $w=65536, d=5$.}
        \label{fig:}
    \end{minipage}
    \hfill
    \begin{minipage}[t]{0.32\linewidth}
        \centering
        \includegraphics[width=\textwidth]{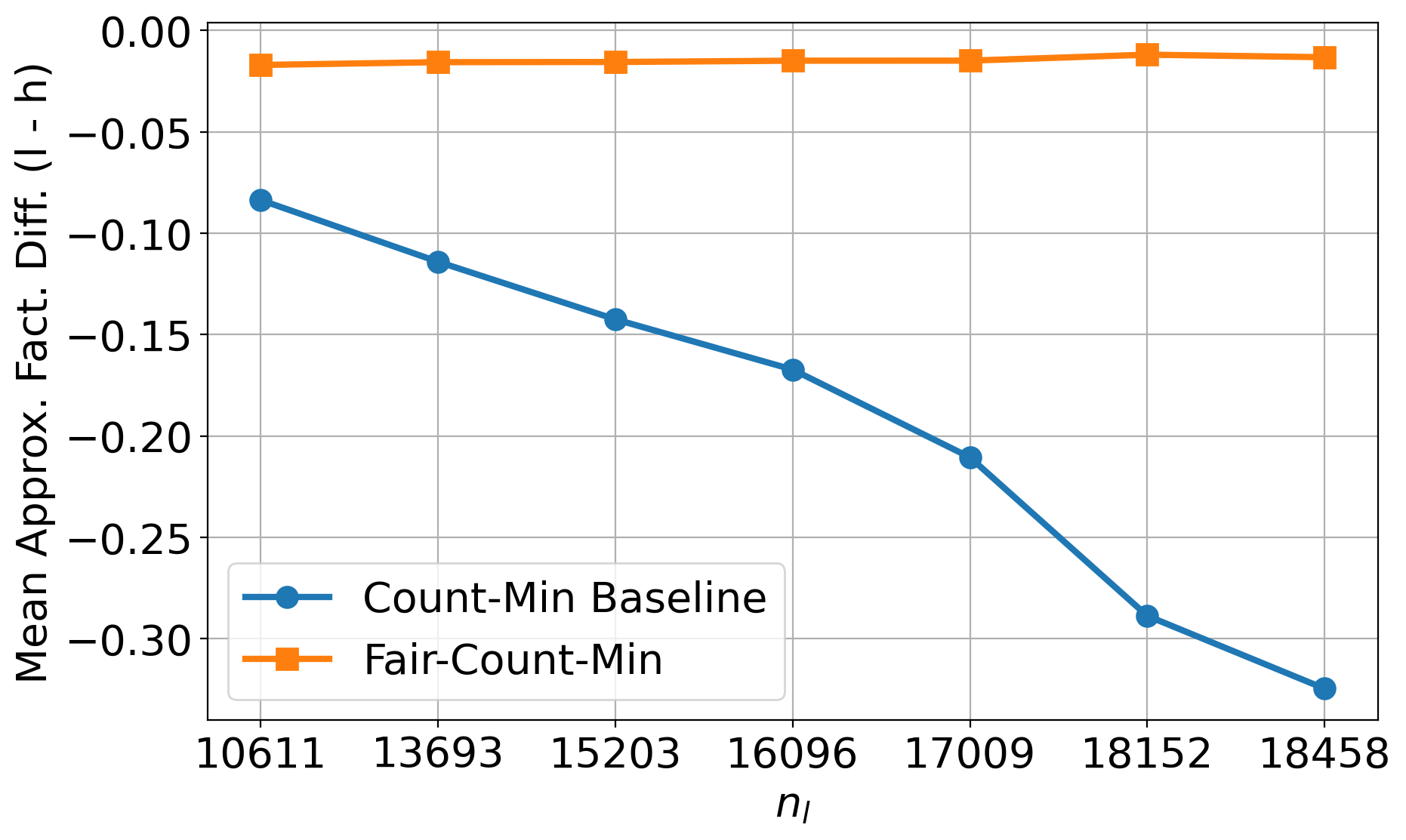}
        \vspace{-2.5em}
        \caption{effect of varying group size $n_l$ on unfairness, \synthetic, $w=512, d=10$.}
        \label{fig:}
    \end{minipage}
    \hfill
    \begin{minipage}[t]{0.32\linewidth}
        \centering
        \includegraphics[width=\textwidth]{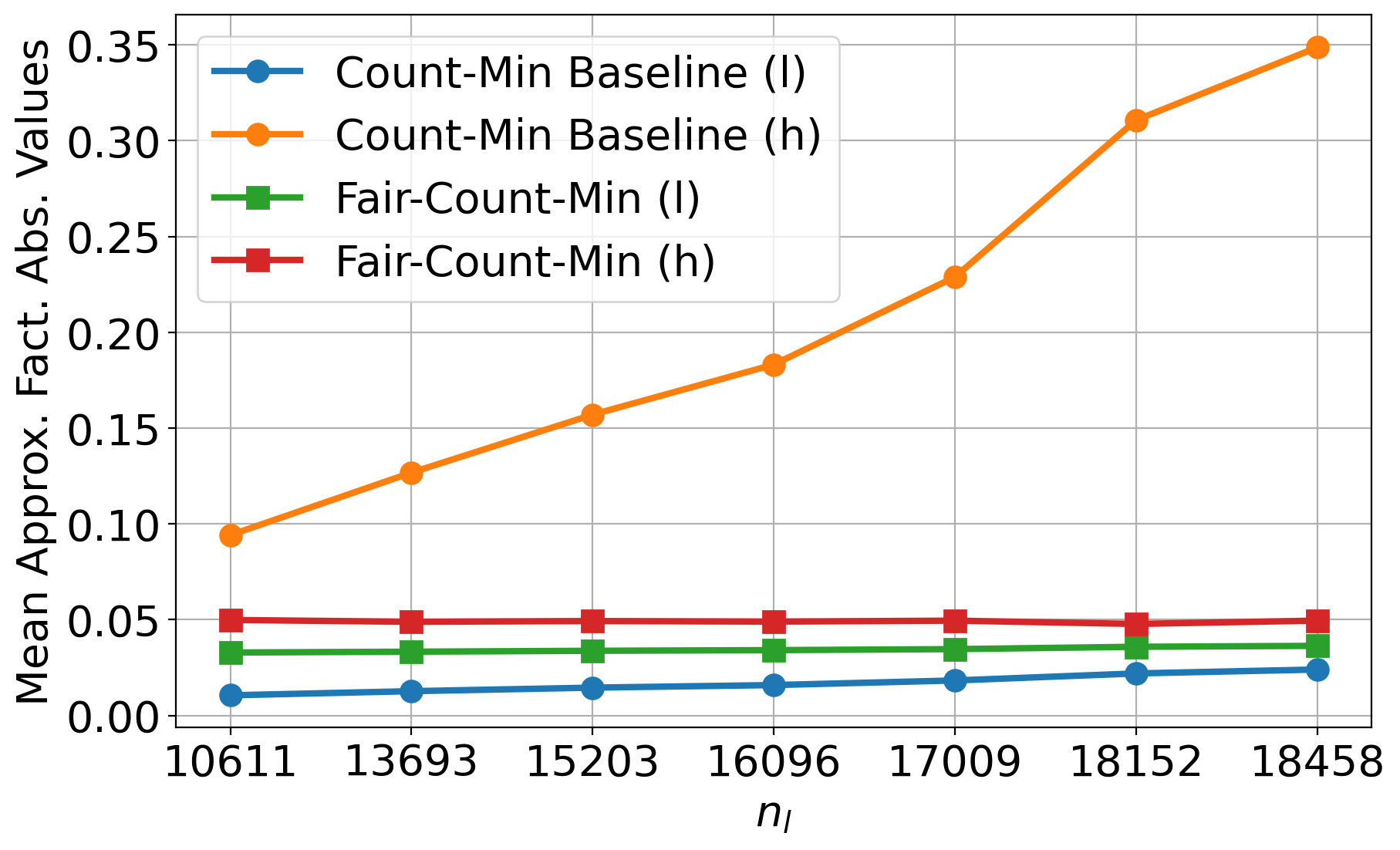}
        \vspace{-2.5em}
        \caption{effect of varying group size $n_l$ on approximation factors, \synthetic, $w=512, d=10$.}
        \label{fig:}
    \end{minipage}
    \hfill
    \begin{minipage}[t]{0.32\linewidth}
        \centering
        \includegraphics[width=\textwidth]{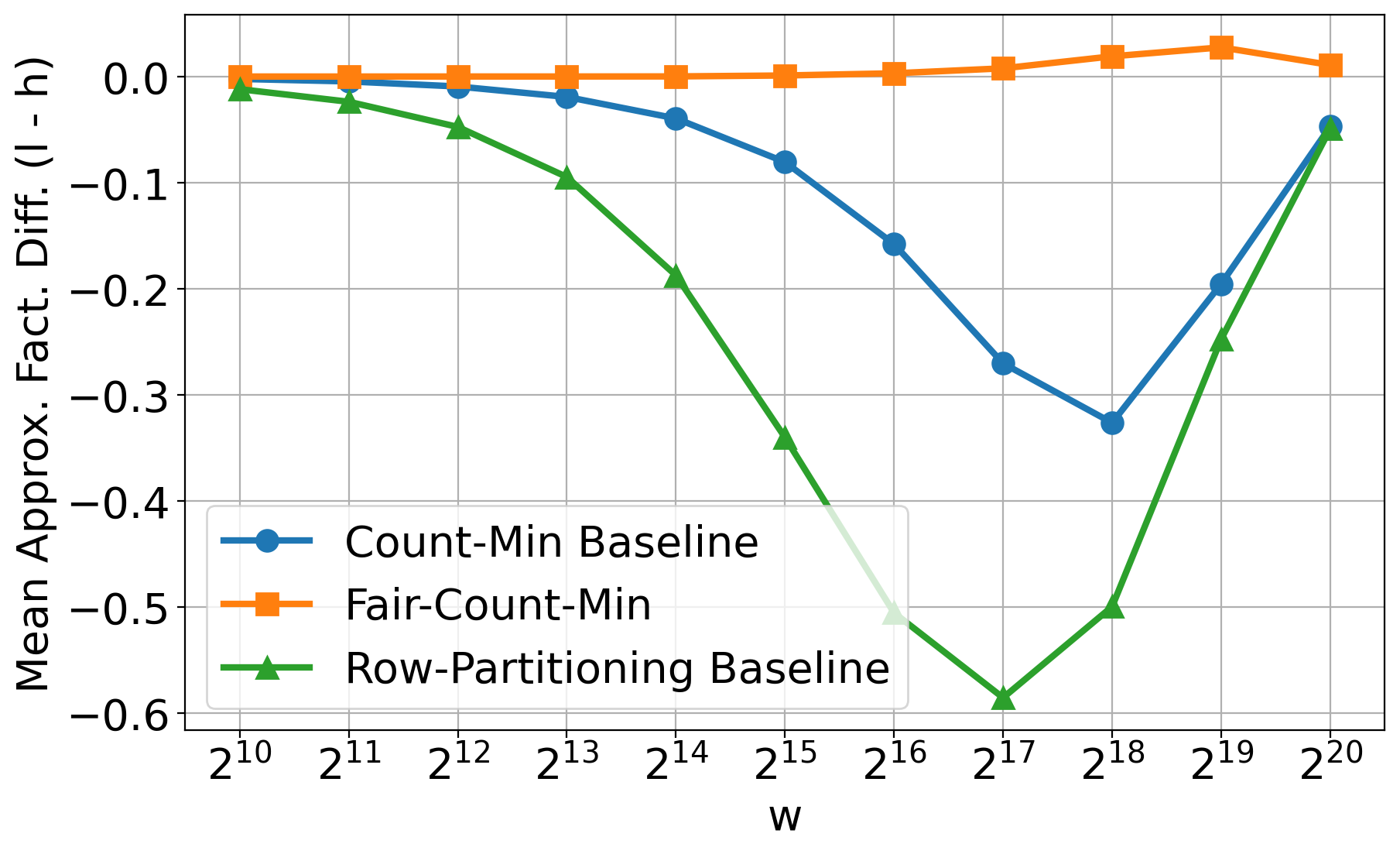}
        \vspace{-2.5em}
        \caption{effect of varying sketch width $w$ on unfairness, \google, $n=1.2M, d=5$.}
        \label{fig:}
    \end{minipage}
    \hfill
    \begin{minipage}[t]{0.32\linewidth}
        \centering
        \includegraphics[width=\textwidth]{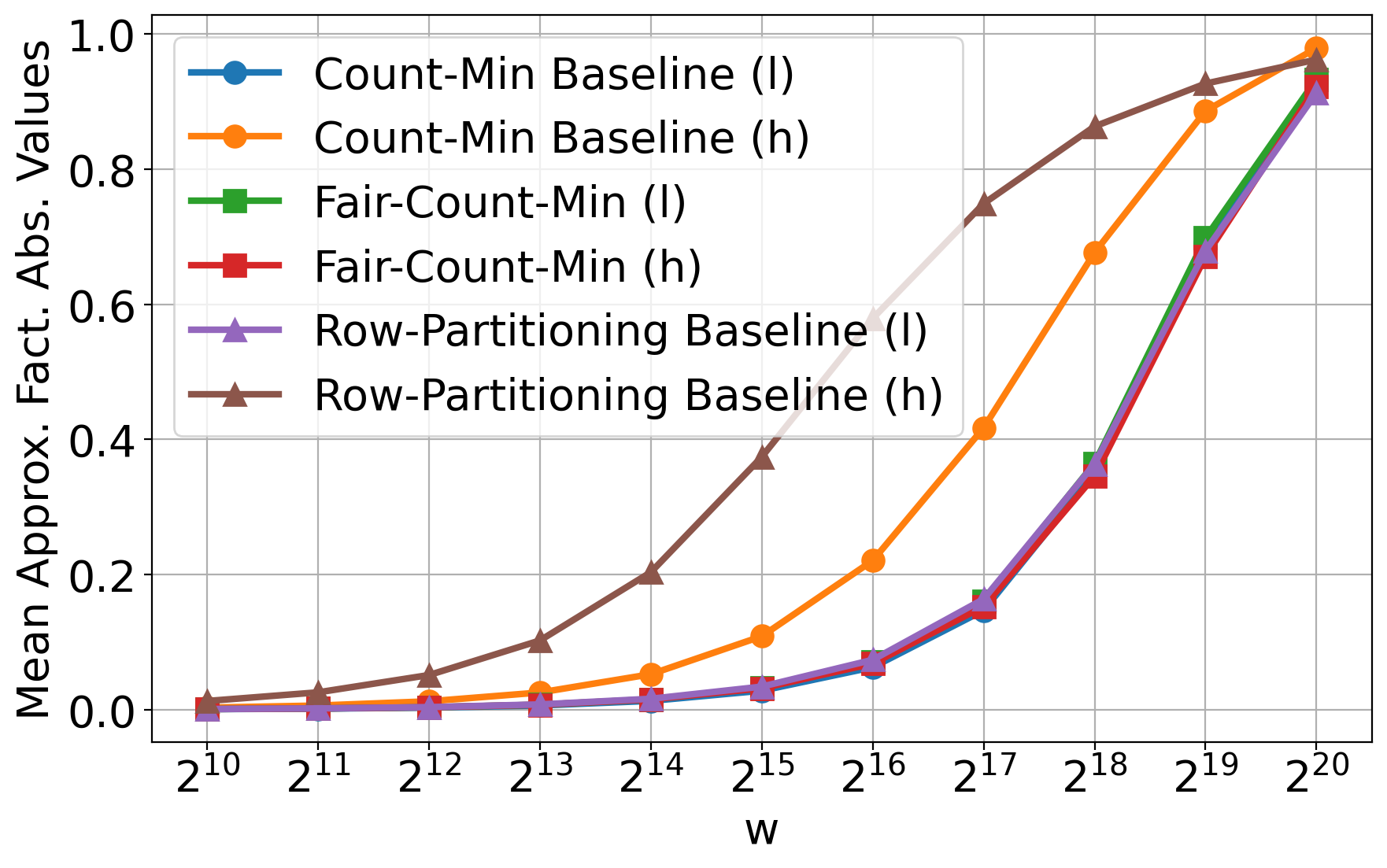}
        \vspace{-2.5em}
        \caption{effect of varying sketch width $w$ on approximation factors, \google, $n=1.2M, d=5$.}
        \label{fig:}
    \end{minipage}
    \hfill
    \begin{minipage}[t]{0.32\linewidth}
        \centering
        \includegraphics[width=\textwidth]{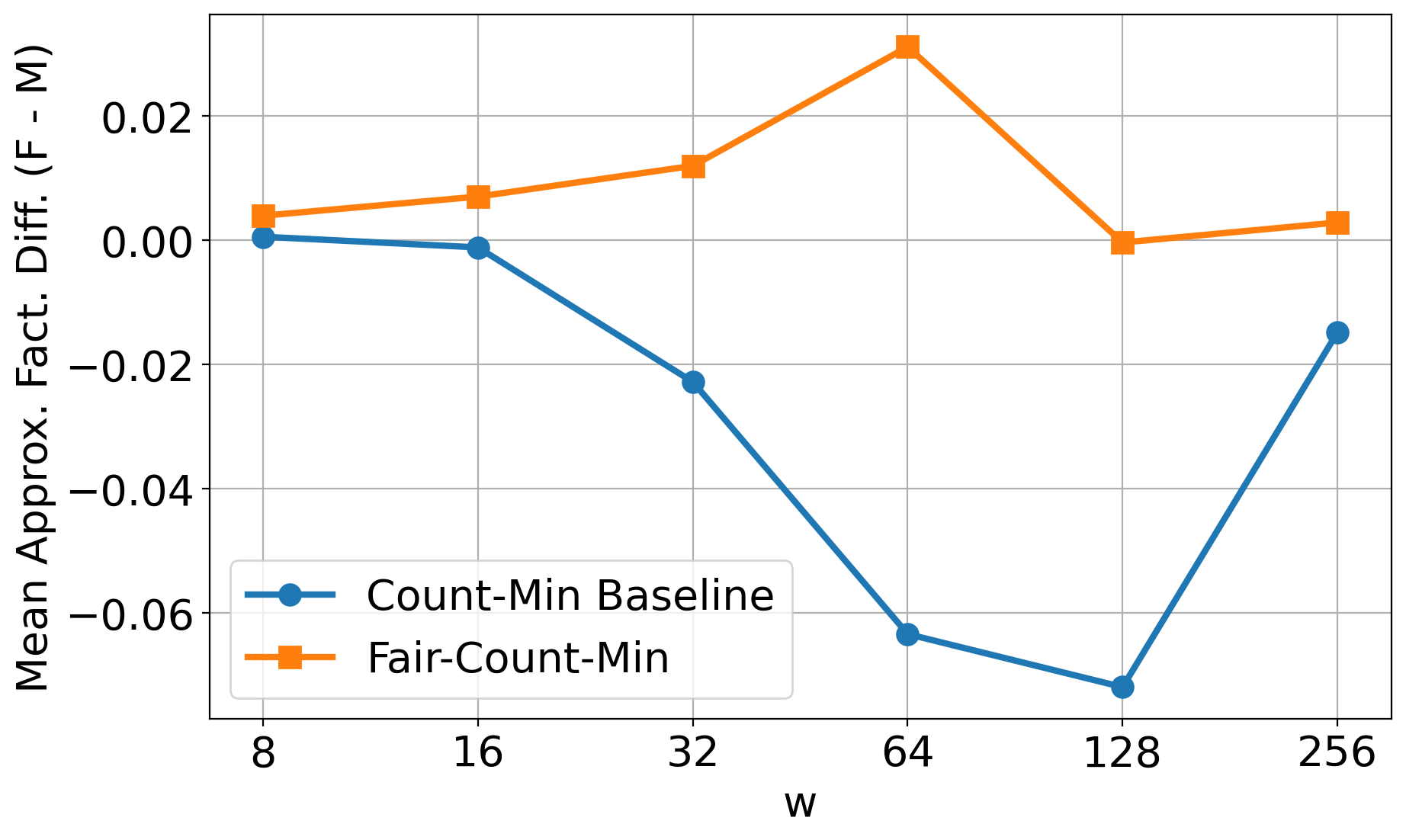}
        \vspace{-2.5em}
        \caption{effect of varying sketch width $w$ on unfairness, \census, $n=430, d=5$.}
        \label{fig:}
    \end{minipage}
    \hfill
    \begin{minipage}[t]{0.32\linewidth}
        \centering
        \includegraphics[width=\textwidth]{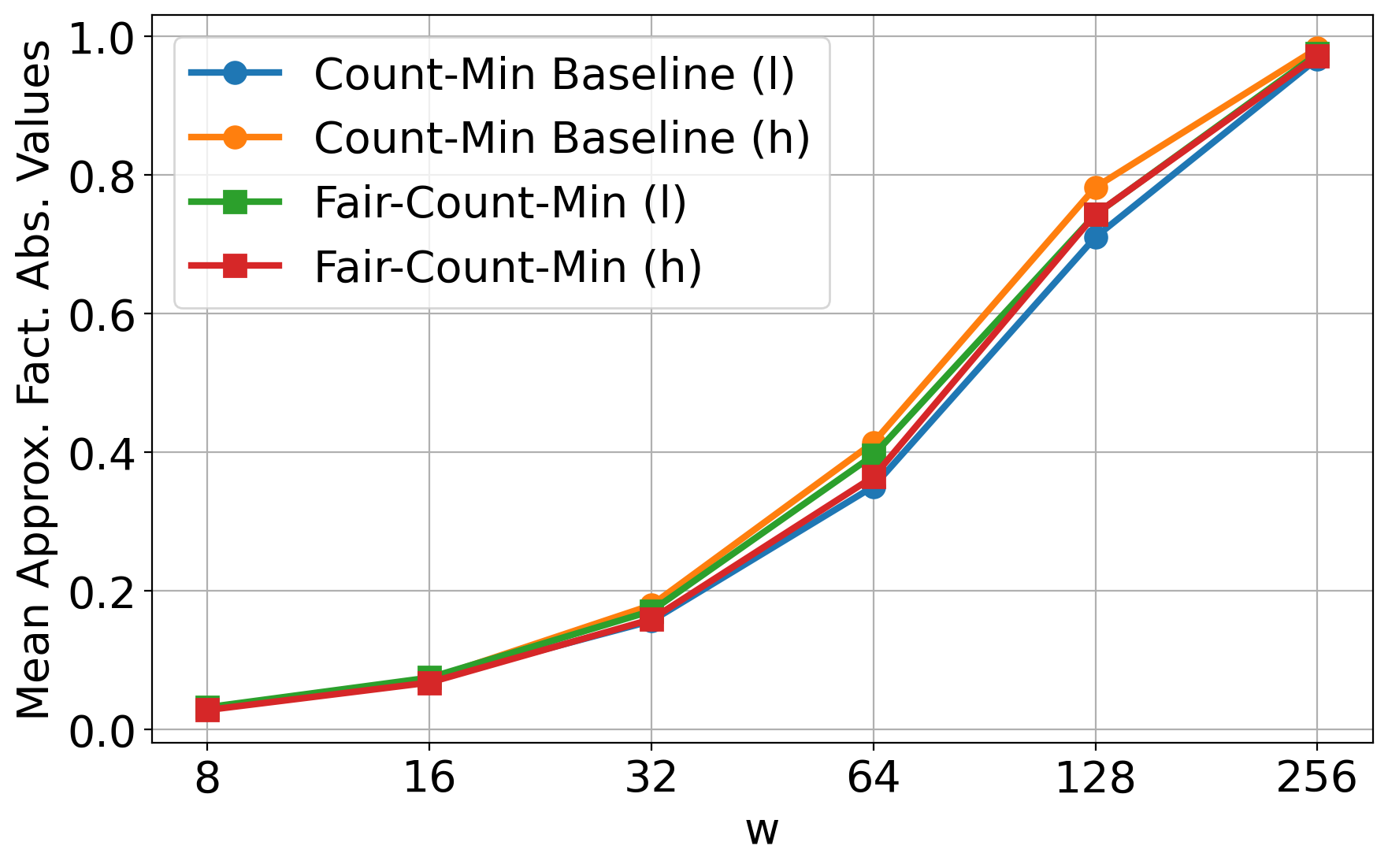}
        \vspace{-2.5em}
        \caption{effect of varying sketch width $w$ on approximation factors, \census, $n=430, d=5$.}
        \label{fig:}
    \end{minipage}
    \hfill
    \begin{minipage}[t]{0.32\linewidth}
        \centering
        \includegraphics[width=\textwidth]{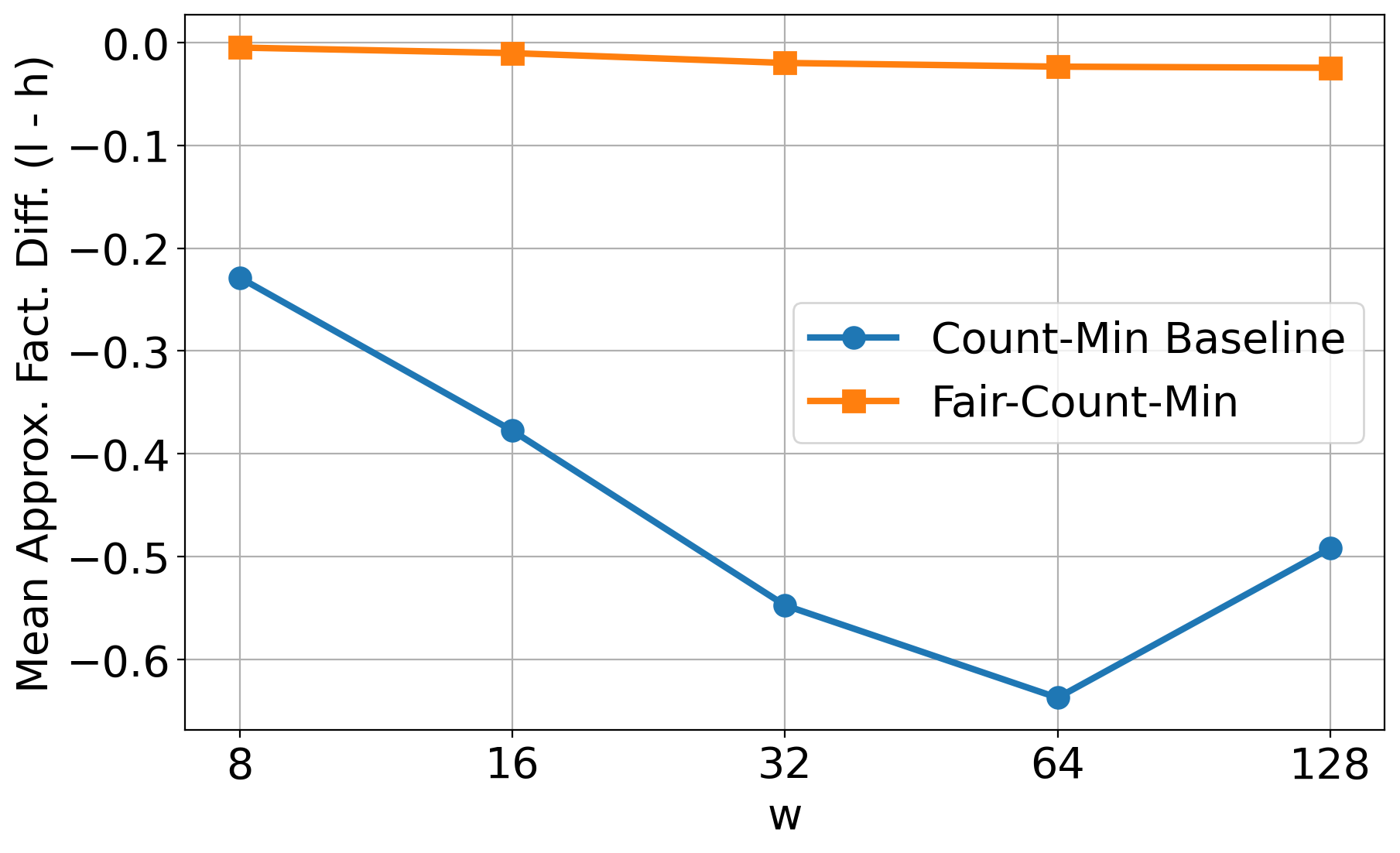}
        \vspace{-2.5em}
        \caption{effect of varying sketch width $w$ on unfairness, \synthetic, $n=20000, d=10$.}
        \label{fig:}
    \end{minipage}
    \hfill
    \begin{minipage}[t]{0.32\linewidth}
        \centering
        \includegraphics[width=\textwidth]{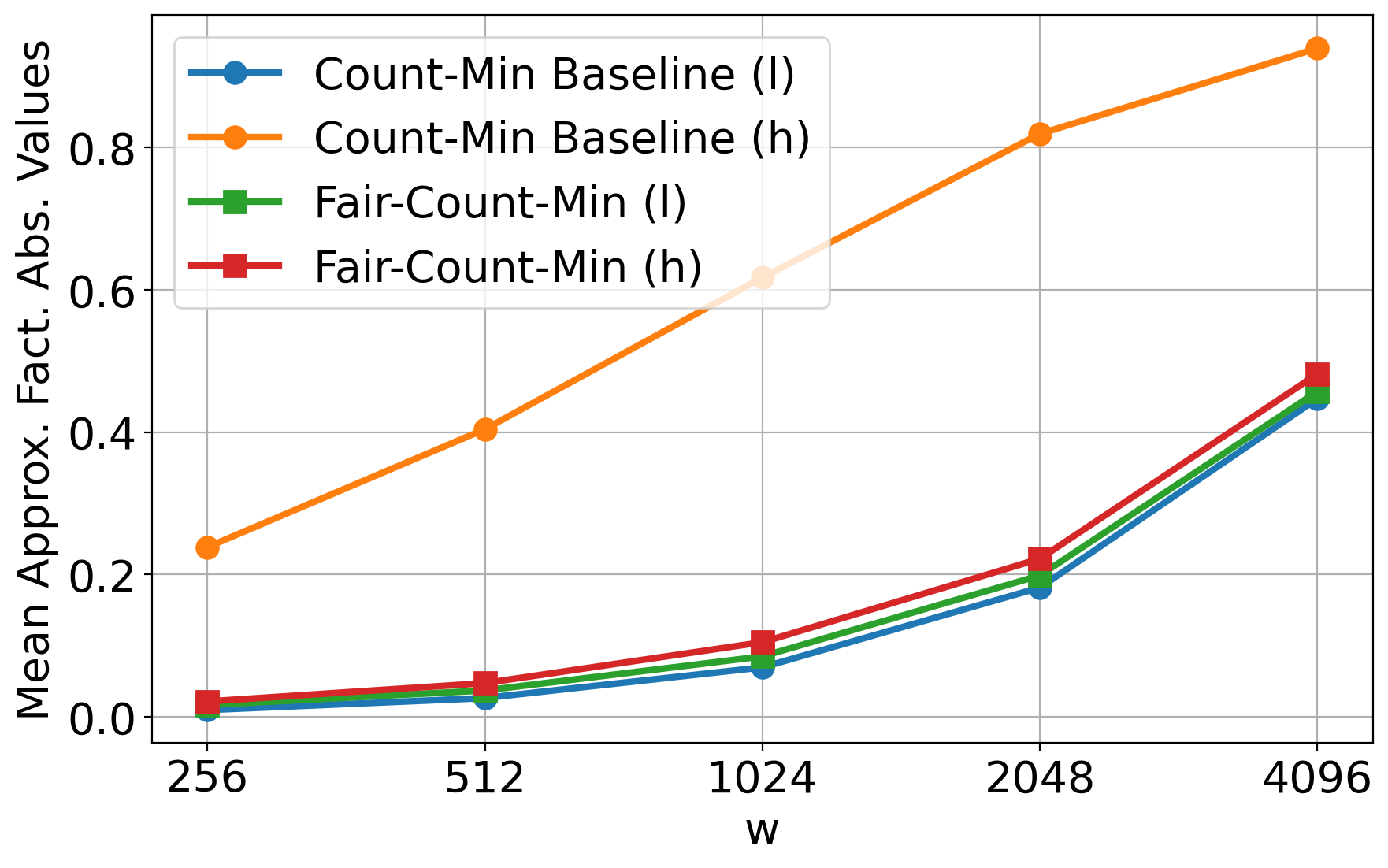}
        \vspace{-2.5em}
        \caption{effect of varying sketch width $w$ on approximation factors, \synthetic, $n=20000, d=10$.}
        \label{fig:}
    \end{minipage}
    \hfill
    \begin{minipage}[t]{0.32\linewidth}
        \centering
        \includegraphics[width=\textwidth]{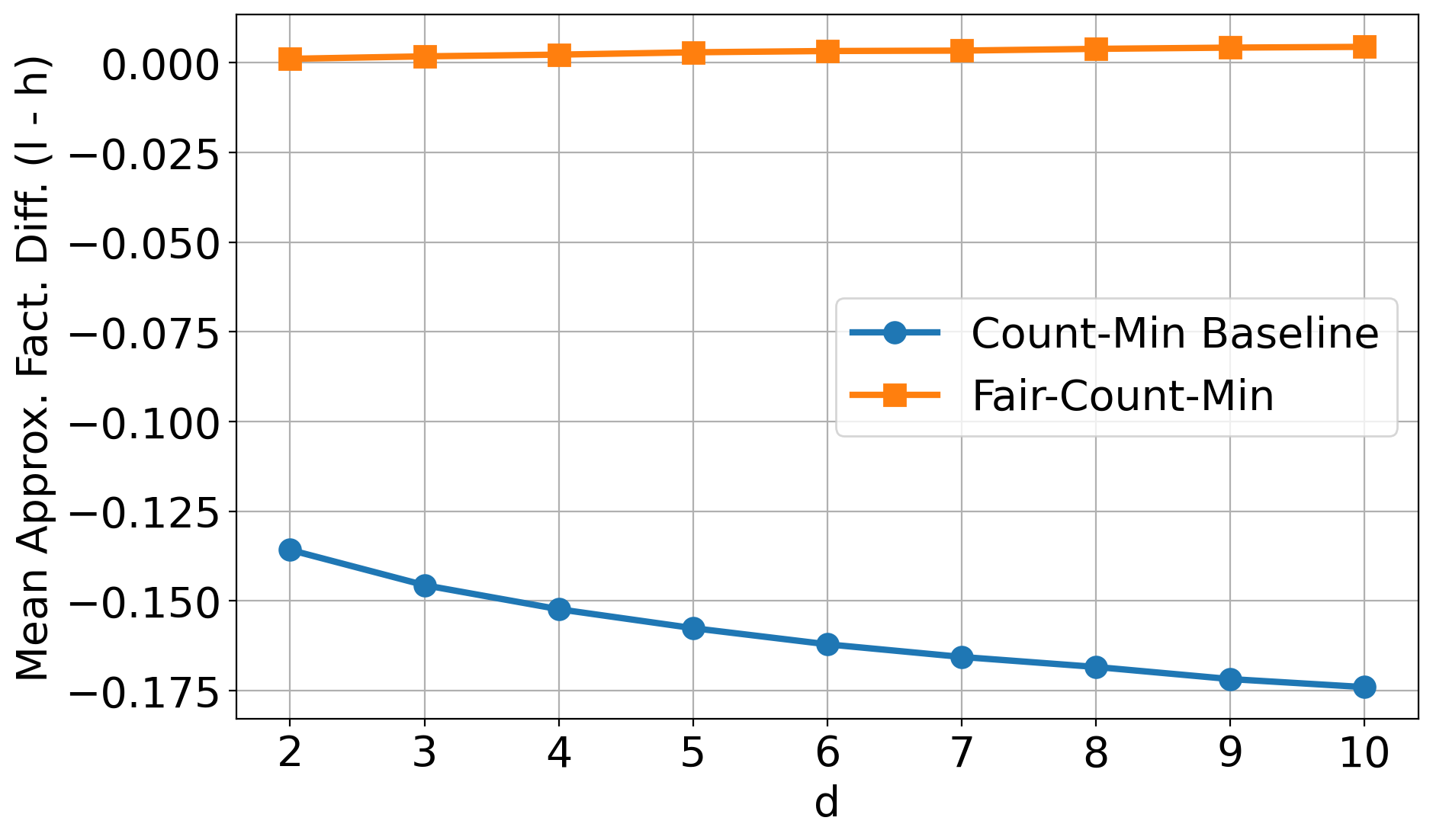}
        \vspace{-2.5em}
        \caption{effect of varying sketch depth $d$ on unfairness, \google, $n=1.2M, w=65536$.}
        \label{fig:}
    \end{minipage}
    \hfill
    \begin{minipage}[t]{0.32\linewidth}
        \centering
        \includegraphics[width=\textwidth]{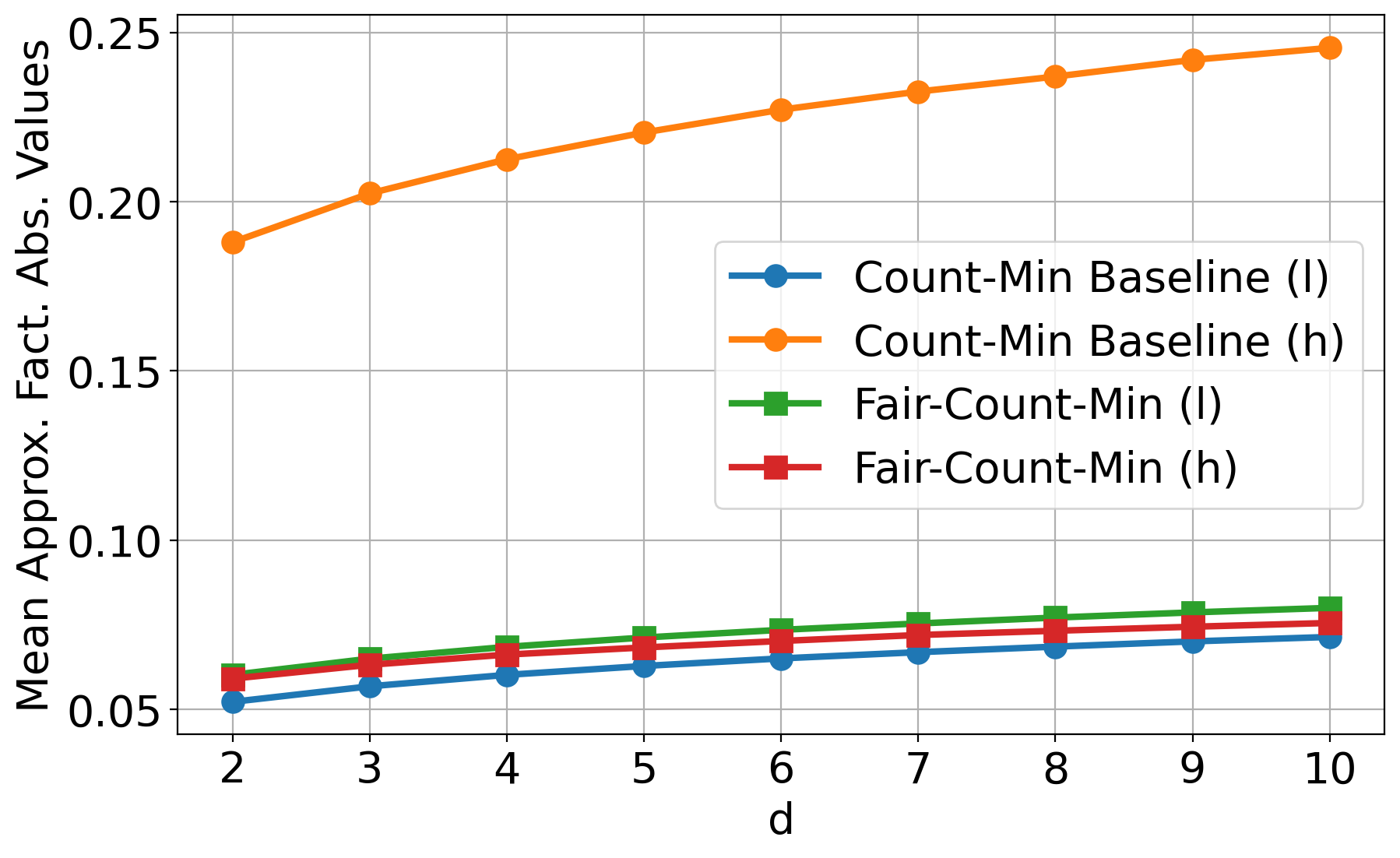}
        \vspace{-2.5em}
        \caption{effect of varying sketch depth $d$ on approximation factors, \google, $n=1.2M, w=65536$.}
        \label{fig:}
    \end{minipage}
\end{figure*}
\begin{figure*}
    \begin{minipage}[t]{0.32\linewidth}
        \centering
        \includegraphics[width=\textwidth]{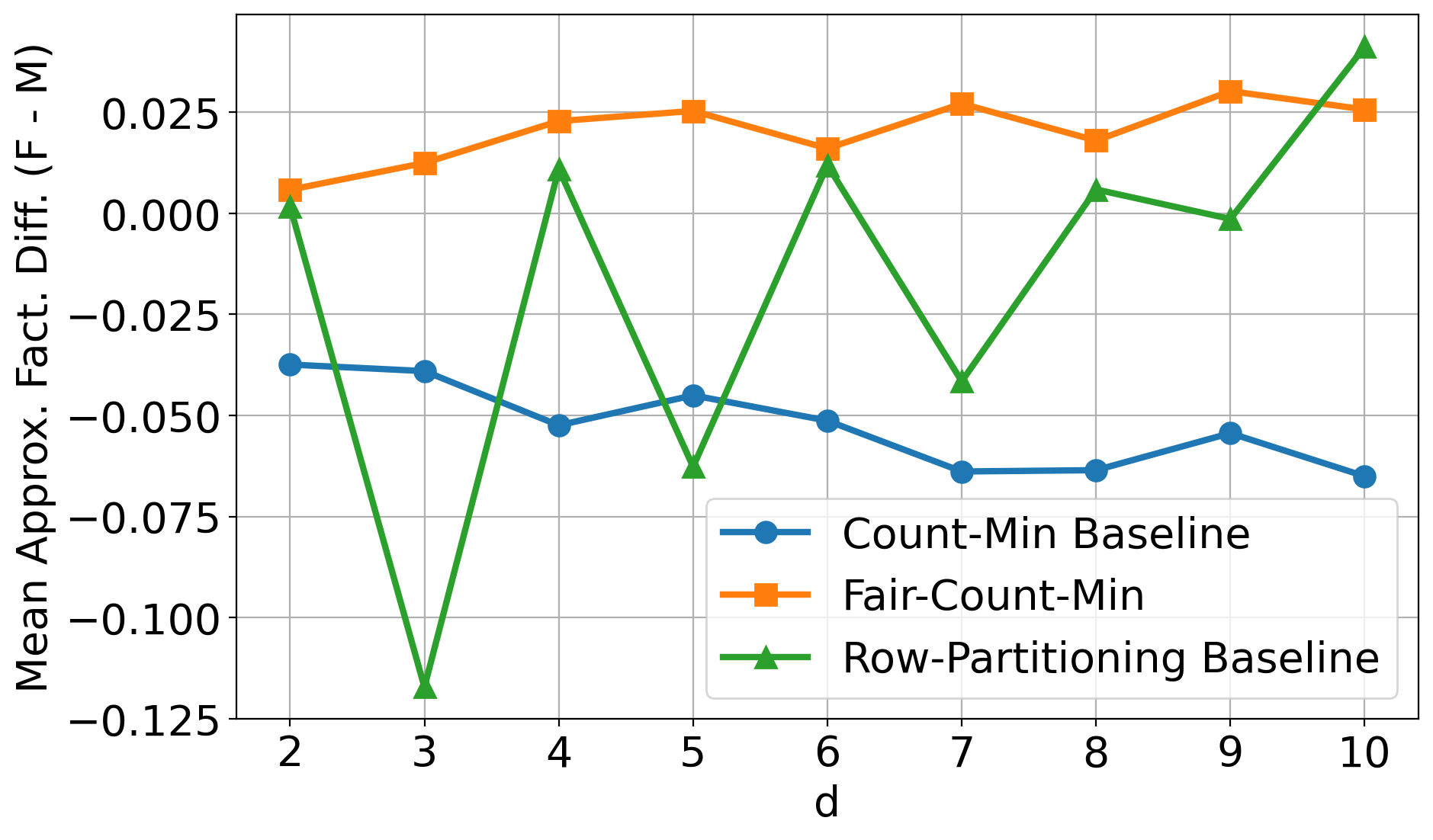}
        \vspace{-2.5em}
        \caption{effect of varying sketch depth $d$ on unfairness, \census, $n=430, w=64$.}
        \label{fig:}
    \end{minipage}
    \hfill
    \begin{minipage}[t]{0.32\linewidth}
        \centering
        \includegraphics[width=\textwidth]{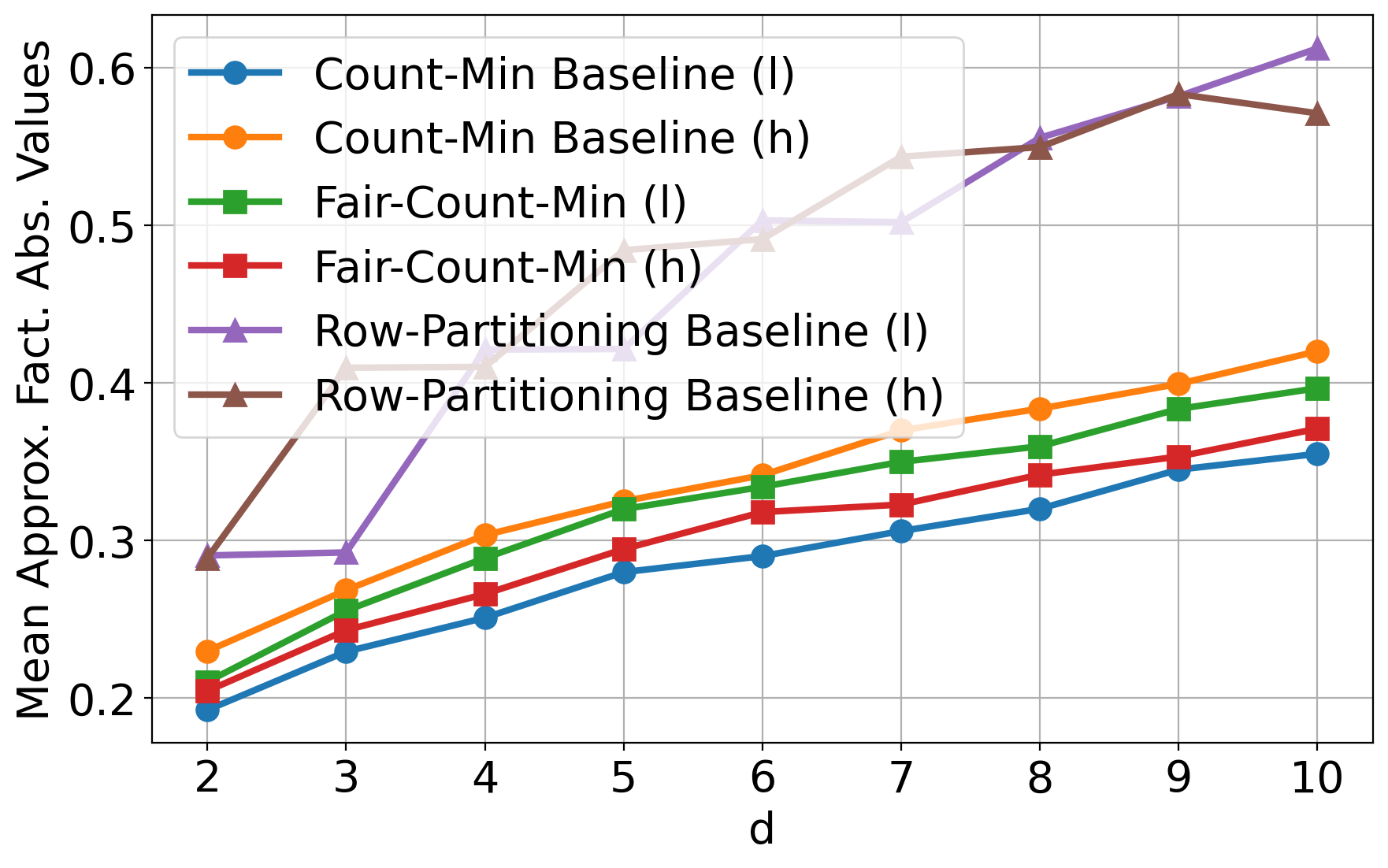}
        \vspace{-2.5em}
        \caption{effect of varying sketch depth $d$ on approximation factors, \census, $n=430, w=64$.}
        \label{fig:}
    \end{minipage}
    \hfill
    \begin{minipage}[t]{0.32\linewidth}
        \centering
        \includegraphics[width=\textwidth]{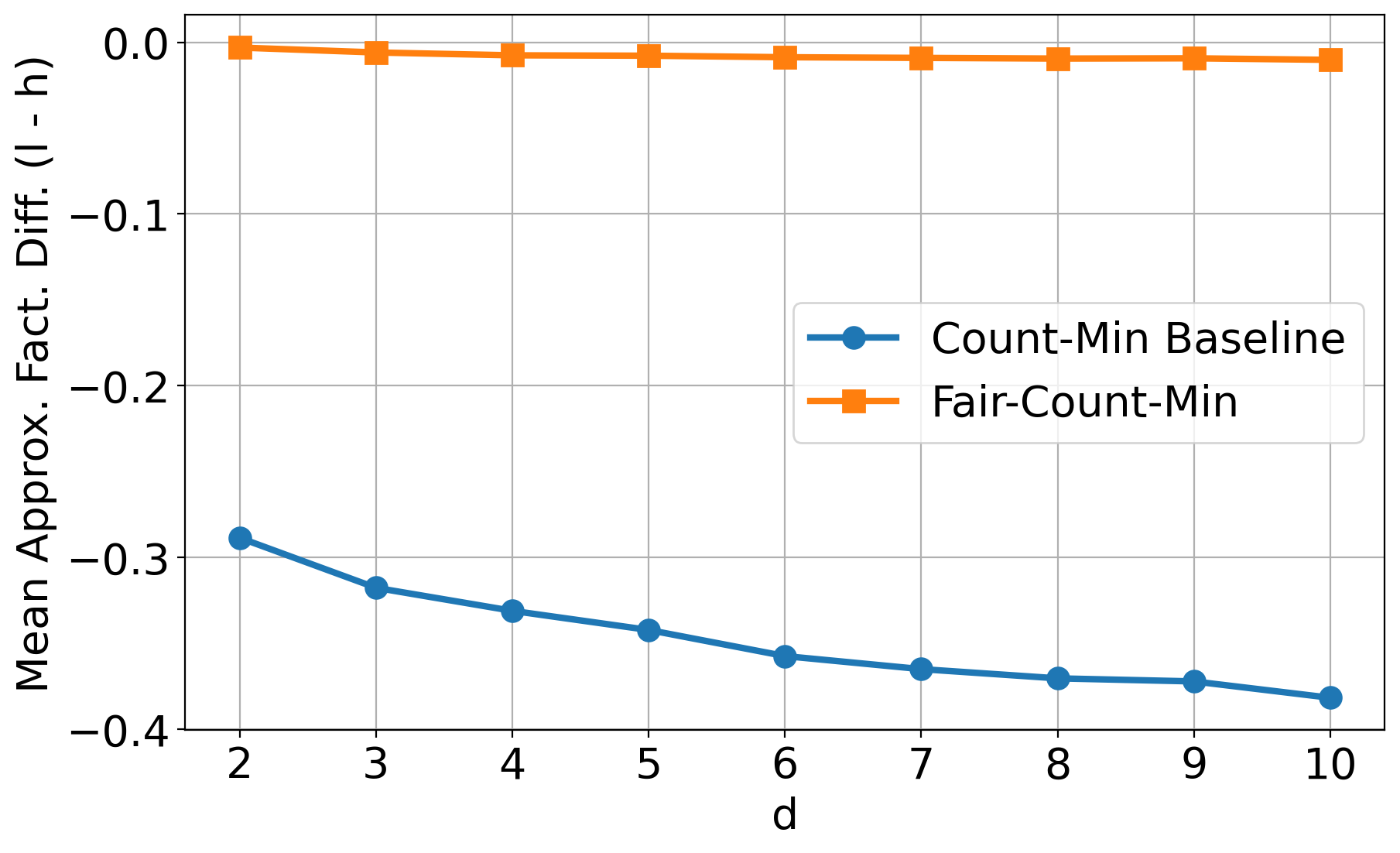}
        \vspace{-2.5em}
        \caption{effect of varying sketch depth $d$ on unfairness, \synthetic, $n=20000, w=512$.}
        \label{fig:}
    \end{minipage}
    \hfill
    \begin{minipage}[t]{0.32\linewidth}
        \centering
        \includegraphics[width=\textwidth]{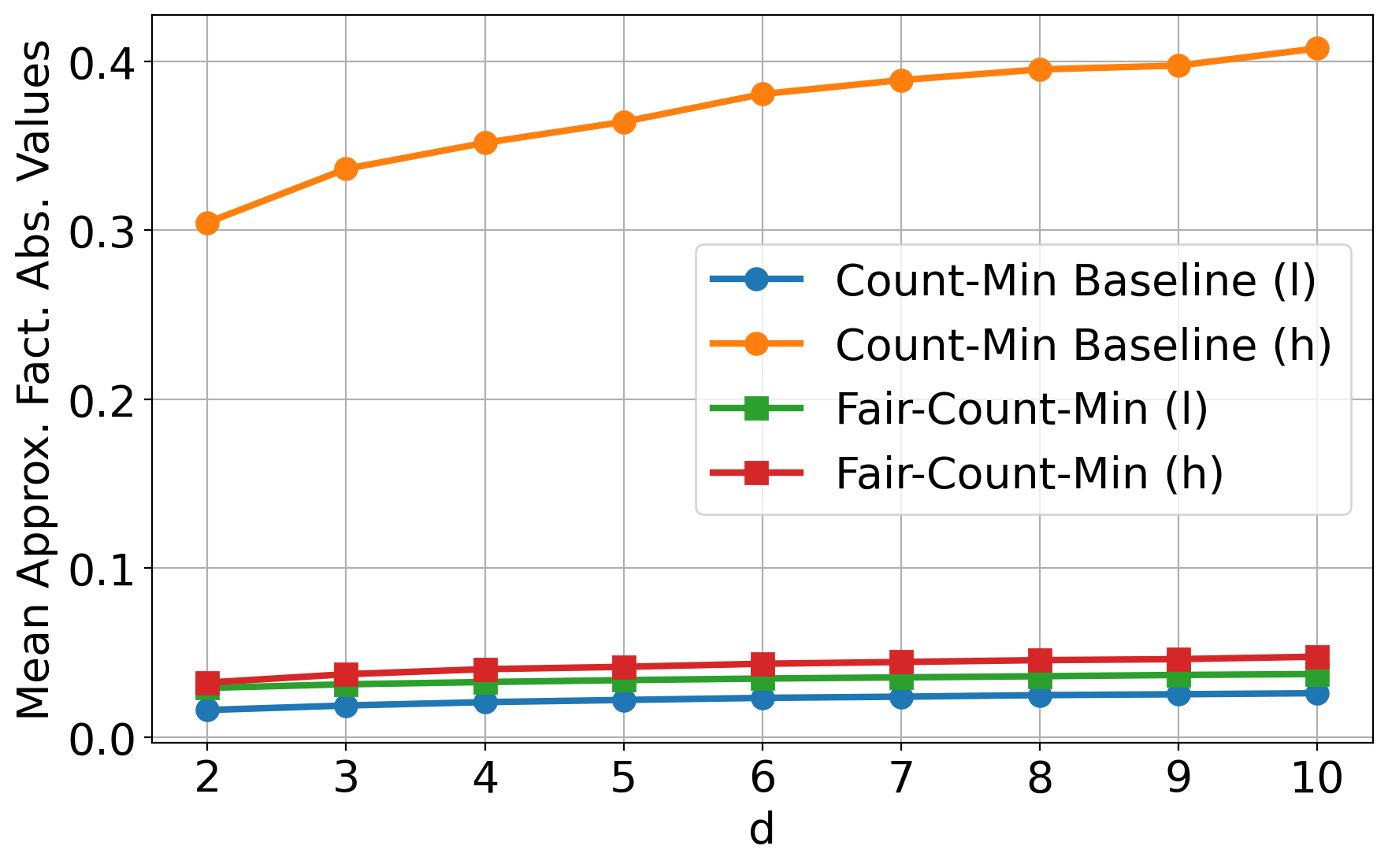}
        \vspace{-2.5em}
        \caption{effect of varying sketch depth $d$ on approximation factors, \synthetic, $n=20000, w=512$.}
        \label{fig:}
    \end{minipage}
    \hfill
    \begin{minipage}[t]{0.32\linewidth}
        \centering
        \includegraphics[width=\textwidth]{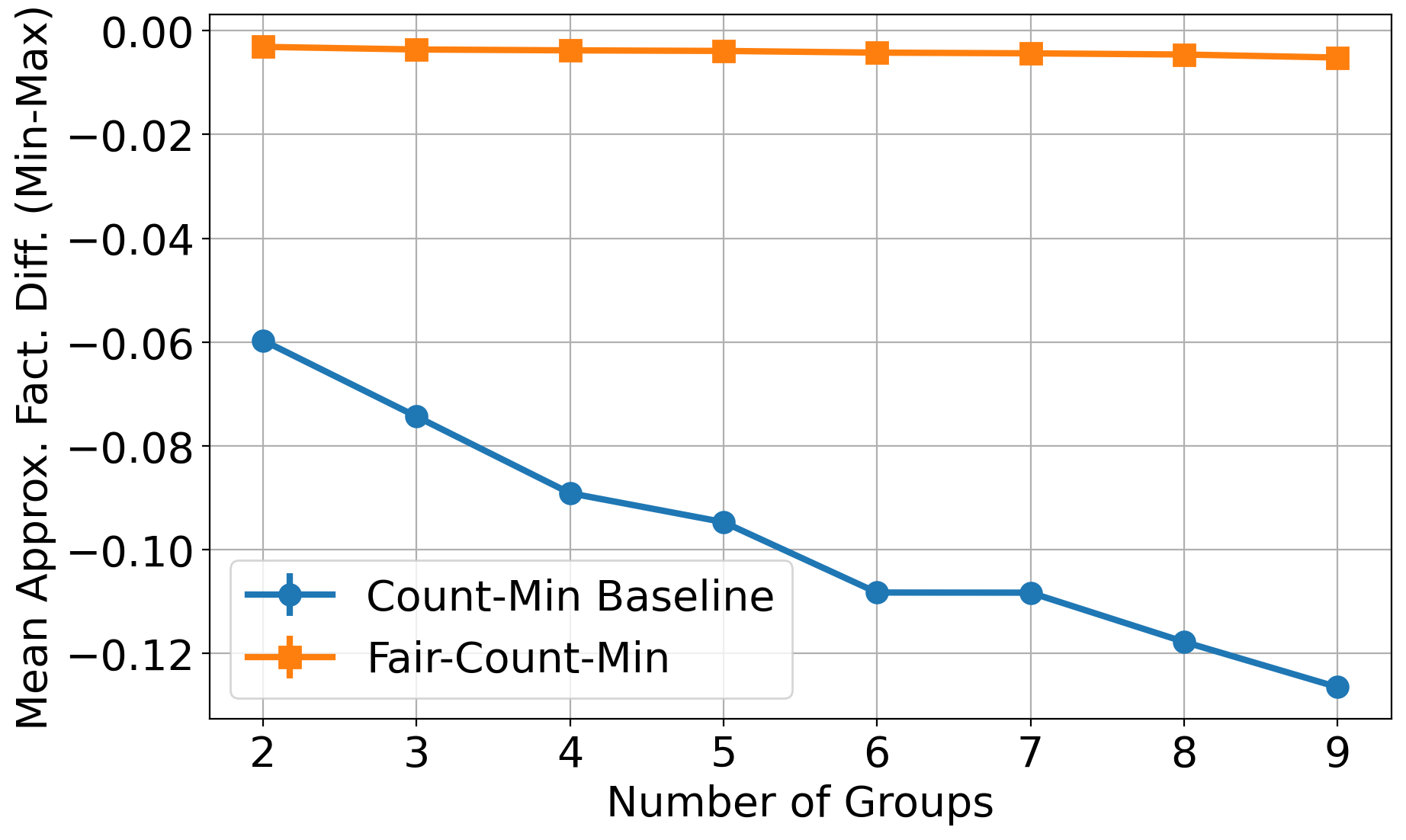}
        \vspace{-2.5em}
        \caption{effect of varying number of groups $\ell$ on unfairness, \google, $n=1.2M, W=65536, d=5$.}
        \label{fig:}
    \end{minipage}
    \hfill
    \begin{minipage}[t]{0.32\linewidth}
        \centering
        \includegraphics[width=\textwidth]{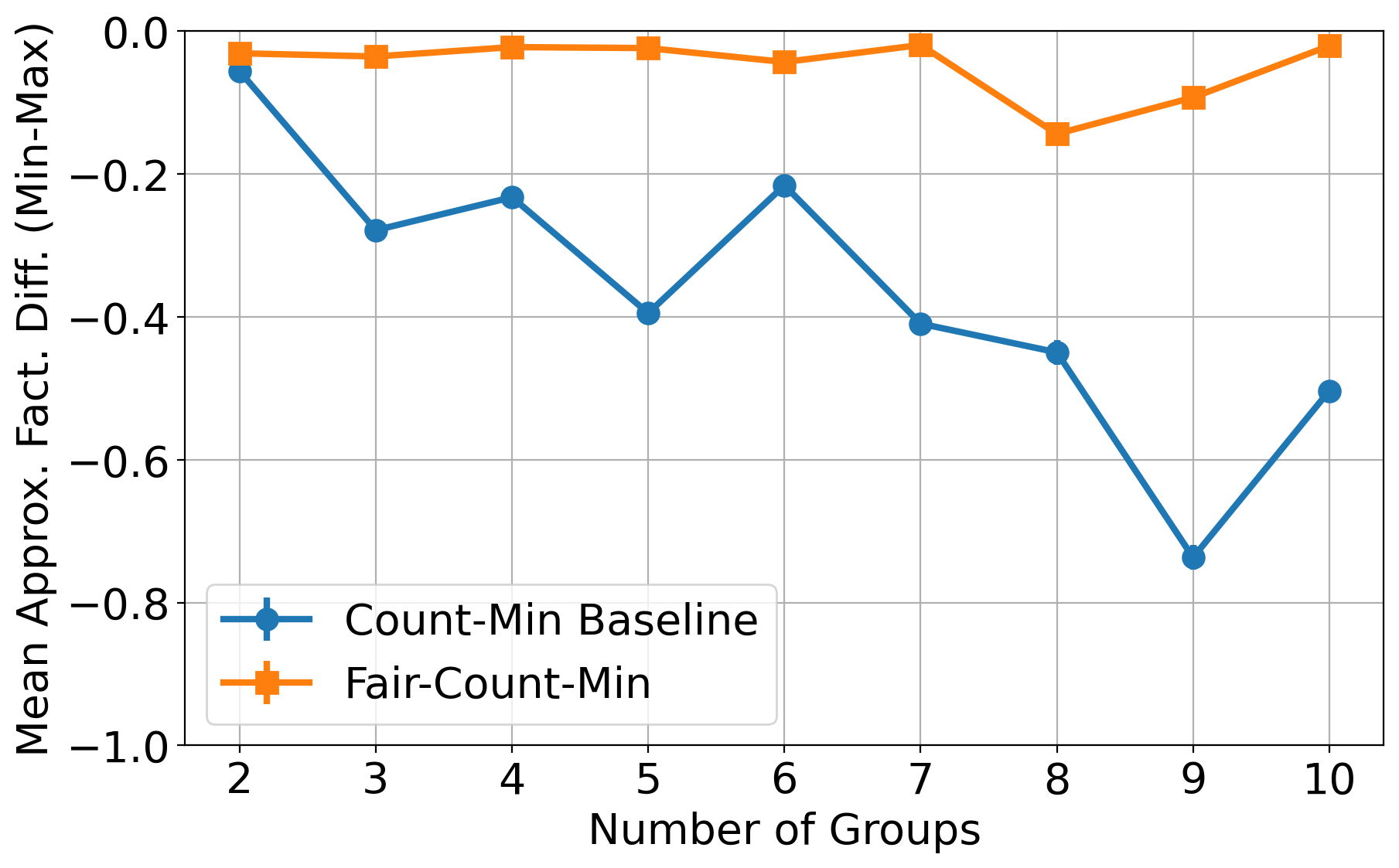}
        \vspace{-2.5em}
        \caption{effect of varying number of groups $\ell$ on unfairness, \census, $n=400, w=64, d=5$.}
        \label{fig:}
    \end{minipage}
    \hfill
    \begin{minipage}[t]{0.32\linewidth}
        \centering
        \includegraphics[width=\textwidth]{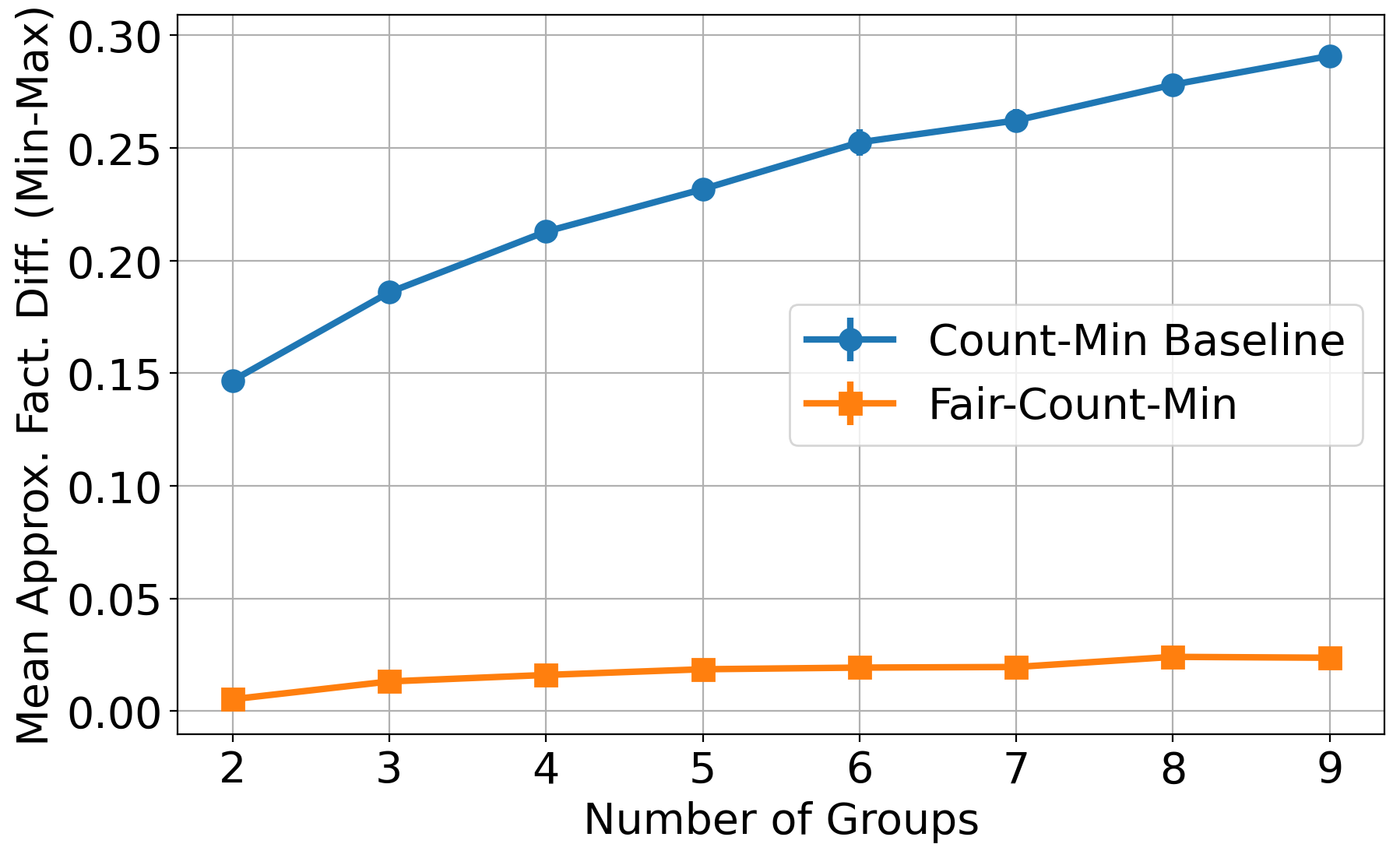}
        \vspace{-2.5em}
        \caption{effect of varying number of groups $\ell$ on unfairness, \synthetic, $n=20000, w=512, d=10$.}
        \label{fig:}
    \end{minipage}
    \hfill
    \begin{minipage}[t]{0.32\linewidth}
        \centering
        \includegraphics[width=\textwidth]{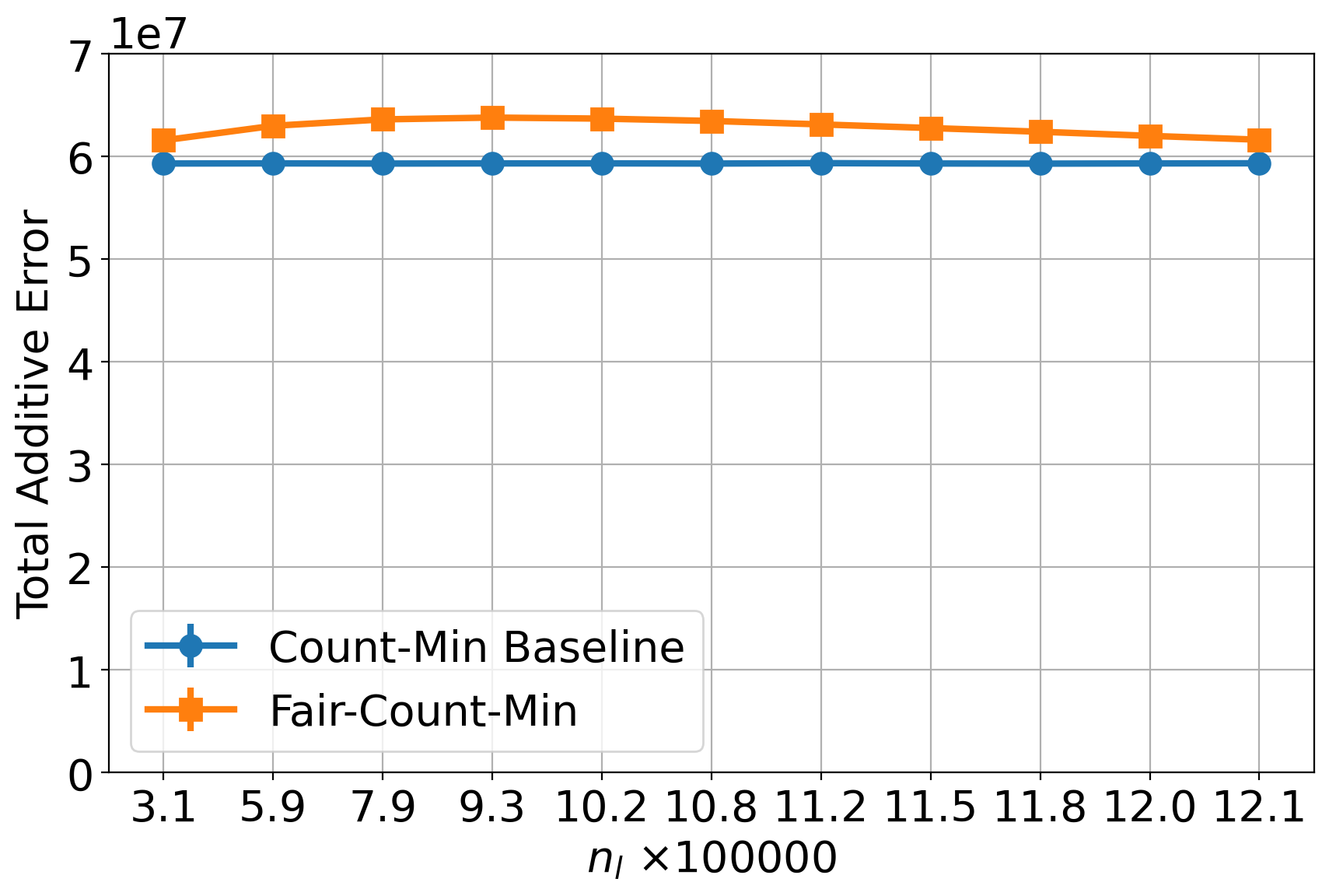}
        \vspace{-2.5em}
        \caption{effect of varying group size $n_l$ on price of fairness, \google, $w=65536, d=5$.}
        \label{fig:}
    \end{minipage}
    \hfill
    \begin{minipage}[t]{0.32\linewidth}
        \centering
        \includegraphics[width=\textwidth]{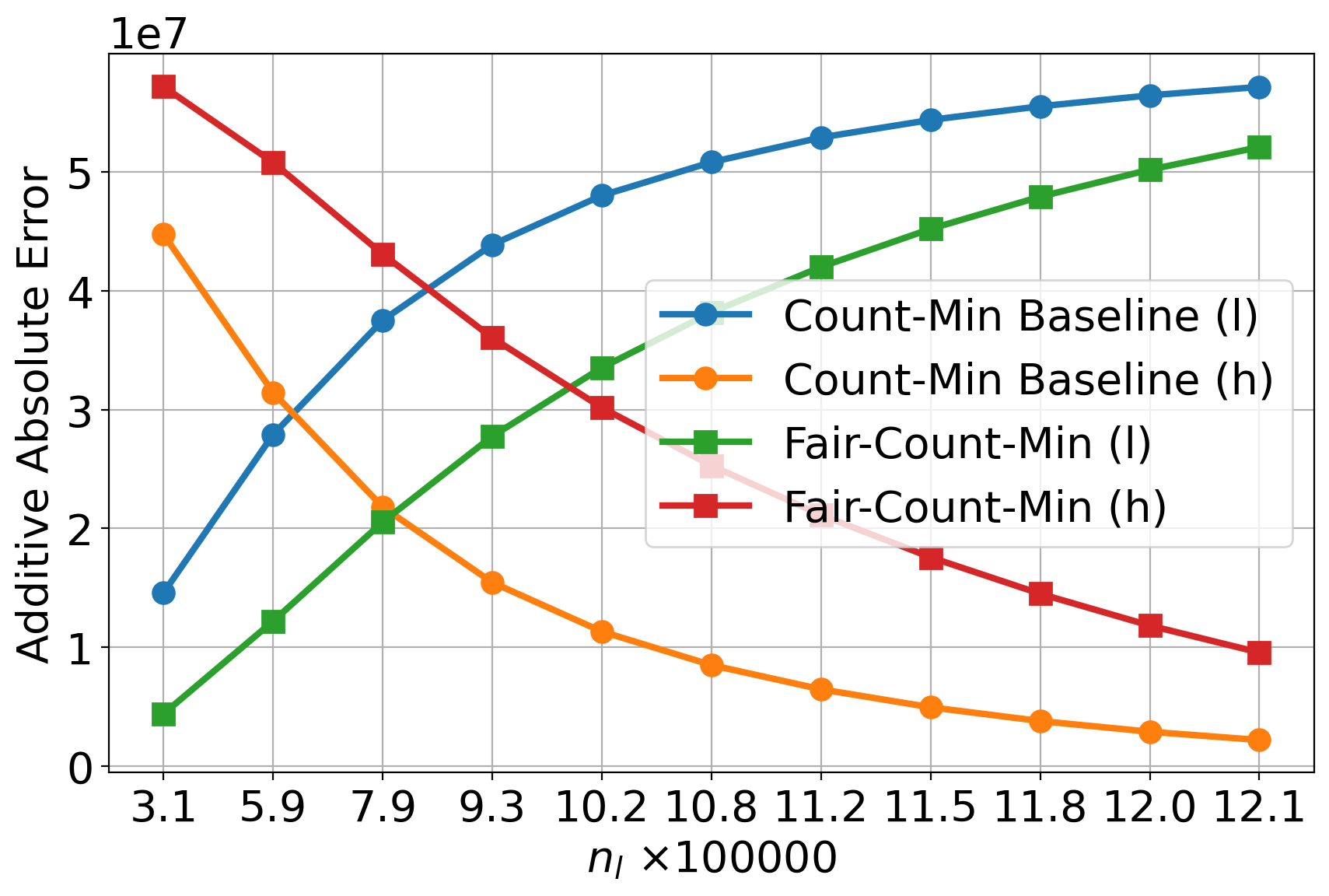}
        \vspace{-2.5em}
        \caption{effect of varying group size $n_l$ on absolute additive errors, \google, $w=65536, d=5$.}
        \label{fig:}
    \end{minipage}
    \hfill
    \begin{minipage}[t]{0.32\linewidth}
        \centering
        \includegraphics[width=\textwidth]{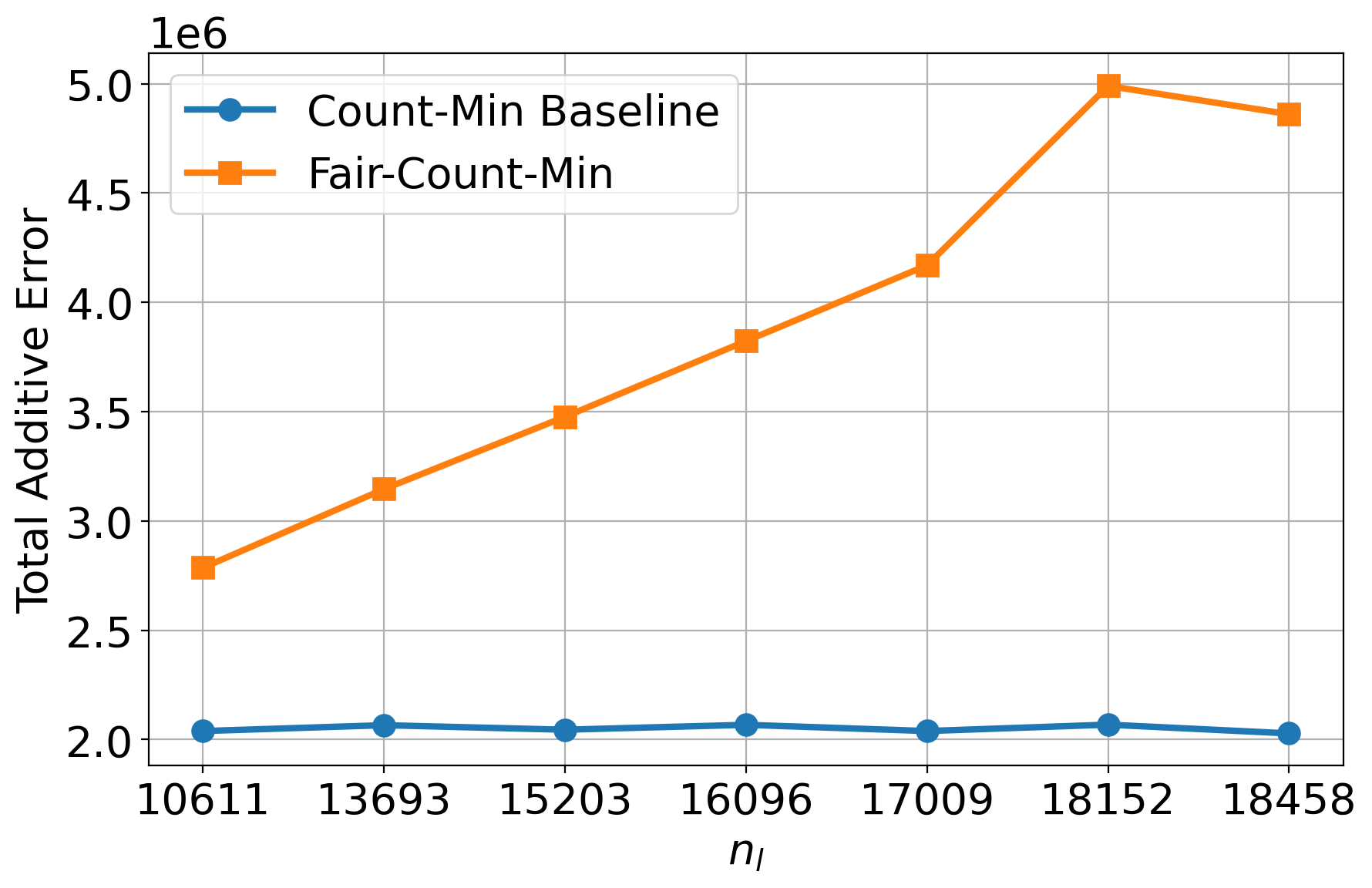}
        \vspace{-2.5em}
        \caption{effect of varying group size $n_l$ on price of fairness, \synthetic, $w=512, d=10$.}
        \label{fig:}
    \end{minipage}
    \hfill
    \begin{minipage}[t]{0.32\linewidth}
        \centering
        \includegraphics[width=\textwidth]{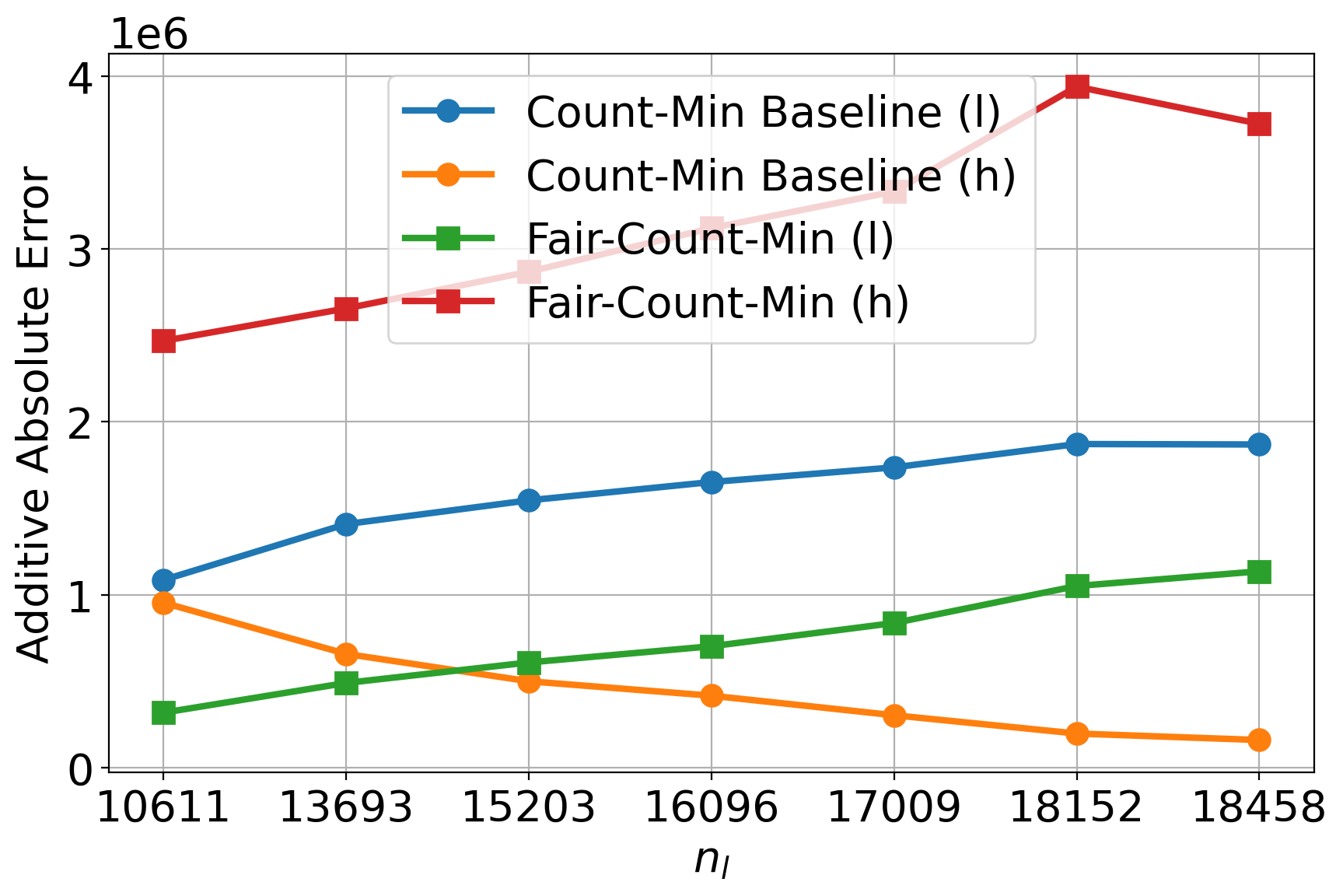}
        \vspace{-2.5em}
        \caption{effect of varying group size $n_l$ on absolute additive errors, \synthetic, $w=512, d=10$.}
        \label{fig:}
    \end{minipage}
    \hfill
    \begin{minipage}[t]{0.32\linewidth}
        \centering
        \includegraphics[width=\textwidth]{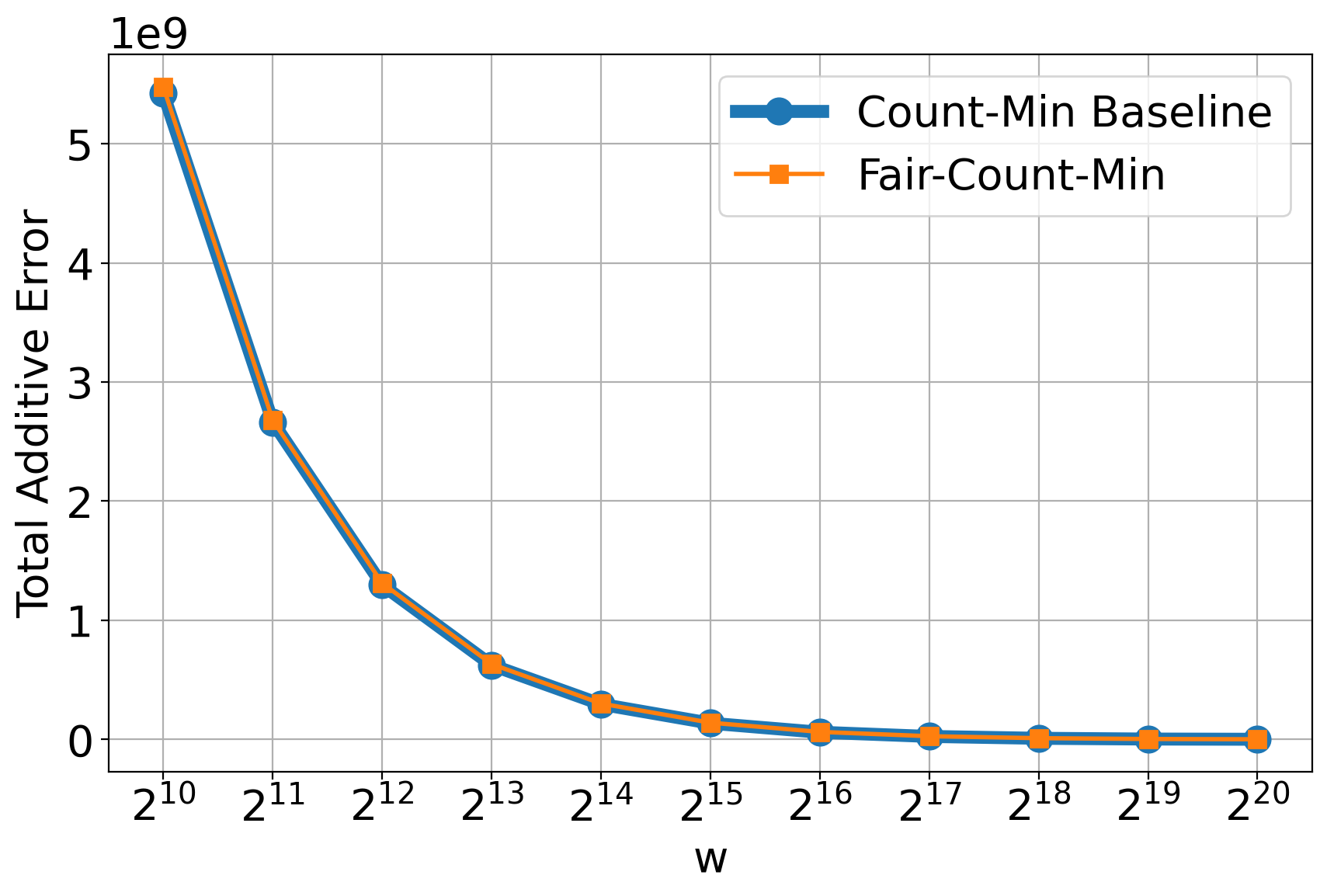}
        \vspace{-2.5em}
        \caption{effect of varying sketch width $w$ on price of fairness, \google, $n=1.2M, d=5$.}
        \label{fig:}
    \end{minipage}
\end{figure*}
\begin{figure*}
    \begin{minipage}[t]{0.32\linewidth}
        \centering
        \includegraphics[width=\textwidth]{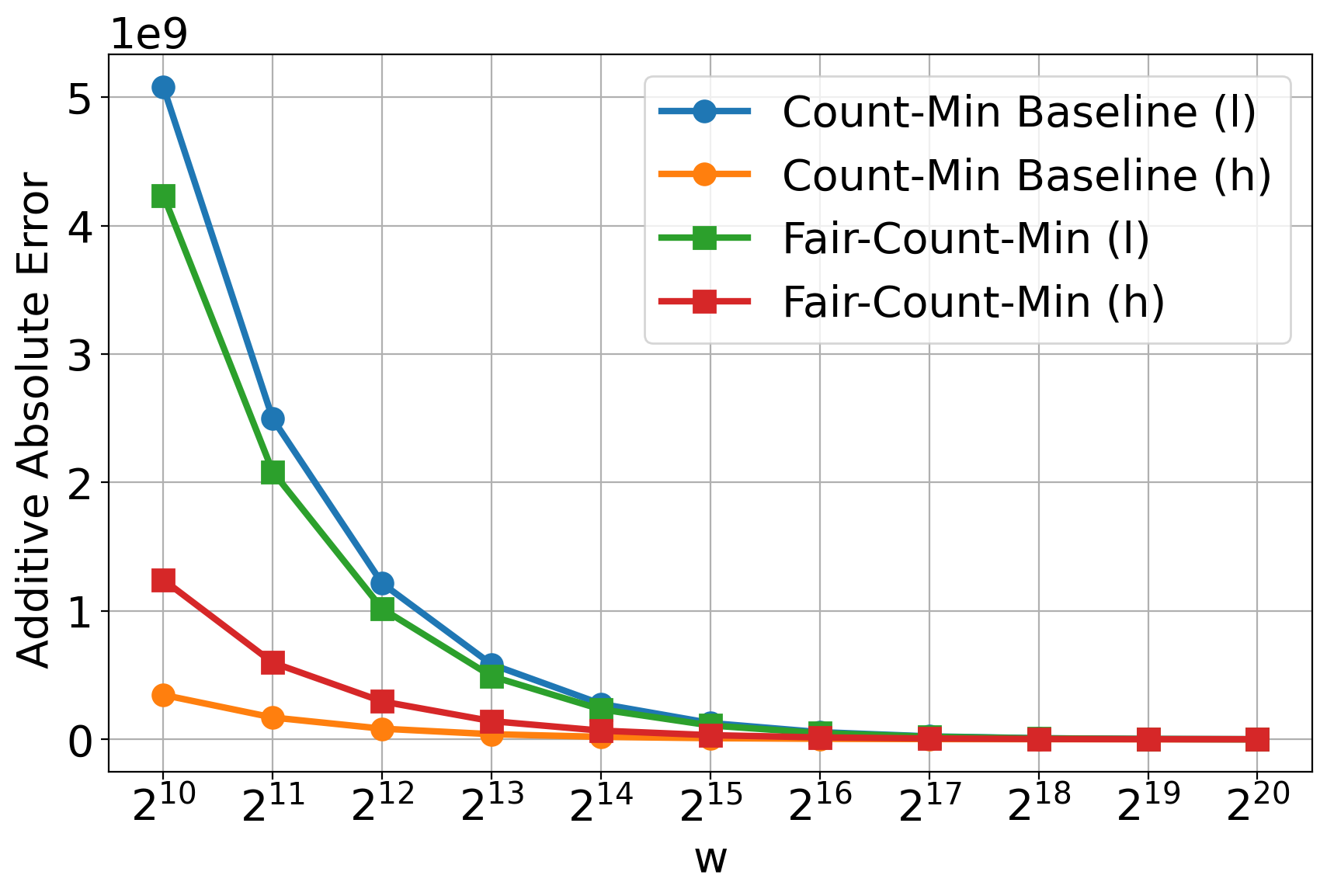}
        \vspace{-2.5em}
        \caption{effect of varying sketch width $w$ on absolute additive errors, \google, $n=1.2M, d=5$.}
        \label{fig:}
    \end{minipage}
    \hfill
    \begin{minipage}[t]{0.32\linewidth}
        \centering
        \includegraphics[width=\textwidth]{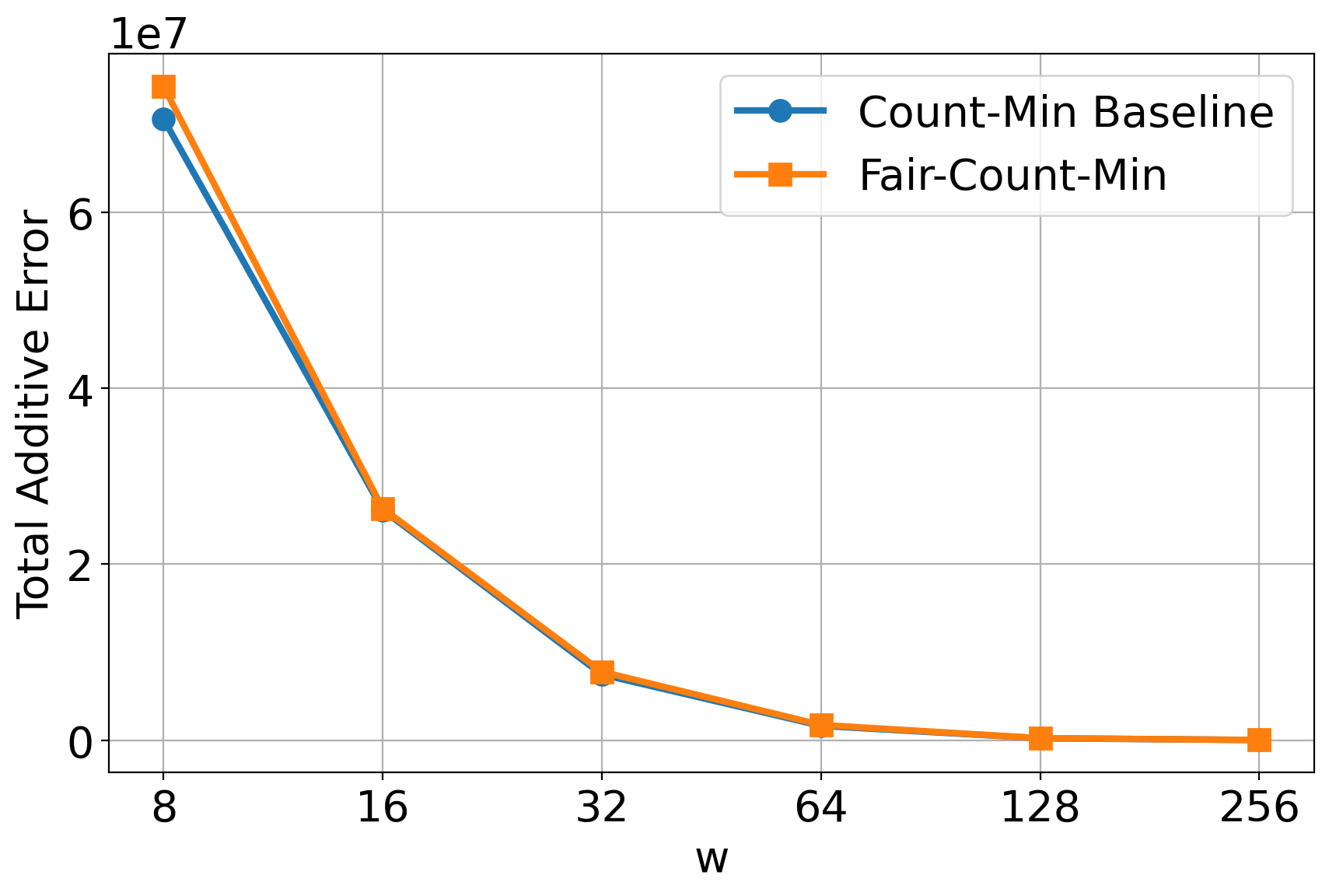}
        \vspace{-2.5em}
        \caption{effect of varying sketch width $w$ on price of fairness, \census, $n=430, d=5$.}
        \label{fig:}
    \end{minipage}
    \hfill
    \begin{minipage}[t]{0.32\linewidth}
        \centering
        \includegraphics[width=\textwidth]{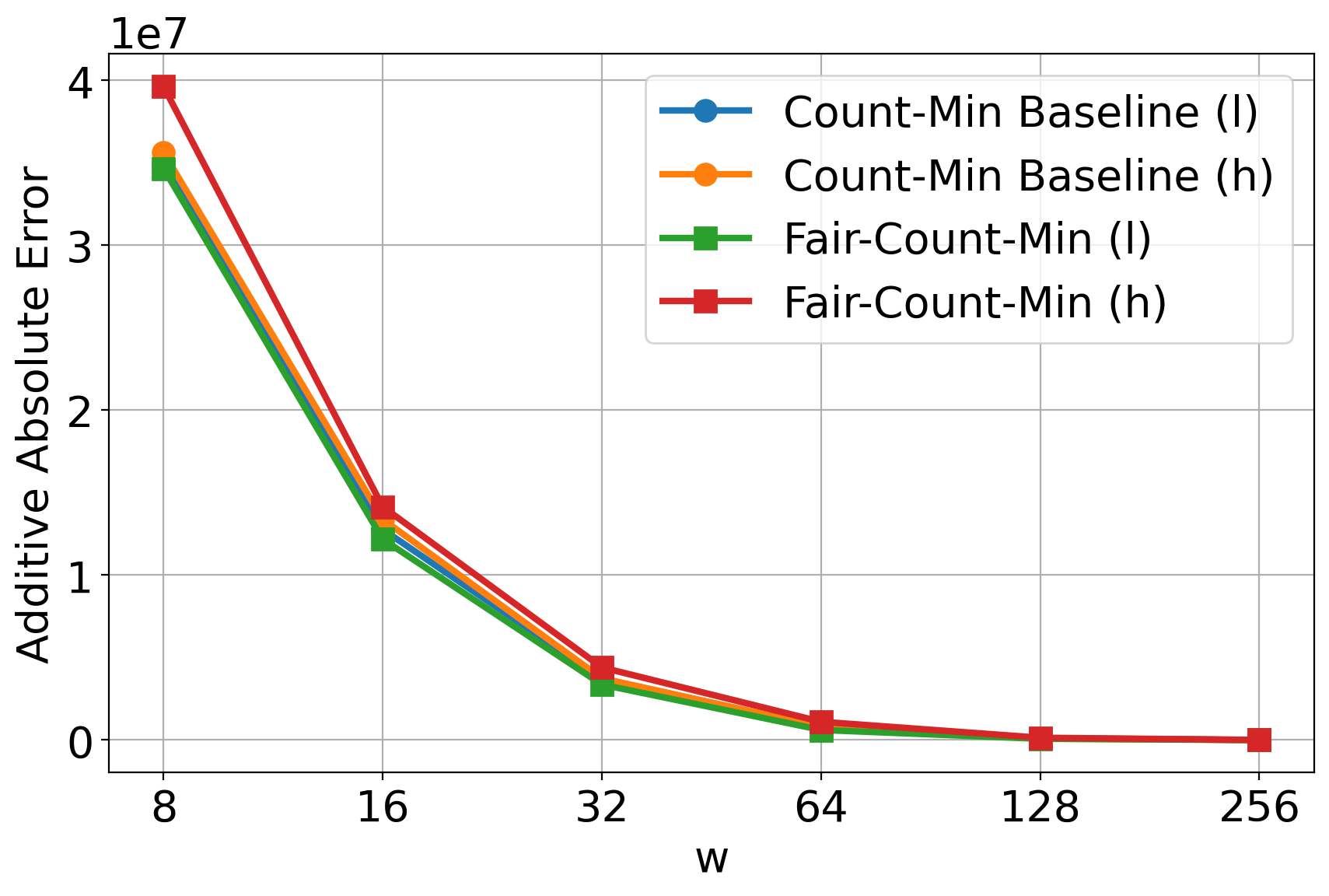}
        \vspace{-2.5em}
        \caption{effect of varying sketch width $w$ on absolute additive errors, \census, $n=430, d=5$.}
        \label{fig:}
    \end{minipage}
    \hfill
    \begin{minipage}[t]{0.32\linewidth}
        \centering
        \includegraphics[width=\textwidth]{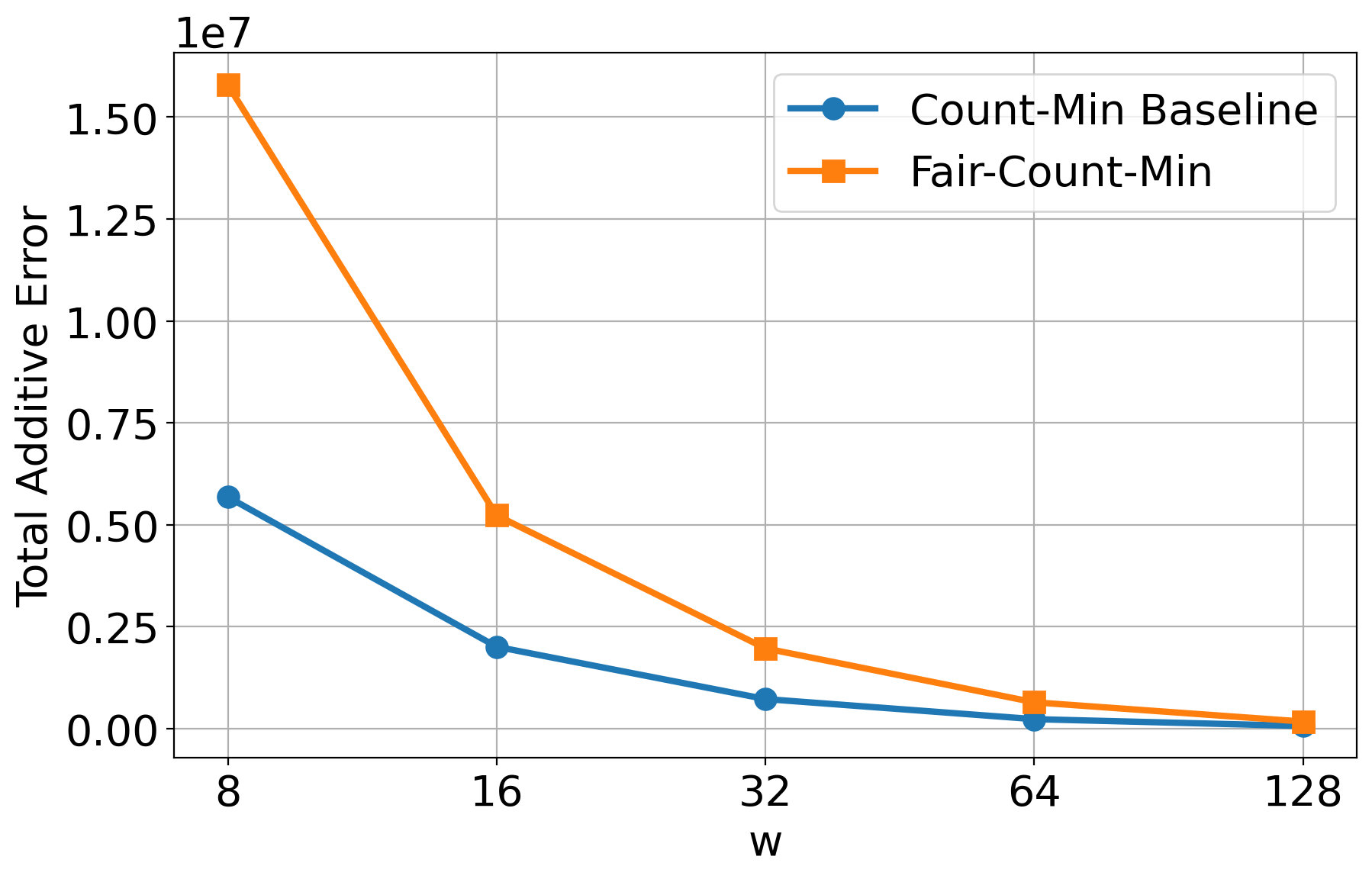}
        \vspace{-2.5em}
        \caption{effect of varying sketch width $w$ on price of fairness, \synthetic, $n=20000, d=10$.}
        \label{fig:}
    \end{minipage}
    \hfill
    \begin{minipage}[t]{0.32\linewidth}
        \centering
        \includegraphics[width=\textwidth]{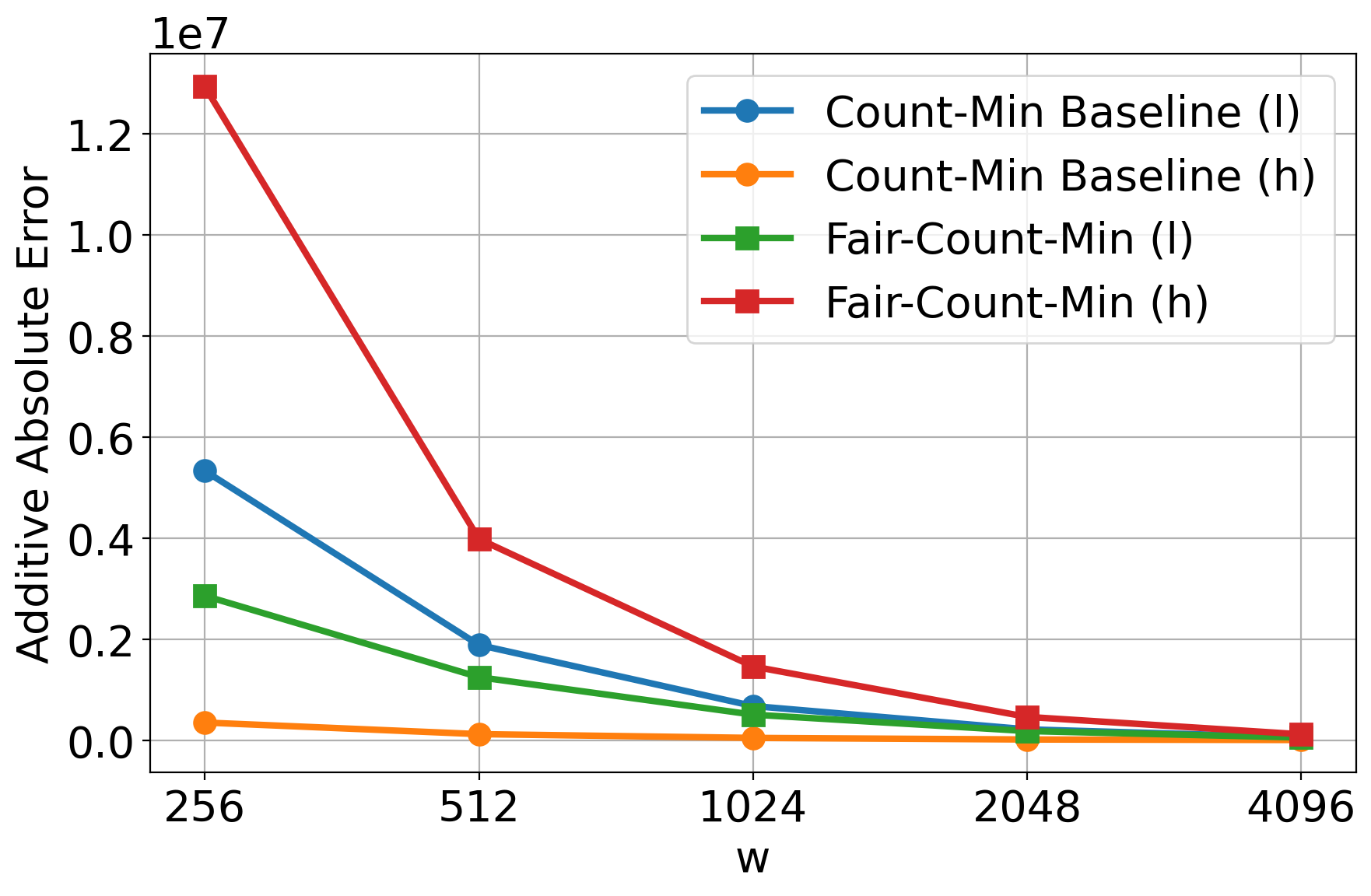}
        \vspace{-2.5em}
        \caption{effect of varying sketch width $w$ on absolute additive errors, \synthetic, $n=20000, d=10$.}
        \label{fig:}
    \end{minipage}
    \hfill
    \begin{minipage}[t]{0.32\linewidth}
        \centering
        \includegraphics[width=\textwidth]{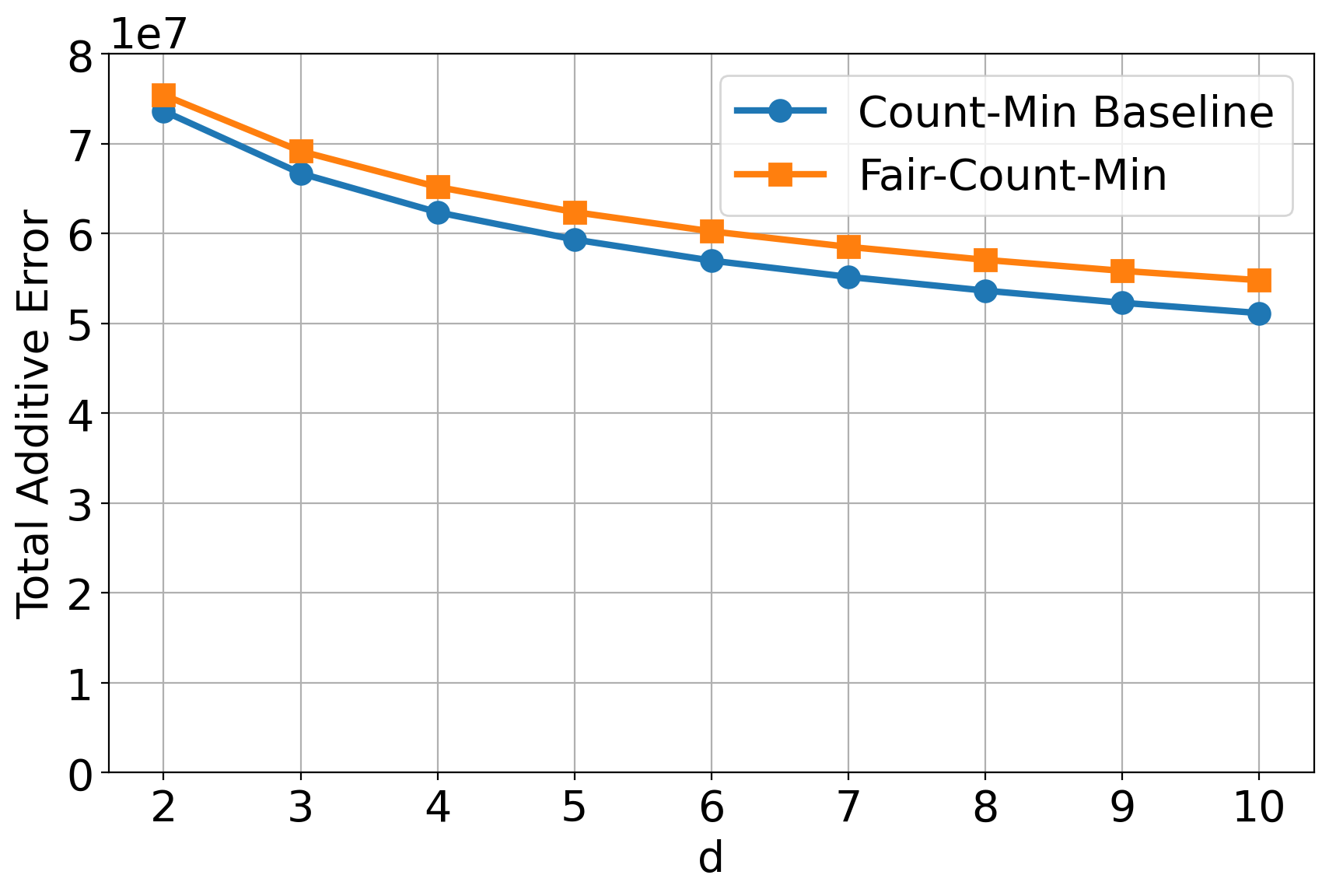}
        \vspace{-2.5em}
        \caption{effect of varying sketch depth $d$ on price of fairness, \google, $n=1.2M, w=65536$.}
        \label{fig:}
    \end{minipage}
    \hfill
    \begin{minipage}[t]{0.32\linewidth}
        \centering
        \includegraphics[width=\textwidth]{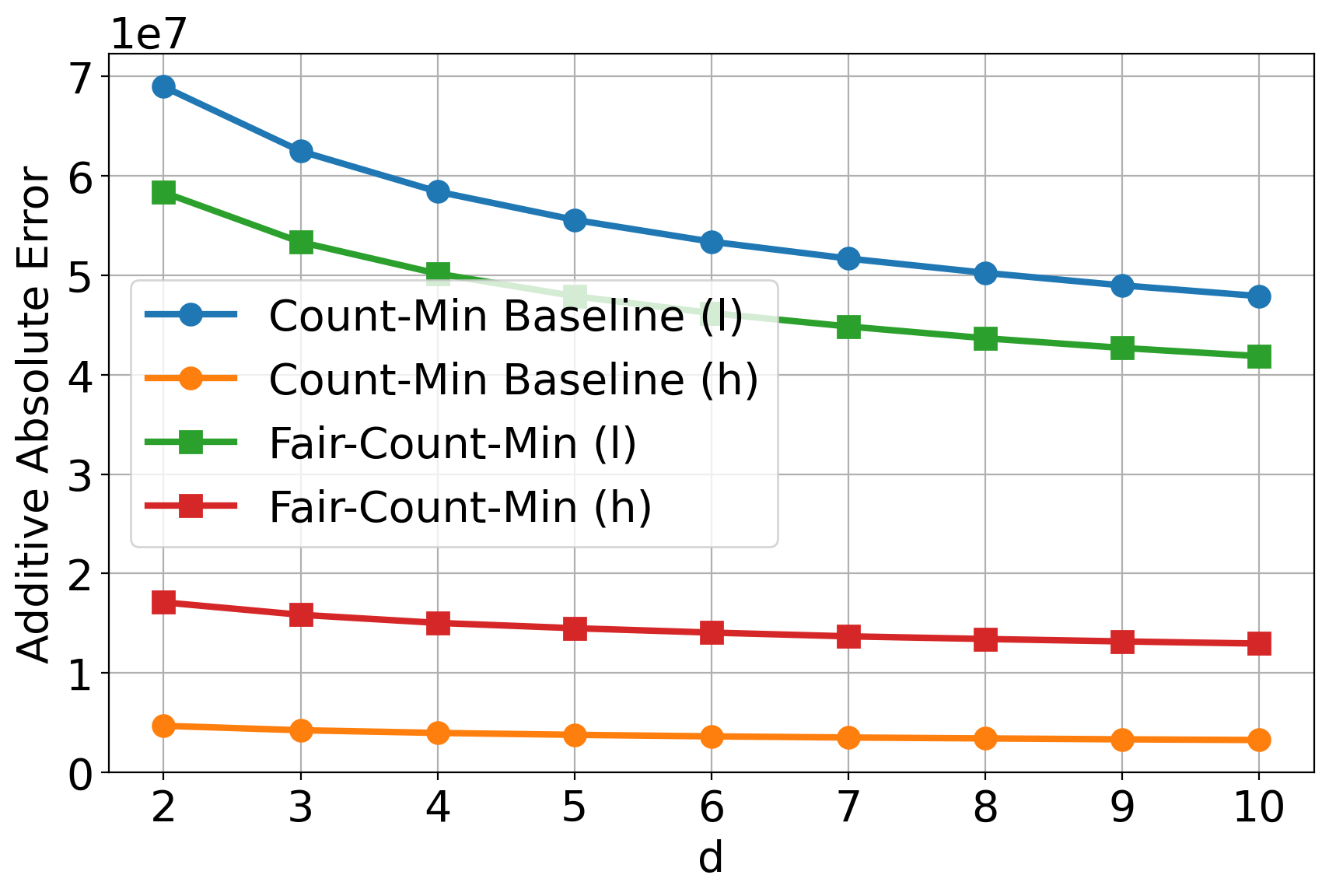}
        \vspace{-2.5em}
        \caption{effect of varying sketch depth $d$ on absolute additive errors, \google, $n=1.2M, w=65536$.}
        \label{fig:}
    \end{minipage}
    \hfill
    \begin{minipage}[t]{0.32\linewidth}
        \centering
        \includegraphics[width=\textwidth]{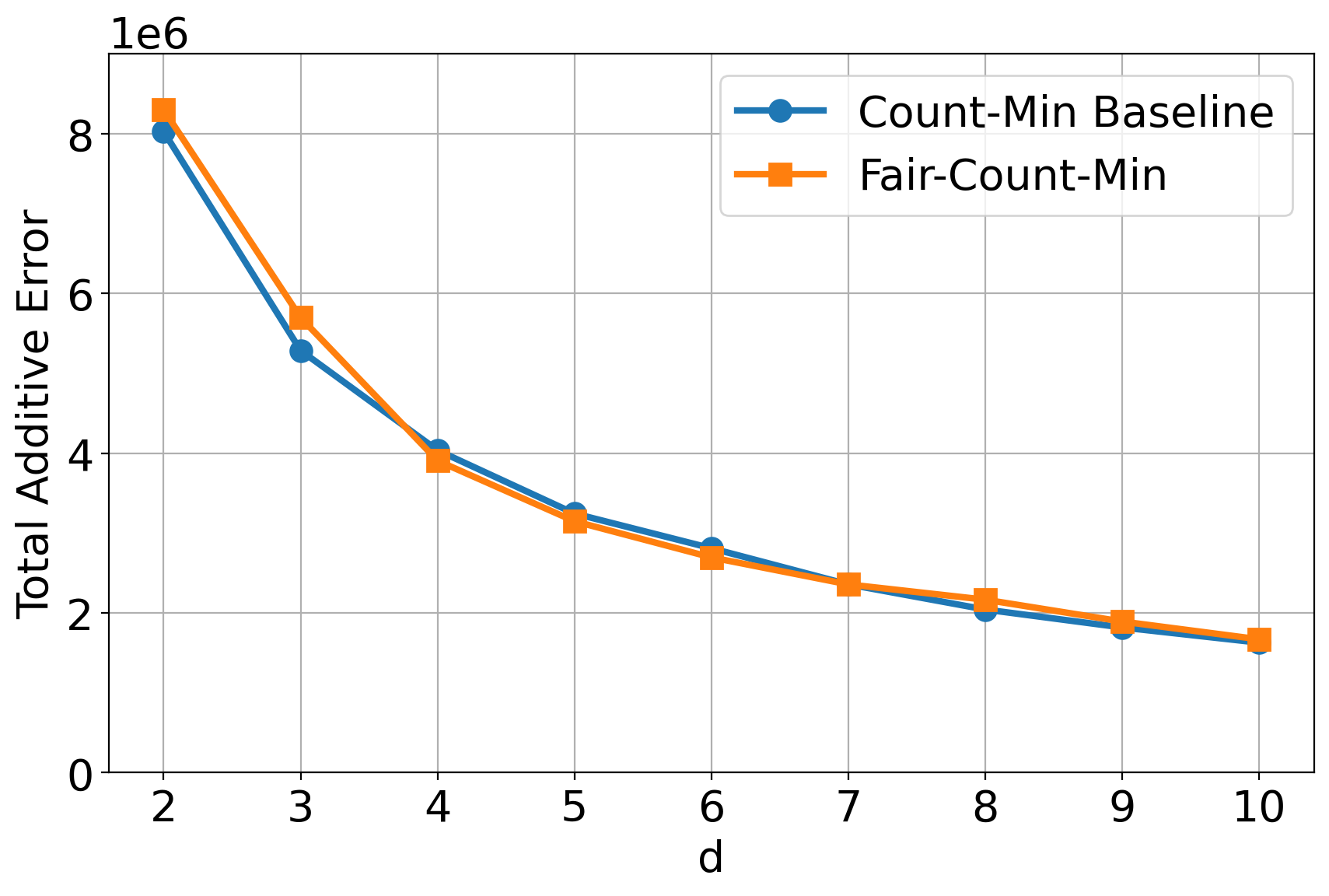}
        \vspace{-2.5em}
        \caption{effect of varying sketch depth $d$ on price of fairness, \census, $n=430, w=64$.}
        \label{fig:}
    \end{minipage}
    \hfill
    \begin{minipage}[t]{0.32\linewidth}
        \centering
        \includegraphics[width=\textwidth]{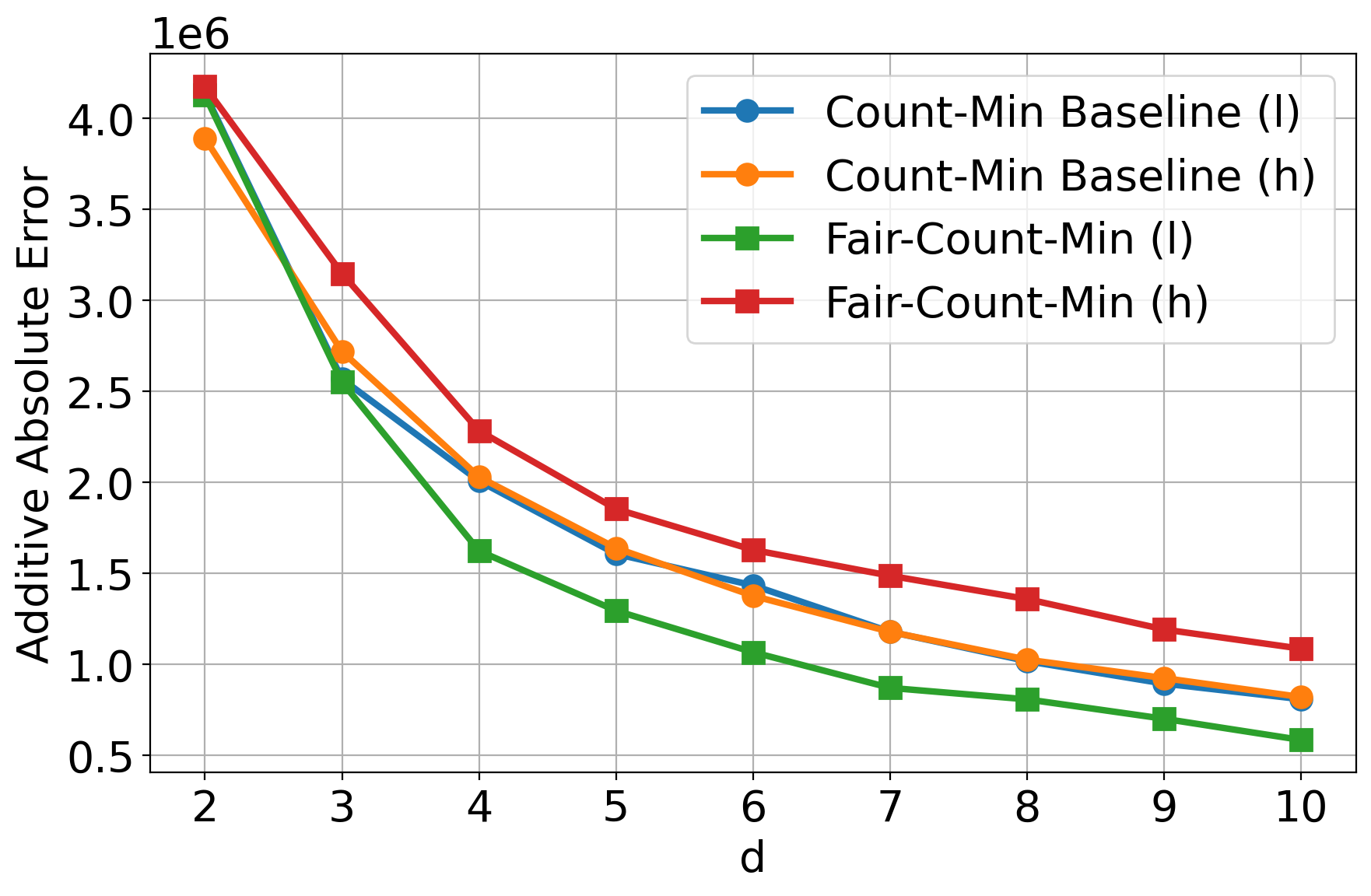}
        \vspace{-2.5em}
        \caption{effect of varying sketch depth $d$ on absolute additive errors, \census, $n=430, w=64$.}
        \label{fig:}
    \end{minipage}
    \hfill
    \begin{minipage}[t]{0.32\linewidth}
        \centering
        \includegraphics[width=\textwidth]{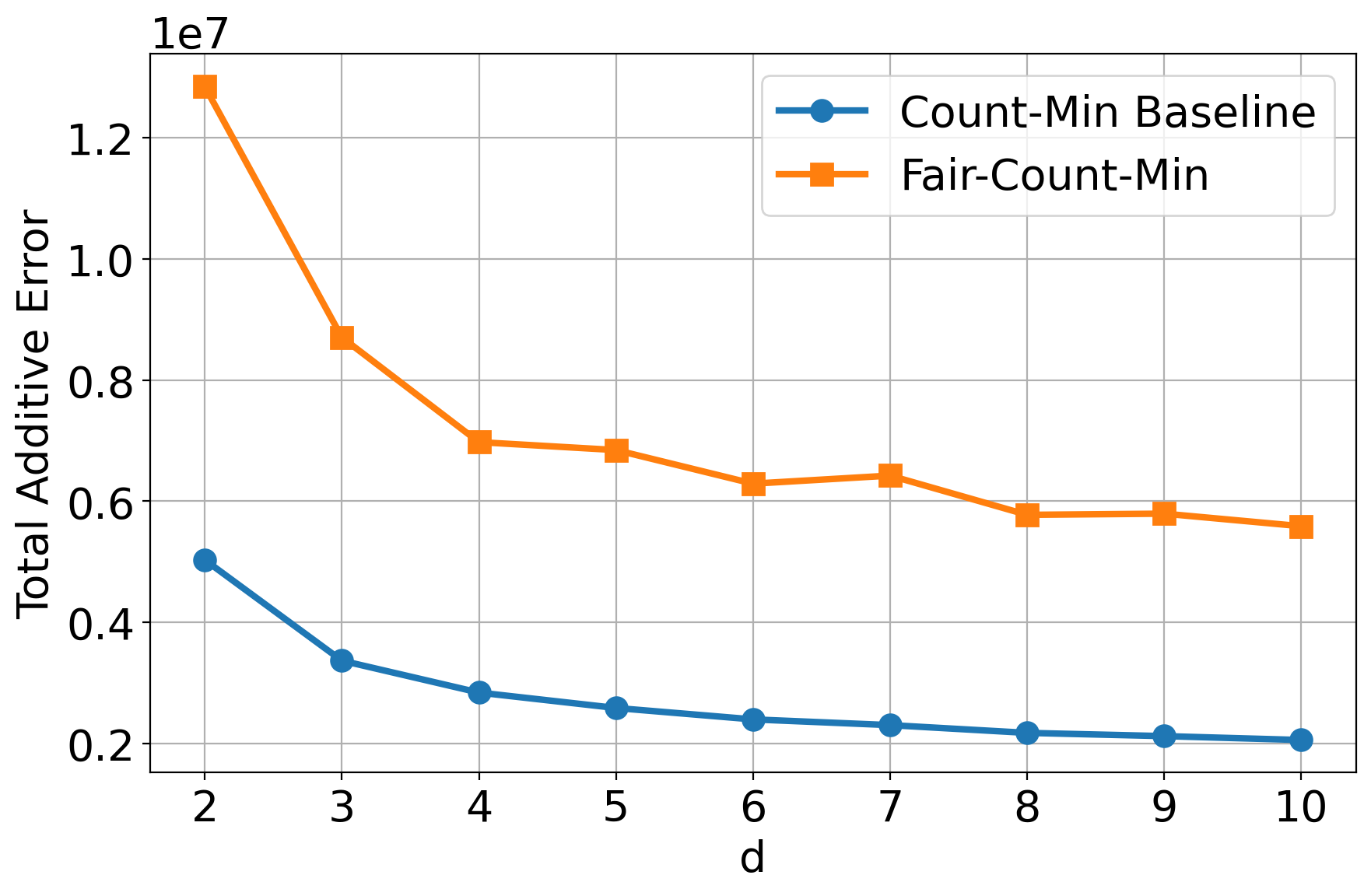}
        \vspace{-2.5em}
        \caption{effect of varying sketch depth $d$ on price of fairness, \synthetic, $n=20000, w=512$.}
        \label{fig:}
    \end{minipage}
    \hfill
    \begin{minipage}[t]{0.32\linewidth}
        \centering
        \includegraphics[width=\textwidth]{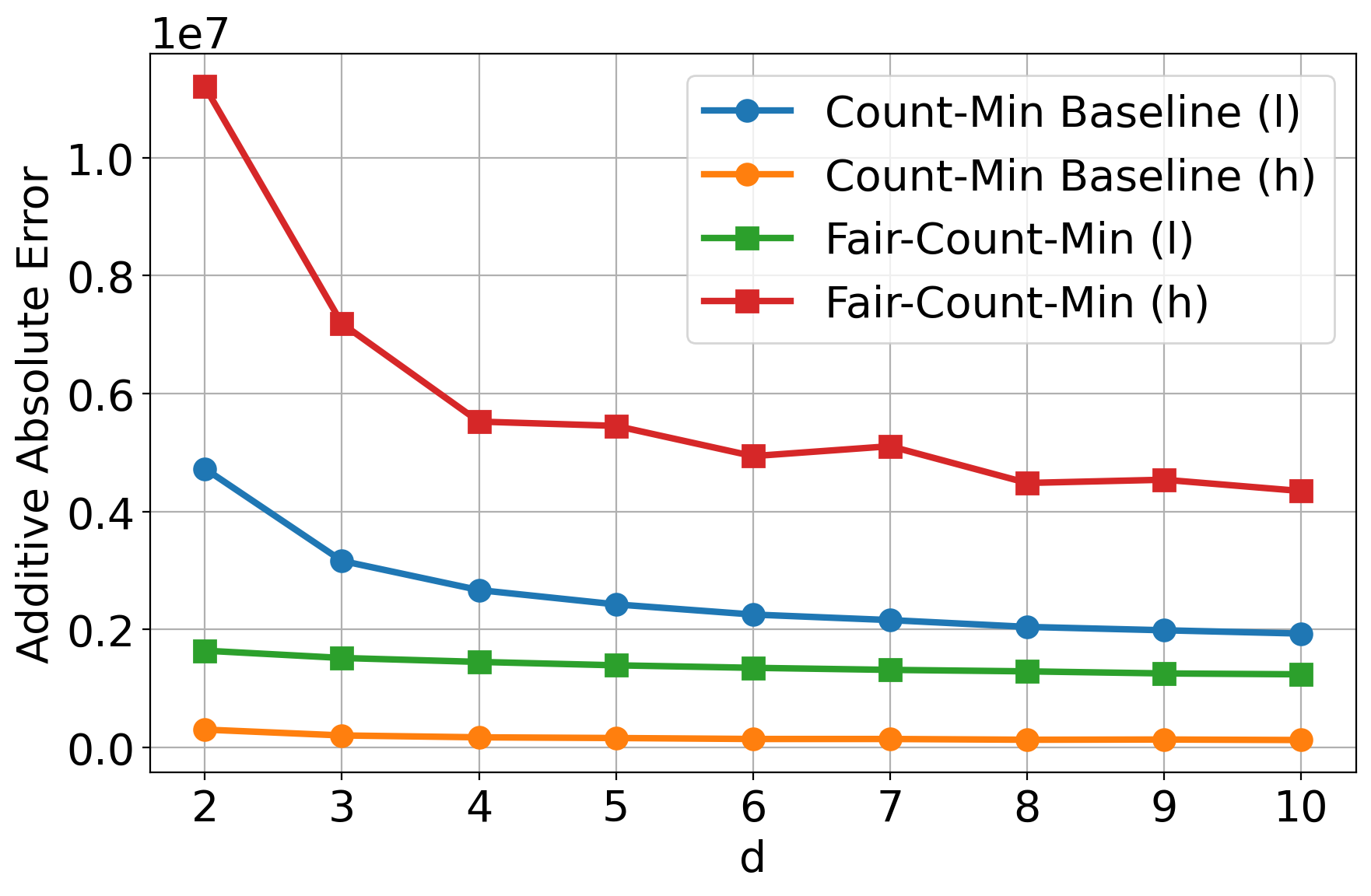}
        \vspace{-2.5em}
        \caption{effect of varying sketch depth $d$ on absolute additive errors, \synthetic, $n=20000, w=512$.}
        \label{fig:}
    \end{minipage}
    \hfill
    \begin{minipage}[t]{0.32\linewidth}
        \centering
        \includegraphics[width=\textwidth]{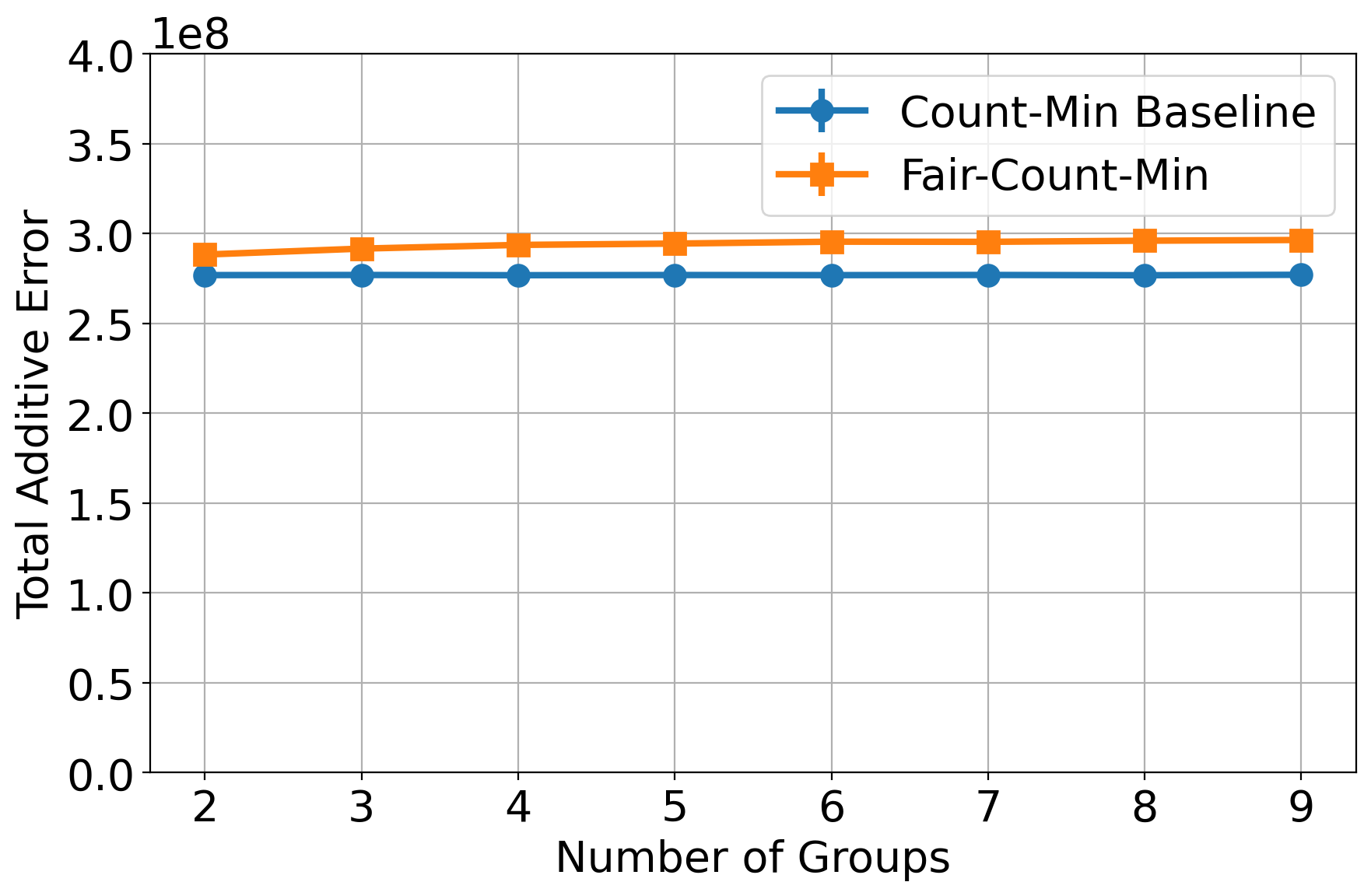}
        \vspace{-2.5em}
        \caption{effect of varying number of groups $\ell$ on price of fairness, \google, $n=1.2M, W=65536, d=5$.}
        \label{fig:}
    \end{minipage}
\end{figure*}

\begin{figure*}
    \begin{minipage}[t]{0.32\linewidth}
        \centering
        \includegraphics[width=\textwidth]{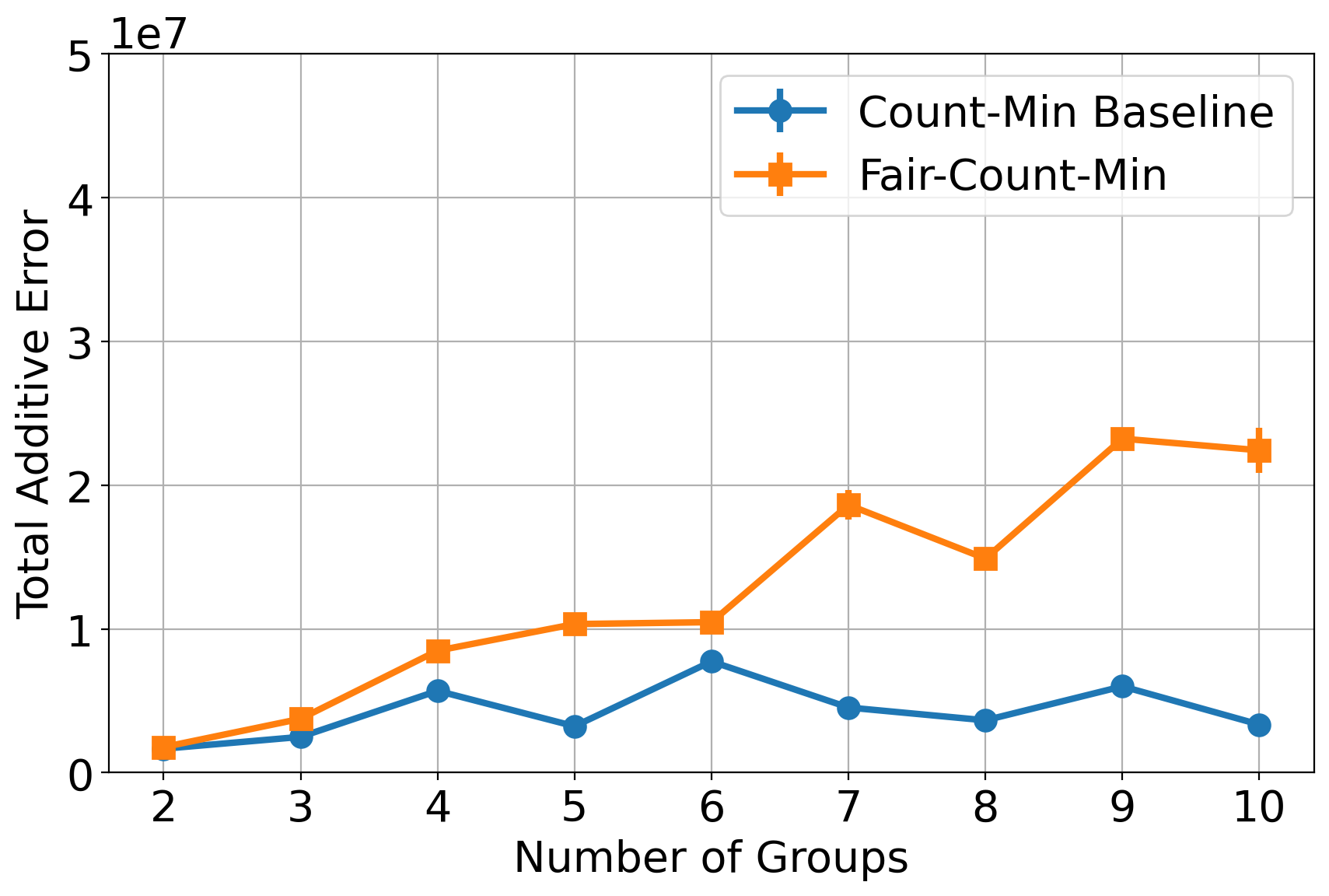}
        \vspace{-2.5em}
        \caption{effect of varying number of groups $\ell$ on price of fairness, \census, $n=400, w=64, d=5$.}
        \label{fig:}
    \end{minipage}
    \hfill
    \begin{minipage}[t]{0.32\linewidth}
        \centering
        \includegraphics[width=\textwidth]{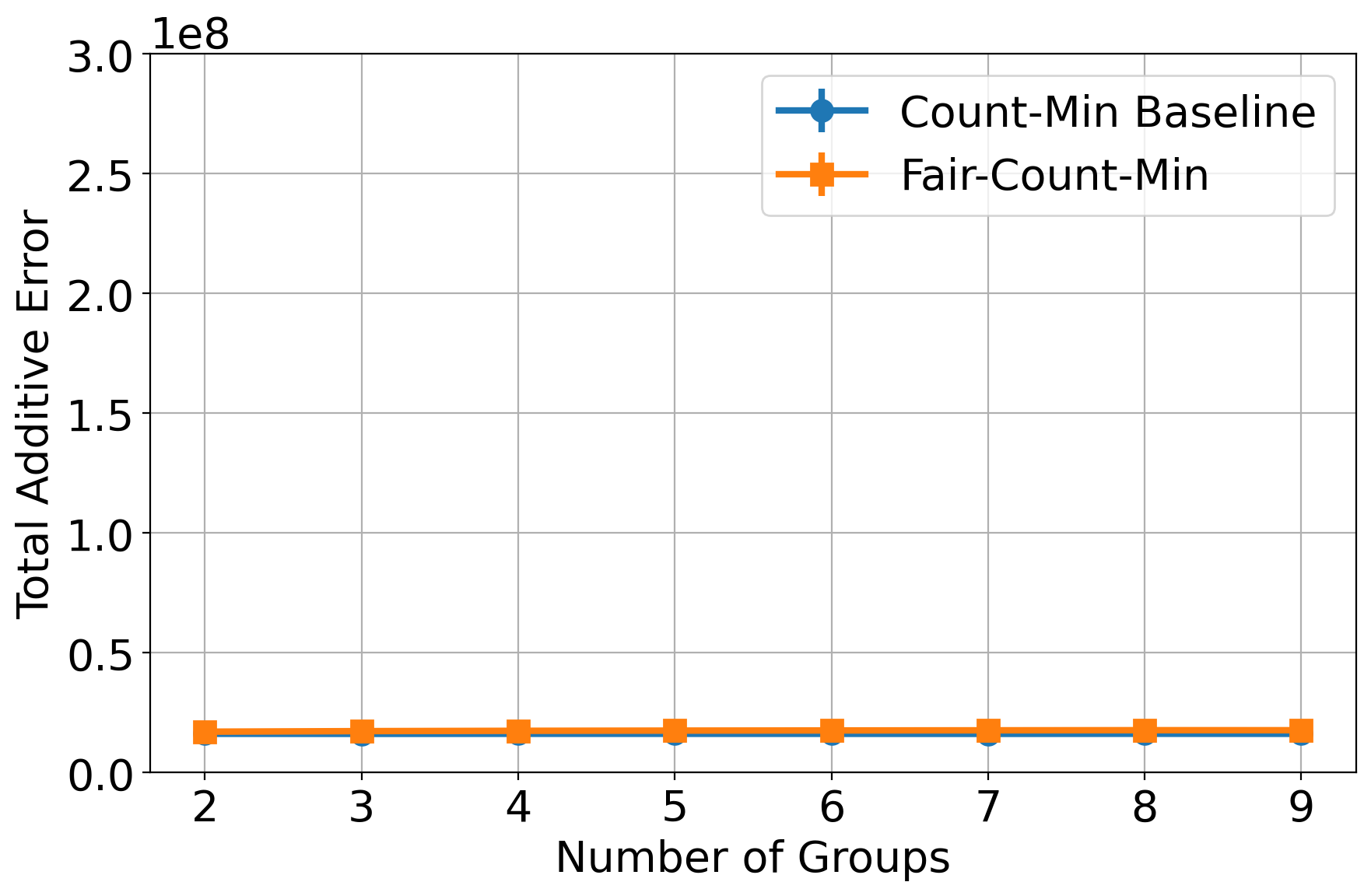}
        \vspace{-2.5em}
        \caption{effect of varying number of groups $\ell$ on price of fairness, \synthetic, $n=20000, w=512, d=10$.}
        \label{fig:}
    \end{minipage}
    \hfill
    \begin{minipage}[t]{0.32\linewidth}
        \centering
        \includegraphics[width=\textwidth]{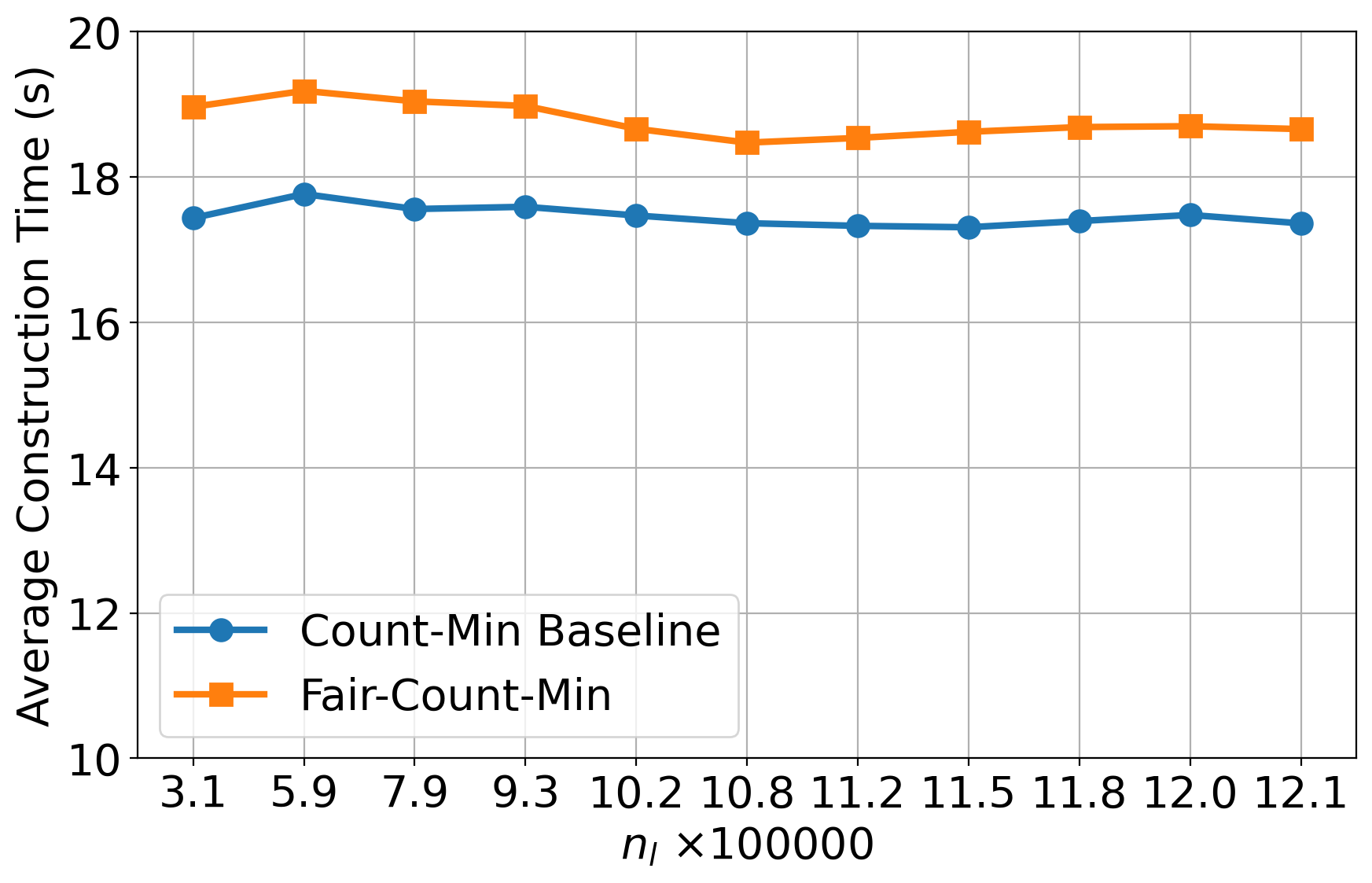}
        \vspace{-2.5em}
        \caption{effect of varying group size $n_l$ on construction time, \google, $w=65536, d=5$.}
        \label{fig:}
    \end{minipage}
    \hfill
    \begin{minipage}[t]{0.32\linewidth}
        \centering
        \includegraphics[width=\textwidth]{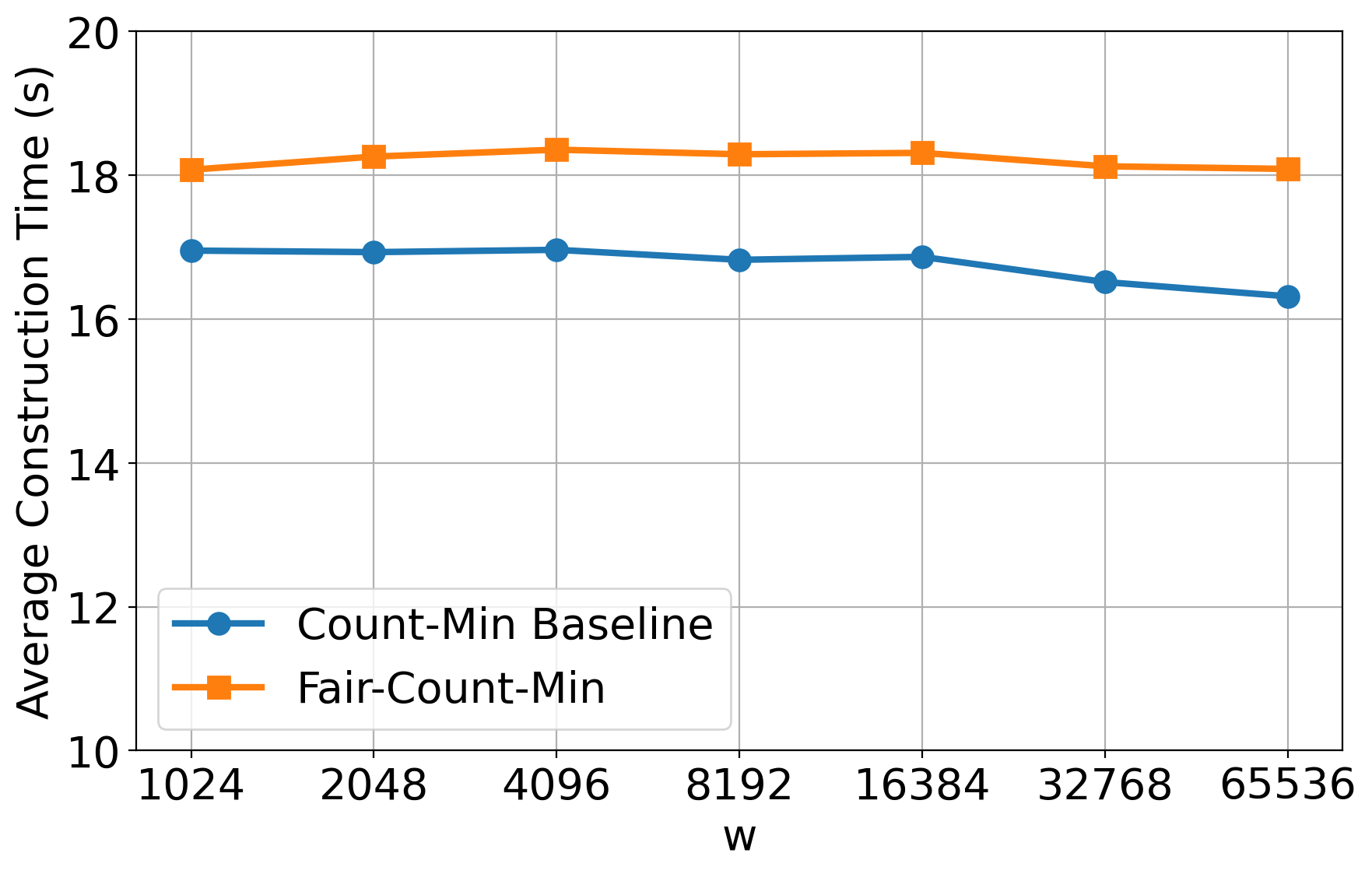}
        \vspace{-2.5em}
        \caption{effect of varying sketch width $w$ on construction time, \google, $n=1.2M, d=5$.}
        \label{fig:}
    \end{minipage}
    \hfill
    \begin{minipage}[t]{0.32\linewidth}
        \centering
        \includegraphics[width=\textwidth]{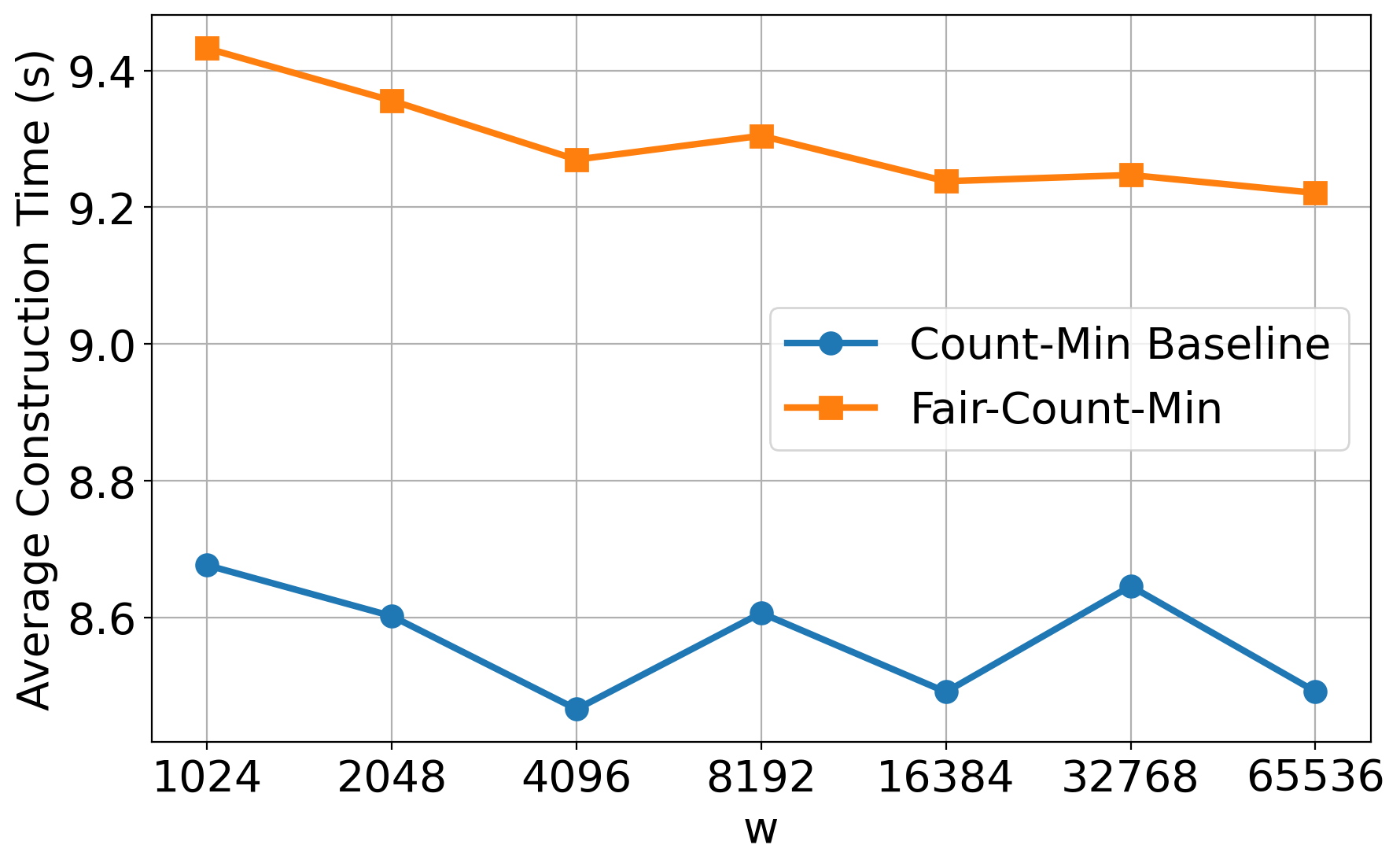}
        \vspace{-2.5em}
        \caption{effect of varying sketch width $w$ on construction time, \census, $n=430, d=5$.}
        \label{fig:}
    \end{minipage}
    \hfill
    \begin{minipage}[t]{0.32\linewidth}
        \centering
        \includegraphics[width=\textwidth]{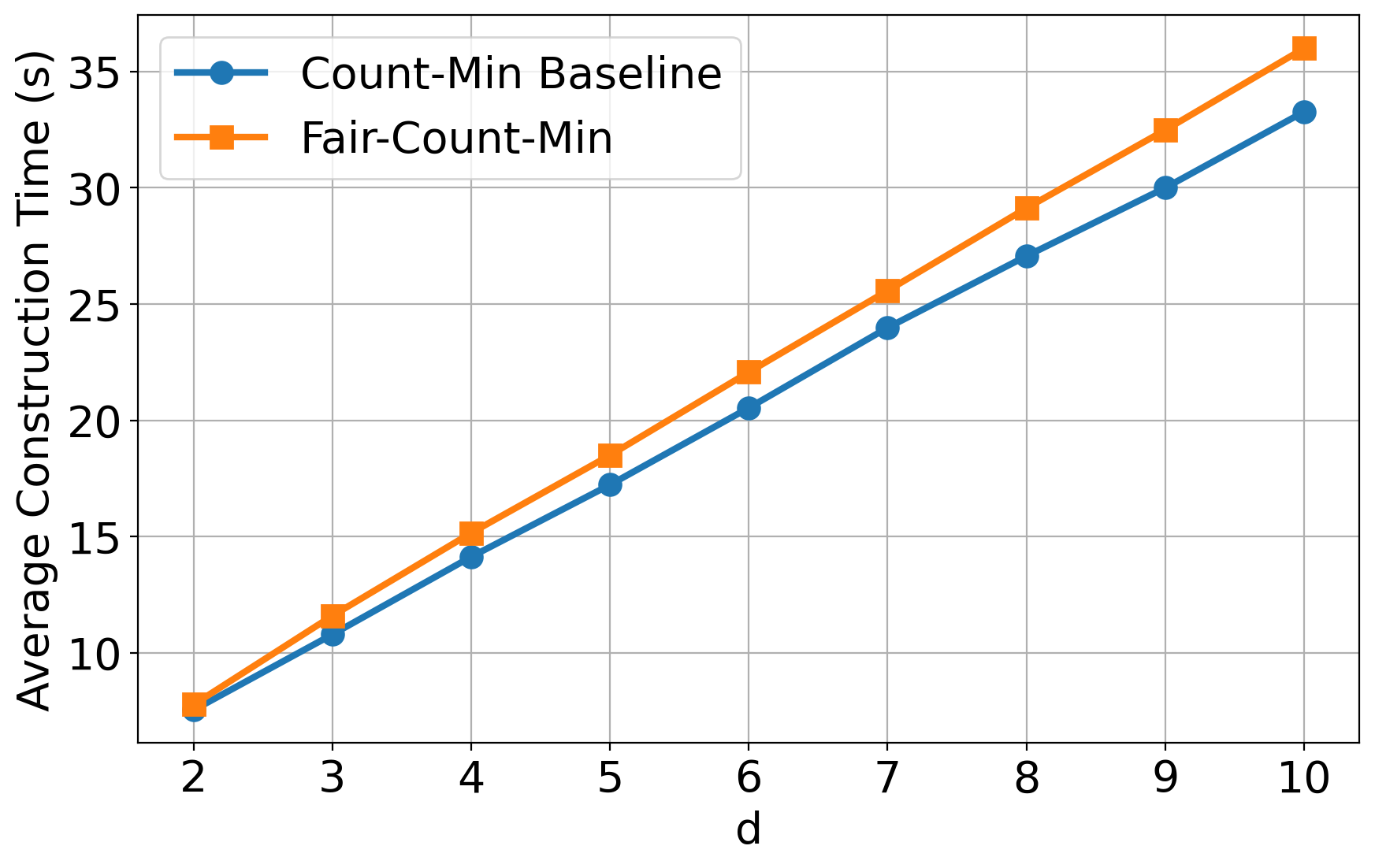}
        \vspace{-2.5em}
        \caption{effect of varying sketch depth $d$ on construction time, \google, $n=1.2M, w=65536$.}
        \label{fig:}
    \end{minipage}
    \hfill
    \begin{minipage}[t]{0.32\linewidth}
        \centering
        \includegraphics[width=\textwidth]{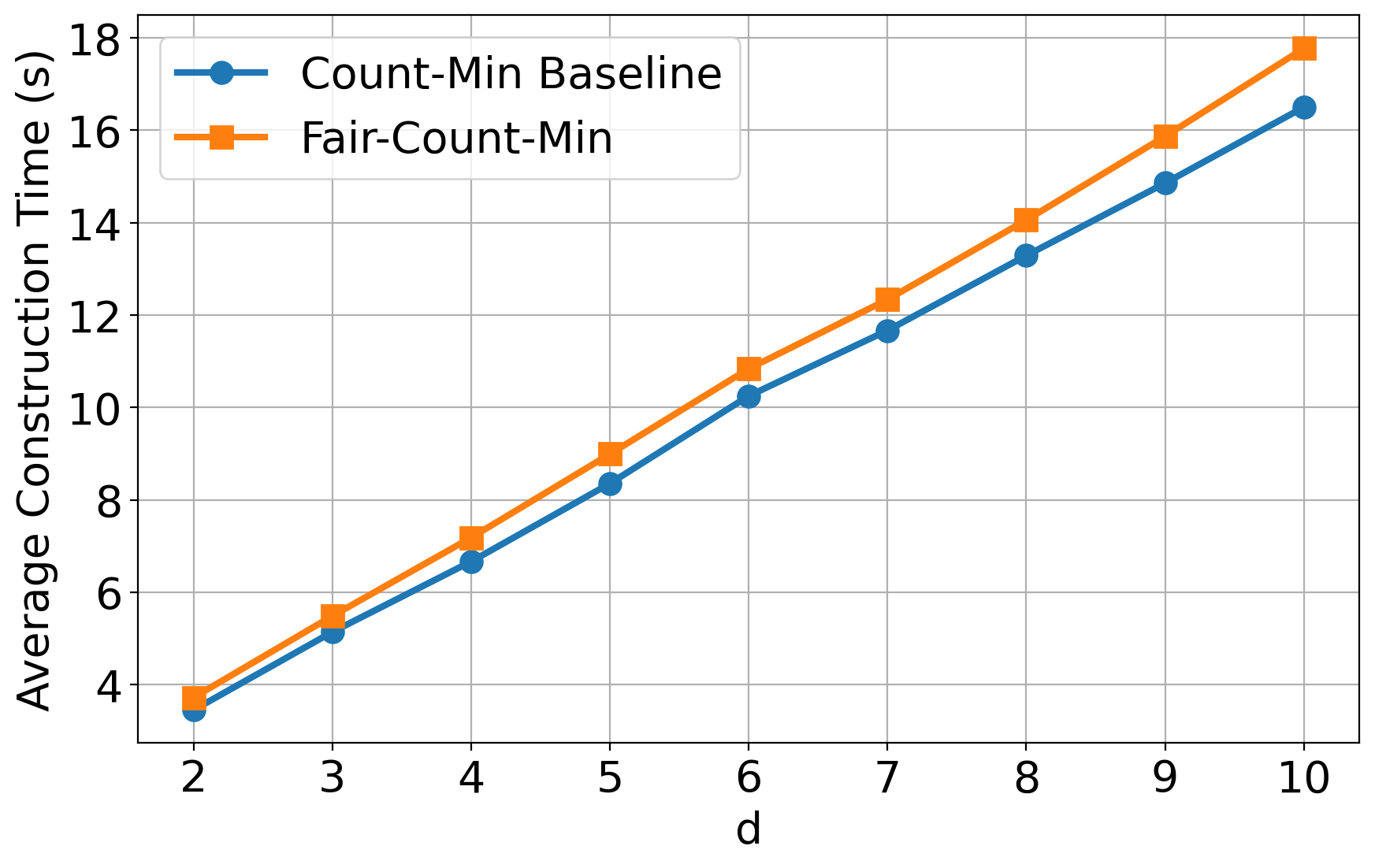}
        \vspace{-2.5em}
        \caption{effect of varying sketch depth $d$ on construction time, \census, $n=430, w=64$.}
        \label{fig:}
    \end{minipage}
    \hfill
    \begin{minipage}[t]{0.32\linewidth}
        \centering
        \includegraphics[width=\textwidth]{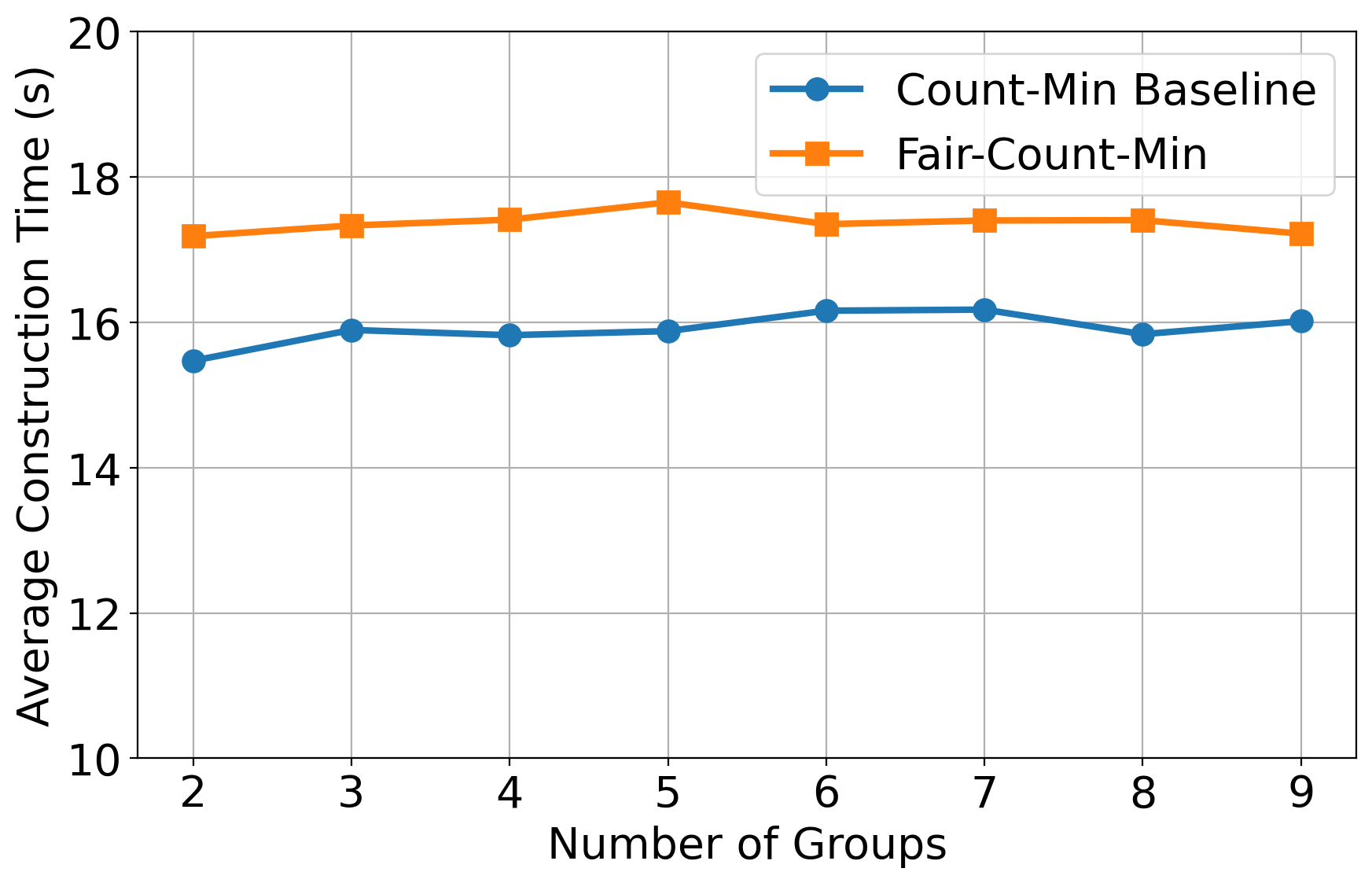}
        \vspace{-2.5em}
        \caption{effect of varying number of groups $\ell$ on construction time, \google, $n=1.2M, w=65536, d=5$.}
        \label{fig:}
    \end{minipage}
    \hfill
    \begin{minipage}[t]{0.32\linewidth}
        \centering
        \includegraphics[width=\textwidth]{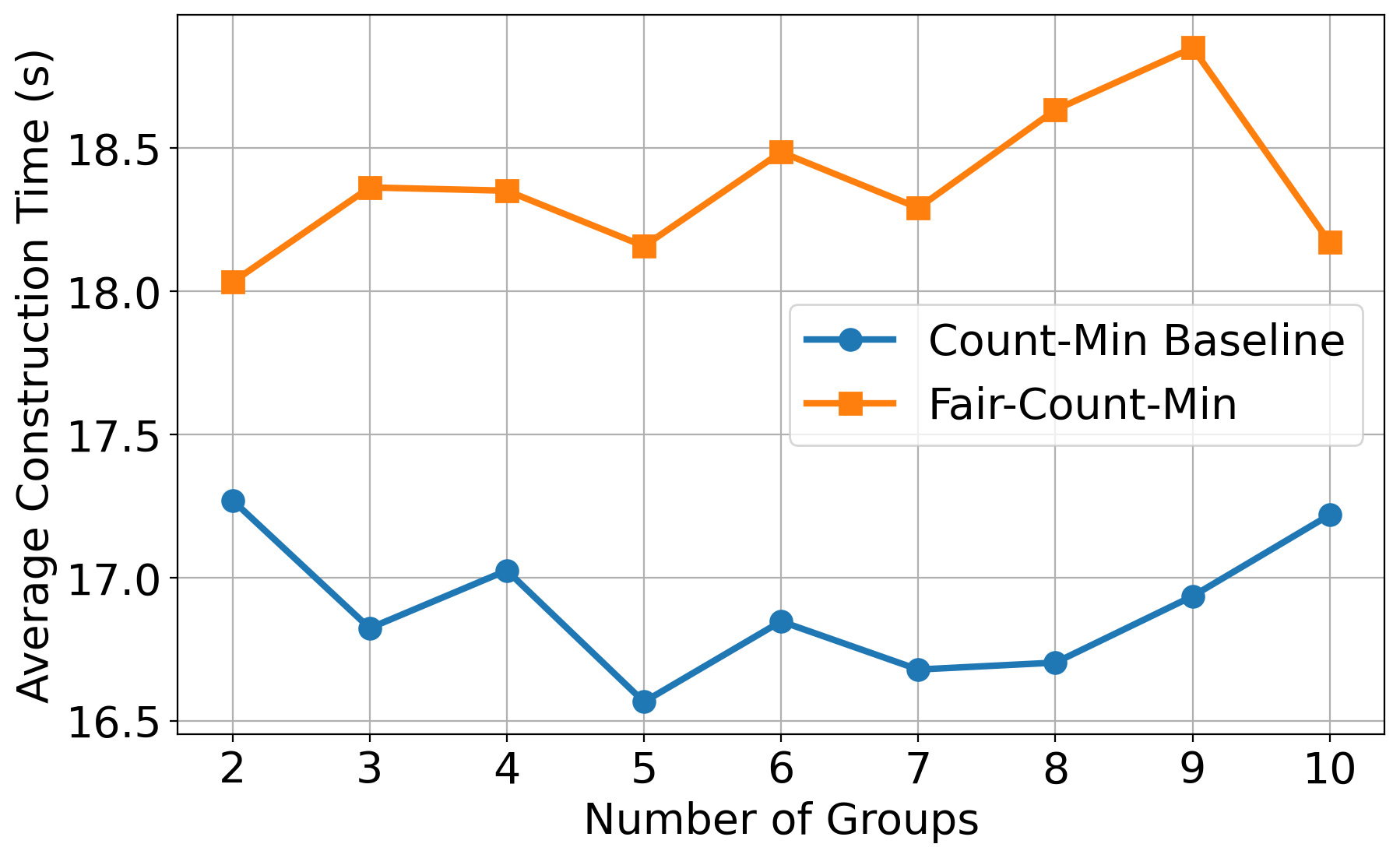}
        \vspace{-2.5em}
        \caption{effect of varying number of groups $\ell$ on construction time, \census, $n=430, w=64, d=5$.}
        \label{fig:}
    \end{minipage}
    \hfill
    \begin{minipage}[t]{0.32\linewidth}
        \centering
        \includegraphics[width=\textwidth]{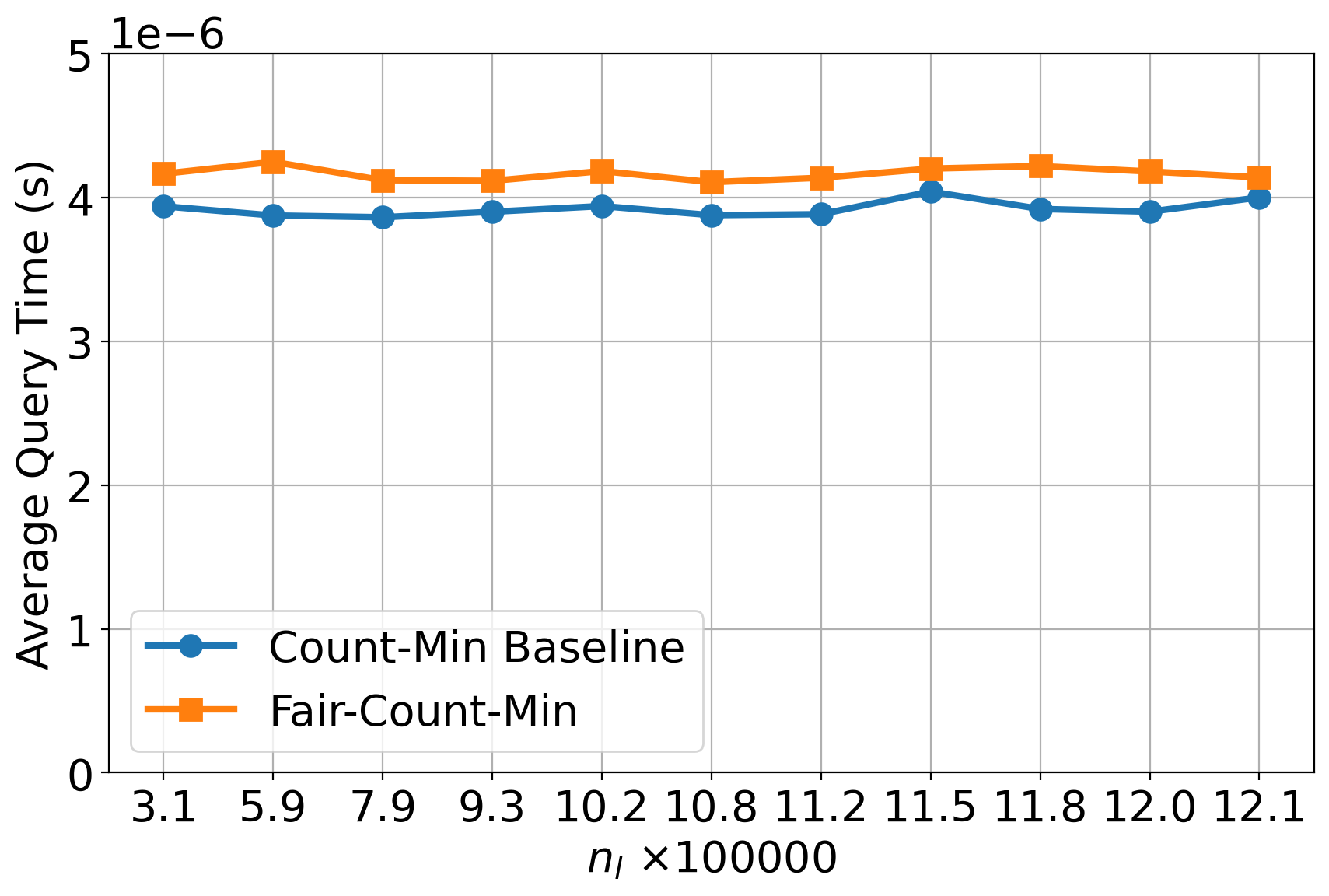}
        \vspace{-2.5em}
        \caption{effect of varying group size $n_l$ on query time, \google, $w=65536, d=5$.}
        \label{fig:}
    \end{minipage}
    \hfill
    \begin{minipage}[t]{0.32\linewidth}
        \centering
        \includegraphics[width=\textwidth]{plots_appendix/google_books/construction_time_plot_varying_w.png}
        \vspace{-2.5em}
        \caption{effect of varying sketch width $w$ on query time, \google, $n=1.2M, d=5$.}
        \label{fig:}
    \end{minipage}
    \hfill
    \begin{minipage}[t]{0.32\linewidth}
        \centering
        \includegraphics[width=\textwidth]{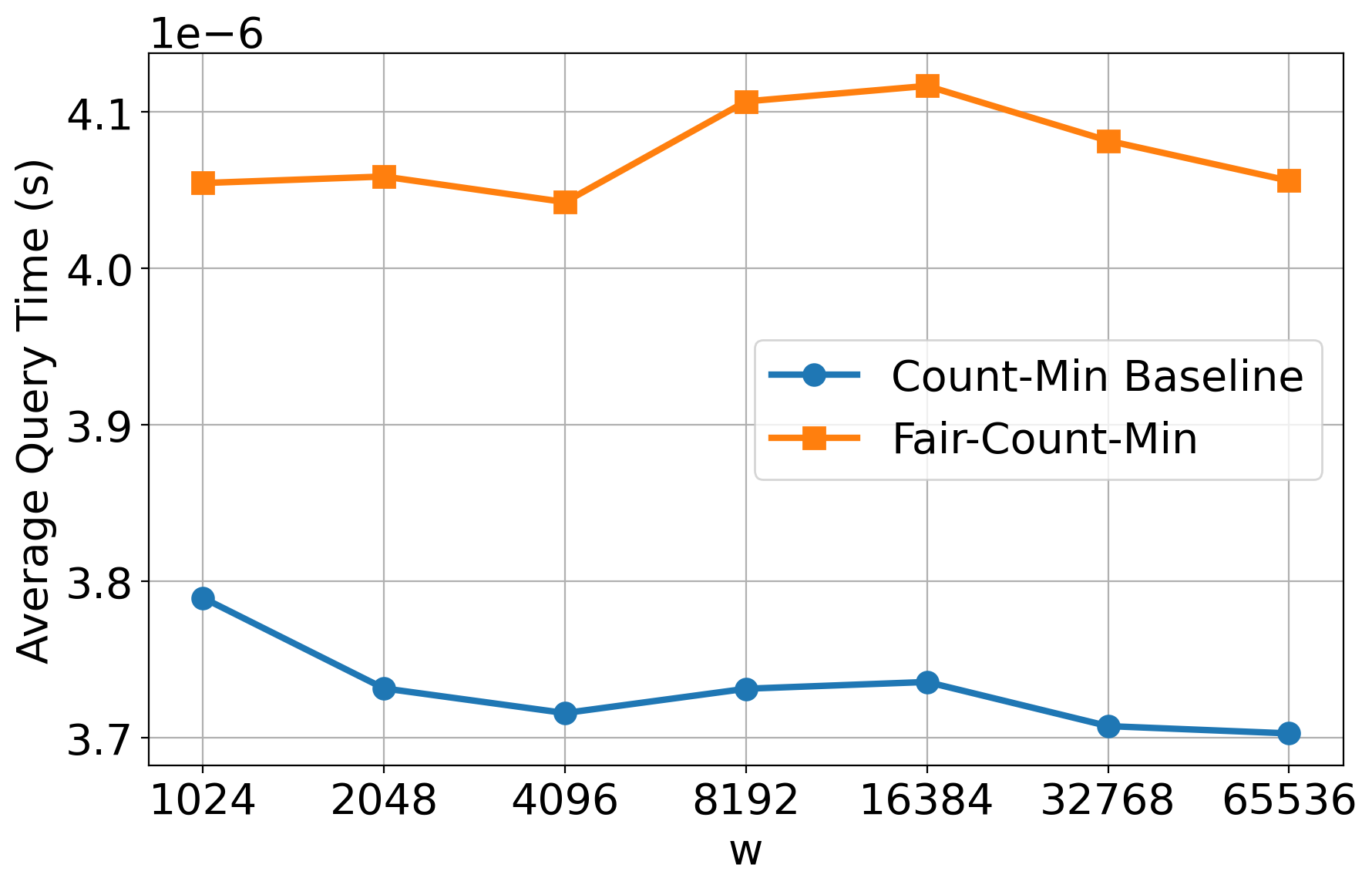}
        \vspace{-2.5em}
        \caption{effect of varying sketch width $w$ on query time, \census, $n=430, d=5$.}
        \label{fig:}
    \end{minipage}
\end{figure*}
\begin{figure*}
    \begin{minipage}[t]{0.32\linewidth}
        \centering
        \includegraphics[width=\textwidth]{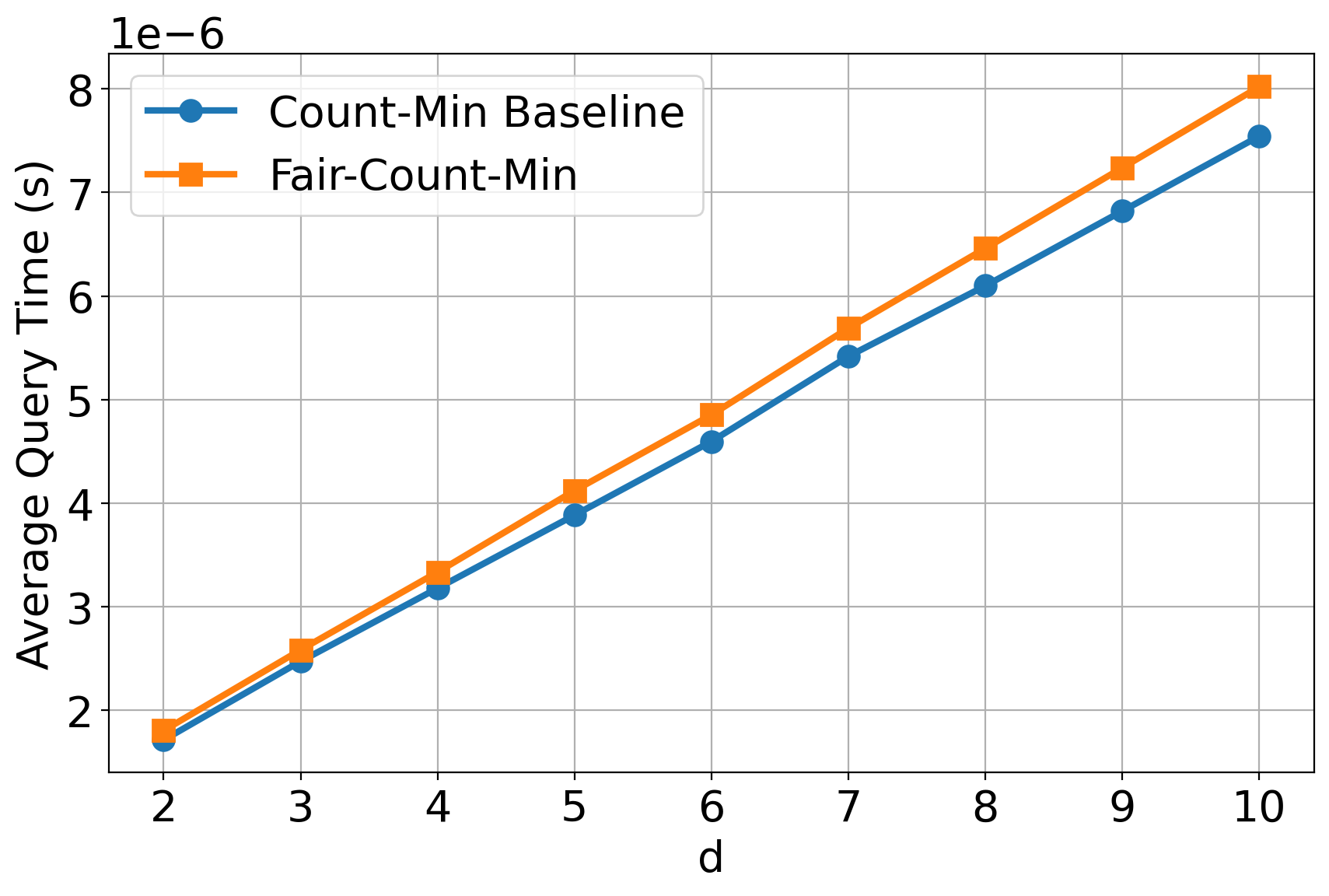}
        \vspace{-2.5em}
        \caption{effect of varying sketch depth $d$ on query time, \google, $n=1.2M, w=65536$.}
        \label{fig:}
    \end{minipage}
    \hfill
    \begin{minipage}[t]{0.32\linewidth}
        \centering
        \includegraphics[width=\textwidth]{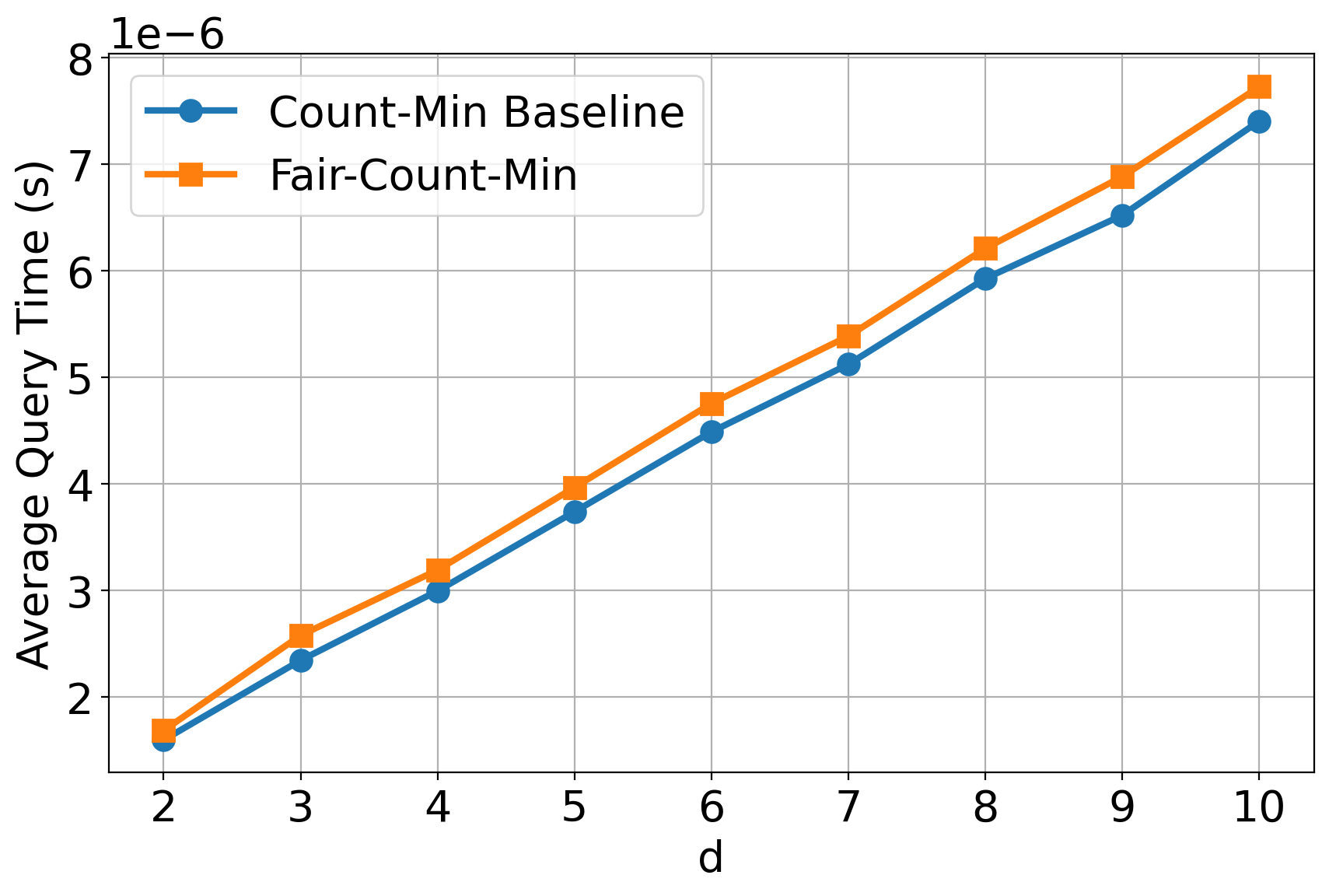}
        \vspace{-2.5em}
        \caption{effect of varying sketch depth $d$ on query time, \census, $n=430, w=64$.}
        \label{fig:}
    \end{minipage}
    \hfill
    \begin{minipage}[t]{0.32\linewidth}
        \centering
        \includegraphics[width=\textwidth]{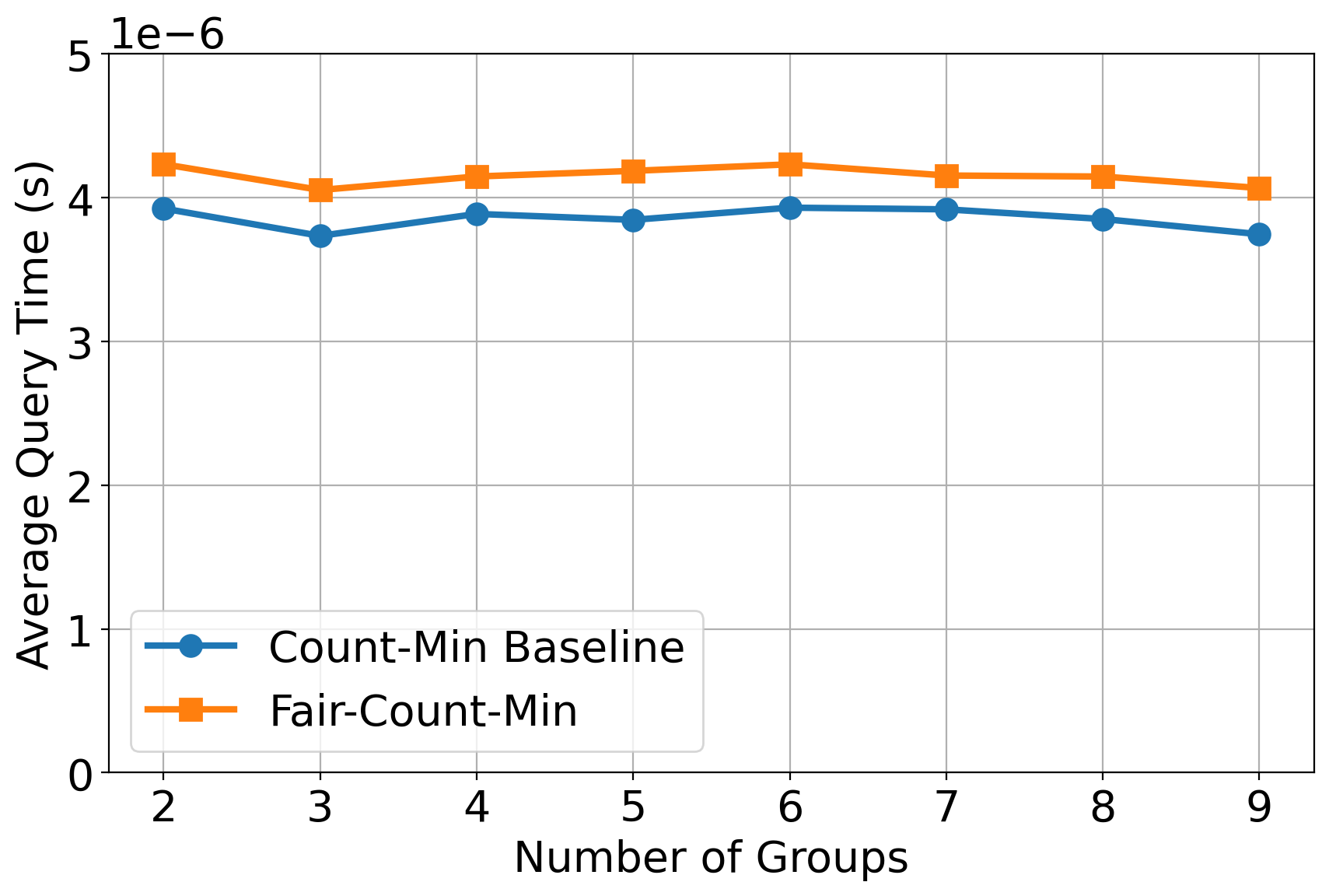}
        \vspace{-2.5em}
        \caption{effect of varying number of groups $\ell$ on query time, \google, $n=1.2M, w=65536, d=5$.}
        \label{fig:}
    \end{minipage}
    \hfill
    \begin{minipage}[t]{0.32\linewidth}
        \centering
        \includegraphics[width=\textwidth]{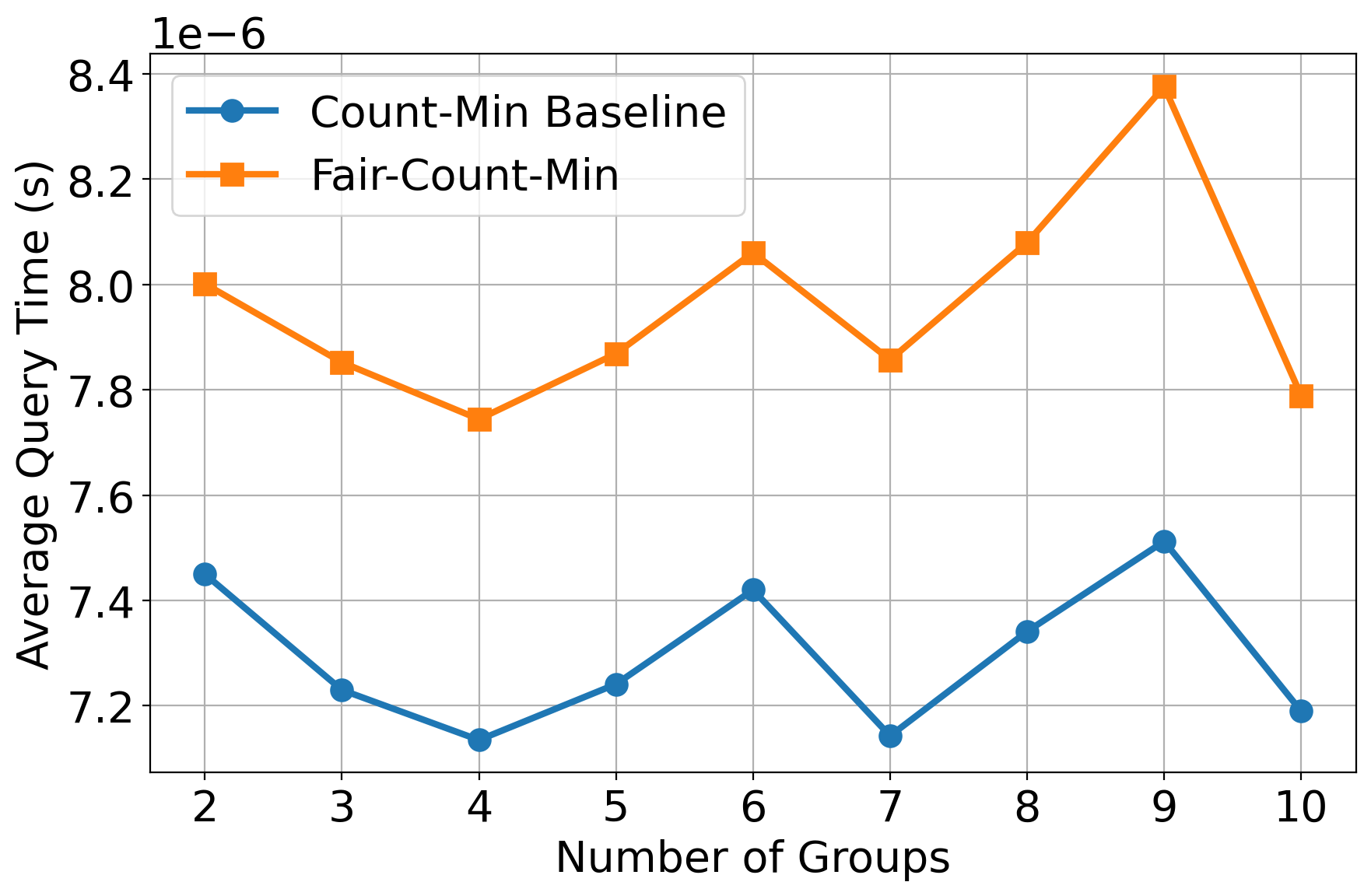}
        \vspace{-2.5em}
        \caption{effect of varying number of groups $\ell$ on query time, \census, $n=430, w=64, d=5$.}
        \label{fig:}
    \end{minipage}
\end{figure*}